\documentclass[sigconf]{acmart}
\acmConference[arXiv '26]{arXiv}{September}{2026}
\renewcommand{\footnotetextcopyrightpermission}[1]{}
\AtBeginDocument{
  }

\usepackage{listings}
\usepackage{longtable}
\usepackage{needspace}
\usepackage{xspace}

\newcommand{\projectName}{\mbox{Generative Tutorial}}
\newcommand{\etal}{et~al.\xspace}
\newcommand{\eg}{e.g.,\xspace}
\newcommand{\ie}{i.\,e.,\xspace}

\providecommand{\studycondition}[1]{\textsc{#1}}

\definecolor{conditionSystem}{HTML}{B35400}
\definecolor{conditionBaseline}{HTML}{0072B2}
\newcommand{\condPreauthored}{\textcolor{conditionBaseline}{\studycondition{Pre-authored}}\xspace}
\newcommand{\condGenerated}{\textcolor{conditionSystem}{\studycondition{Generated}}\xspace}

\newcommand{\statsum}[3]{{$M=#1\,#3$, $SD=#2\,#3$}}

\usepackage{xcolor}
\usepackage{array}
\usepackage{soul}

\begin{document}

\title{\projectName: Towards Live Contextualized Visual Instructions for Physical Tasks}

\settopmatter{
  authorsperrow=4,
  printacmref=false,
  printccs=false,
  printfolios=true
}

\author{Muzhe Wu}
\authornote{Both authors contributed equally to this research.}
\email{henrw@umich.edu}
\affiliation{%
  \institution{University of Michigan}
  \city{Ann Arbor}
  \state{MI}
  \country{USA}
}

\author{Zuchen Li}
\authornotemark[1]
\email{zuchenli@umich.edu}
\affiliation{%
  \institution{University of Michigan}
  \city{Ann Arbor}
  \state{MI}
  \country{USA}
}

\author{Xu Wang}
\email{xwanghci@umich.edu}
\affiliation{%
  \institution{University of Michigan}
  \city{Ann Arbor}
  \state{MI}
  \country{USA}
}

\author{Anhong Guo}
\email{anhong@umich.edu}
\affiliation{%
  \institution{University of Michigan}
  \city{Ann Arbor}
  \state{MI}
  \country{USA}
}

\begin{abstract}
Visual instructions for physical tasks are typically authored in one context and followed in another, requiring users to translate demonstrated tools, materials, and spatial relationships into their own environment.
We introduce Generative Tutorial, a conceptual framework for live visual instruction that depicts intended outcomes and actions within the user's environment and task flow.
A formative evaluation of state-of-the-art image and video generation identifies failures and potential benefits across 15 physical tasks.
Drawing on these findings, we build an augmented-reality prototype system that proactively generates goal images and demonstration videos using observed workspace context and predicted visual outcomes of preceding actions.
A 24-participant lab study found higher task performance quality, greater perceived workspace correspondence, and shorter step-confirmation intervals with the system than with pre-authored guidance.
Qualitative findings highlighted how contextual resemblance shapes trust, how generation errors affect interpretation, and how guidance delivery should adapt to users’ needs, informing future designs.

\end{abstract}



\begin{teaserfigure}
  \includegraphics[width=\textwidth]{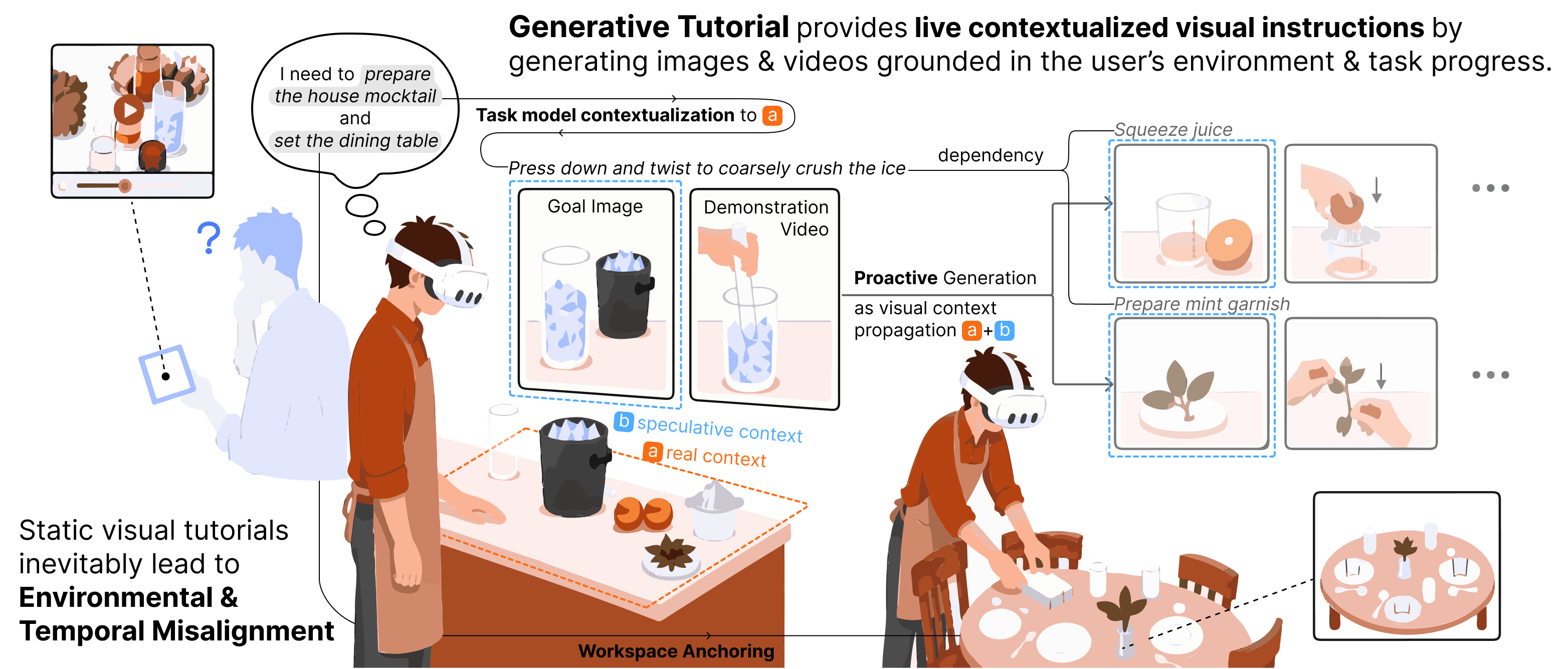}
  \caption{\textbf{Generative Tutorial reimagines tutorials as live, contextualized visual instructions adapted to a user’s environment and evolving task progress.}
  The prototype system continuously captures real-world context and grounds the task model in the user’s environment.
  For each step, it aims to generate a workspace-grounded goal image and demonstration video (\eg\ showing how to press and twist a muddler to crush ice).
  By propagating real and speculative visual context forward, the system proactively generates guidance for upcoming steps (\eg\ squeezing juice and preparing a mint garnish).
  The generated guidance is presented and situated in its associated task workspace, allowing instructions to coexist across activities and remain available as the user switches between them (\eg\ moving from preparing a house mocktail to setting the dining table).}
  \Description{
  Composite overview of Generative Tutorial's contextualized visual instruction capabilities and system workflow. The left section shows the environmental and temporal misalignment inherent in static visual tutorials, paired with an illustration of a user struggling to map a floating 2D video to their physical environment. The center section shows task model contextualization for a user wearing a mixed-reality headset at a kitchen counter, utilizing real and speculative context to dynamically generate workspace-grounded goal images and demonstration videos for their immediate task. The right section shows proactive generation for future dependent steps, such as squeezing juice and preparing a mint garnish, alongside workspace distribution scenarios where generated instructions remain spatially situated as the user switches activities to set the dining table.
  }
  \label{fig:teaser}
\end{teaserfigure}

\maketitle

\section{Introduction}
Visual instructions---ranging from online (\eg\ YouTube) tutorial videos to diagrams and illustrated guides---are a primary means for guiding physical tasks\footnote{In this work, we define ``physical tasks'' broadly as activities involving bodily movement to achieve physical, rather than digital, goals. We use ``instruction'' and ``guidance'' interchangeably to refer to information intended to directly support task performance, as distinct from information for training/learning.},
especially when task execution relies on spatial relationships, material states, or tacit techniques~\cite{smith2018many,mayer2024past}.
Despite their ubiquity, these media are mostly pre-authored within fixed contexts.
Users must mentally translate generic demonstrations to accommodate variations in their own tools, viewpoints, and workspace layouts~\cite{li2026substantial}, while constantly realigning them to their actual task flow~\cite{sweller1988cognitive,hegarty2004mechanical}.
Prior research in HCI has explored augmented reality (AR) interfaces to reduce the burden of spatial mapping by registering visual cues directly onto the physical environment~\cite{kong2021tutoriallens,liuInstruMentARAutoGenerationAugmented2023,zhaoGuidedRealityGenerating2025}.
Beyond instructional efficacy, a persistent challenge for these systems is the inherent trade-off between authoring effort and visual fidelity: highly realistic overlays demand intensive manual instrumentation, resulting guidance typically defaults to abstract, low-fidelity visual assets.

Recent advances in visual generative models, such as image and video generation models~\cite{rombach2022high,blattmann2023align}, have enabled high-fidelity visual synthesis at ever-decreasing latency.
This capability raises a new possibility for physical task guidance:
conditioned on a live view and task description, these models may be prompted to depict how a step should be performed with the user's own objects and scene (\eg\ what a fabric would look like aligned with a seam, where a soldering iron should contact a circuit pad, or how a mechanical gear should seat in its housing).
In this work, we conceptualize this paradigm of on-the-fly, environment-aware visual guidance as \textbf{Generative Tutorial} (see Figure~\ref{fig:teaser}).

Yet realizing this vision presents two substantial challenges.
First, the ability to generate realistic media does not inherently guarantee effective instruction.
While current models can produce visually plausible output, they are prone to errors in object identity, target configuration, action execution, or temporal continuity~\cite{souvcek2025showhowto};
how these errors manifest and the extent to which they may degrade instructional utility remain unclear.
Furthermore, steering visual generative models for \emph{live} physical task guidance raises both design and technical challenges.
Rather than simply generating isolated images or videos,
effective live guidance requires identifying and capturing task-relevant context, scheduling generation to accommodate changing contexts and substantial model latency, and presenting generated output that maintains clear spatial correspondence with its physical referents without inducing excessive perceptual load.

To answer the first question, we conduct a formative evaluation of state-of-the-art off-the-shelf image and video generation models, qualitatively analyzing 176 artifacts generated for 45 sampled unit actions\footnote{We use ``unit action'' in this work to refer to a bounded segment of a task organized around one immediate physical objective.} across 15 physical tasks.
We identify 27 observable issue codes spanning action interpretation, target-state accuracy, contextual coherence, and presentation, with at least one issue in 67 of 86 images and 82 of 90 videos.
These recurring failures motivate prioritizing low-consequence, reversible actions with visually verifiable outcomes and validating both the requested change and its source context.
Failures involving obscured details or unclear motion further motivate complementary goal images and demonstration videos for inspecting task-relevant details.

We then instantiate \projectName\ in an AR prototype system.
The system first contextualizes high-level task intents from live observations into unit actions assigned to registered workspaces.
To prepare guidance ahead of the user's actions, it propagates generated goal states along action dependencies, using observed or speculative visual context to generate goal images and demonstration videos.
The system presents available media side-by-side anchored to the physical workspace, allowing users to compare intended outcomes and demonstrated motion with their own actions and explicitly confirm step completion.
In a counterbalanced lab study with 24 participants across four everyday tasks, the prototype system demonstrated improved task quality, higher perceived workspace correspondence, and shortened step-confirmation intervals compared with a pre-authored guidance baseline.
Qualitative findings highlighted how local resemblance shaped interpretation and trust, generated inconsistencies required judgment, and complementary media, pacing, and controls should personalize to users' needs, pace, and physical actions.

In summary, this work contributes:
\begin{itemize}
    \item \projectName{}, a conceptual framework that connects environmental observations, action grounding, and visual generation to support live instructions aligned with users' physical surroundings and task progress.
    \item A formative evaluation of state-of-the-art visual generative models on context-conditioned unit-action generation across 15 physical tasks, identifying recurring failures and design implications for generated visual guidance.
    \item An AR prototype system that proactively generates goal images and demonstration videos from observed and speculative visual context and presents them beside the physical workspace.
    \item Findings from a comparative study with 24 participants across four everyday tasks, showing the prototype system's benefits for task quality and workspace correspondence, and revealing how contextual resemblance, generation errors, and media timing shape users' interpretation of guidance.
\end{itemize}

\section{Related Work}
Our work builds on context-adaptive physical task guidance, visual instructions and their creation process, and recent HCI literature on steering visual generative models.

\subsection{Context-Adaptive Physical Task Guidance}
Physical tasks involve rich contextual factors that lead to task difficulty~\cite{liu2012task}.
We categorize how prior systems adapt guidance to accommodate these factors along three dimensions: the \emph{environment}, \emph{task}, and \emph{user}.

Environment-adaptive systems register instructions to local objects and configurations.
A common strategy is to spatialize guidance by embedding prerecorded instructional steps with respect to specific elements in the workspace, such as kitchen items~\cite{chenPaperToPlaceTransformingInstruction2023}, circuit boards~\cite{chatterjeeARDWAugmentedReality2022}, and origami paper~\cite{chen2025origamisensei}, allowing users to consume instructions in place.
For example, CARING-AI~\cite{shiCARINGAIAuthoringContextaware2025} introduced a pipeline that authors humanoid-avatar instructional animations from expert demonstrations and natural speech, then spatially and temporally blends them into learners' practice environment.

Task-adaptive systems align or reconfigure predefined procedures.
Prior work has segmented and aligned procedural content with users' progress, such as assembly-video sequencing and tracking~\cite{yamaguchiVideoAnnotatedAugmentedReality2020}, or repurposed it to support metacognitive activities such as self-exploration~\cite{wangEXplainMRGeneratingRealtime2025} and reflection~\cite{zhangFollowingUnderstandingInvestigating2025}.
The form of guidance would depend on task characteristics.
Activities emphasizing coordinated body movements often benefit from visualizations of kinetic-chain motion (\eg\ basketball~\cite{wengBridgingCoachingKnowledge2025}, table tennis~\cite{maAvaTTARTableTennis2024}, and welding~\cite{xu2026weldar}) or muscle activation intensity (\eg\ physical rehabilitation~\cite{zhuMuscleRehabImprovingUnsupervised2022},  massage~\cite{jiangDesigningLLMPoweredMultimodal2025}).
For procedural tasks, which often comprise interdependent actions, recent systems seek to anticipate progress and provide proactive assistance~\cite{arakawa2022prismtracker, arakawa2024prismqa, arakawa2025scaling}.
For example, PrISM-Observer~\cite{arakawaPrISMObserverInterventionAgent2024} intervenes when inferred progress suggests omissions or ordering errors, while Satori~\cite{li2025satori} forecasts task progression to provide multimodal guidance.

User-adaptive systems personalize guidance to accommodate individual capabilities and real-time performance.
Prior systems have tailored instructions to address users' errors~\cite{arakawaPrISMObserverInterventionAgent2024}, inferred knowledge mastery~\cite{ren2025rubikon,huang2021adaptutar}, posture~\cite{joFlowARHowDifferent2023}, and perceptual abilities~\cite{huhVid2CoachTransformingHowTo2025,ningAROMAMixedInitiativeAI2025,liuHumanUnifiedApproach2024,leeCookARAffordanceAugmentations2024}.
For example, Vid2Coach~\cite{huhVid2CoachTransformingHowTo2025} and AROMA~\cite{ningAROMAMixedInitiativeAI2025} enable blind and low-vision people to access cooking instructions through audio narration powered by vision-language model (VLM)~\cite{radford2021learning} agents.

Our work primarily focuses on adapting physical task guidance to the user's environment and task flow:
we employ visual generative models to produce high-fidelity instructional media that is visually grounded in the user's actual workspace and current action.
Like much prior research~\cite{chenPaperToPlaceTransformingInstruction2023,shiCARINGAIAuthoringContextaware2025,zhangFollowingUnderstandingInvestigating2025,maAvaTTARTableTennis2024}, we situate our exploration in augmented reality for its situated guidance capability.
While recent model-driven systems primarily generate language-based guidance (\eg\ audio narration~\cite{ningAROMAMixedInitiativeAI2025,huhVid2CoachTransformingHowTo2025}) or translate text into other modalities with predefined templates (\eg\ visual overlays~\cite{zhaoGuidedRealityGenerating2025}, haptics~\cite{ho2026generative}), we investigate the unique value of directly synthesizing rich visual instructions via model generation.

\subsection{Visual Instructions and Creation Process}
Cognitive psychology has long established the significance of visual instructions in reducing the cognitive effort to translate abstract descriptions into physical manipulations~\cite{wilson2002six}.
The representations of visual instructions vary in abstraction: symbolic overlays (\eg\ arrows, highlights) direct attention~\cite{ren2025rubikon,zhaoGuidedRealityGenerating2025}, diagrams suppress incidental details to emphasize structure~\cite{hegarty1993constructing}, and photographs preserve realistic physical configurations~\cite{skulmowski2022preference}.
Temporally, static images efficiently communicate discrete procedural states while video conveys continuous motion~\cite{pongnumkul2011pause}.

Traditionally, visual instruction creation has centered on capturing and parsing human expert demonstrations.
A large body of HCI research has sought to reduce this authoring burden by supporting ad hoc tracking of expert actions during demonstrations, such as interactions with key objects or task elements~\cite{kong2021tutoriallens,liuInstruMentARAutoGenerationAugmented2023,stoverTAGGARGeneralPurposeTask2024,gupta2012duplotrack}, hand movements~\cite{huang2021adaptutar}, or by applying AI-assisted annotation during post-hoc processing~\cite{fengVideo2ActionReducingHuman2023}.
More recent work has taken generative approaches.
Often powered by vision-language reasoning systems~\cite{zhongHelpVizAutomaticGeneration2021,jo2025generative} and simulation toolkits~\cite{gunturuAugmentedPhysicsCreating2024,chengVisMimicIntegratingMotion2025}, this line of work generates structured visual representations either from scratch or from inputs not readily consumable as instructions~\cite{chiAutomaticInstructionalVideo2021,yangVideoMixAggregatingHowTo2025}.

Following the latter paradigm, our work seeks to answer whether high-fidelity visual instructions can be meaningfully generated during live use with minimum expert intervention.
Alongside systems such as Guided Reality~\cite{zhaoGuidedRealityGenerating2025} which map live VLM outputs to abstract visual primitives, we investigate an alternative strategy by treating generative image and video models as direct engines for high-fidelity runtime instructional content.
Our empirical findings complement recent advances in computer vision~\cite{menon2024generating,souvcek2025showhowto} by highlighting interaction needs beyond visual generation, including accommodating changing task contexts, supporting users' judgment, and adapting representations to their needs, as discussed in Section~\ref{sec:discussion-effectiveness}.

\subsection{Steering Visual Generative Models in HCI}
Recent HCI research has explored a growing range of interactions with visual generation models.
Much of this work focuses on image generation to support open-ended creative workflows and prompt iteration~\cite{zhouStyleFactoryBetterStyle2024,gangulyShadowMagicDesigningHumanAI2024,rajaramBlendScapeEnablingEndUser2024}.
To improve controllability, these systems often introduce multimodal inputs like sketches~\cite{sarukkaiBlockDetailScaffolding2024} and region inpainting~\cite{dangWorldSmithIterativeExpressive2023,rajaramBlendScapeEnablingEndUser2024}.
Comparatively, video generation remains sparsely explored, with a narrower focus on concept exploration and evaluation~\cite{liVideoCraftMixedRealityempowered2025,wang2025boundary}.
Issues of high latency and persistent motion errors restrict current applications to asynchronous creative production rather than real-time utility.

Our work shifts attention from open-ended creative workflows to live physical-task guidance, where generated media must reflect the current physical scene, faithfully depict a bounded action, and arrive while its context remains relevant.
We orchestrate off-the-shelf image and video generation models and proactively schedule generation according to dependencies among task steps.
To improve physical grounding and task fidelity, we condition generation on visual evidence from the workspace, constrain prompts to the intended unit action, and apply additional visual processing to emphasize action-relevant changes.
Together, these techniques examine how visual generative models can function not only as creative tools, but also as timely and context-sensitive sources of practical instruction.

\section{Conceptual Framework for Live Contextualized Visual Instruction}
\label{sec:design-framework}

\begin{figure*}[t]
    \centering
    \includegraphics[width=\linewidth]{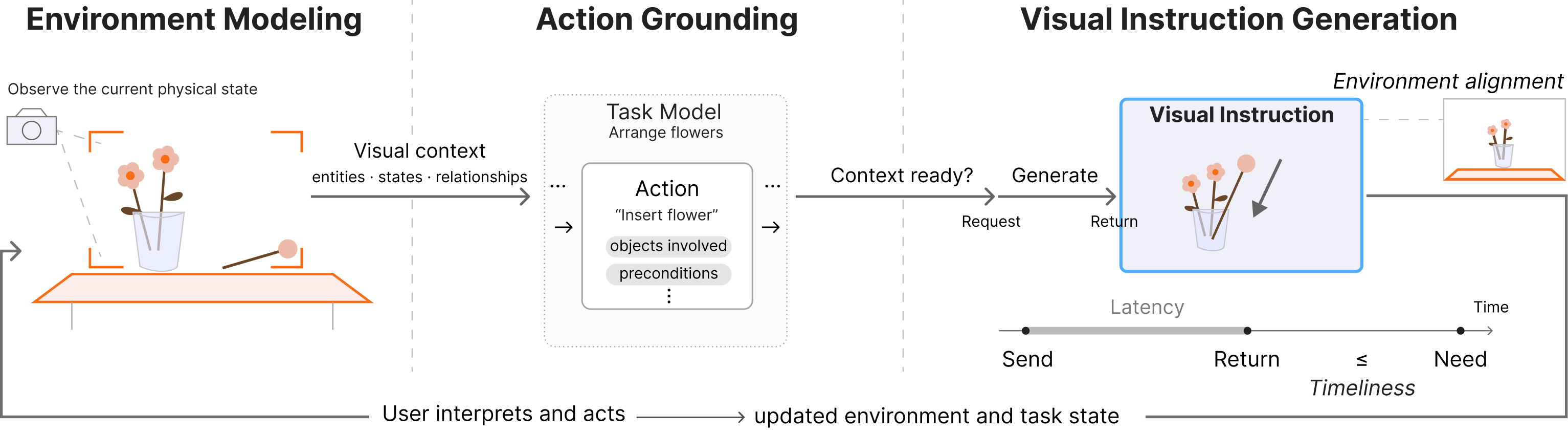}
    \caption{Conceptual framework for \projectName{}, illustrated through flower arranging.
    Environment Modeling represents observed entities, states, and relationships as visual context.
    Action Grounding relates an action in the task model, including its objects and preconditions, to the current scene.
    Once sufficient context is available, Visual Instruction Generation depicts how to act or what the action should achieve.
    Environment alignment connects the depiction to the physical scene, while timeliness concerns its availability at the time of need.
    User interpretation and action lead to changes in the environment and task state that inform subsequent guidance.}
    \Description{
    Composite overview of Generative Tutorial's conceptual framework, illustrated through a flower arranging task. The Environment Modeling section shows the system observing the current physical state to extract visual context, including entities, states, and relationships. The Action Grounding section shows a task model relating a specific action, such as inserting a flower, along with its involved objects and preconditions, to the current scene. The Visual Instruction Generation section shows the creation of a visual instruction once context is ready, paired with an environment alignment step that maps the depiction to the physical scene, and a timeline demonstrating that the system's latency must satisfy the user's timeliness need before the user interprets, acts, and updates the environment state.
    }
    \label{fig:conceptual-framework}
\end{figure*}

We propose \projectName{} as a form of physical task guidance that uses observations of the user's physical environment to generate visual guidance for actions as a task unfolds.
Figure~\ref{fig:conceptual-framework} organizes this process into three connected stages: environment modeling, action grounding, and visual instruction generation.
Together, these stages connect what is observed, what the user intends to do, and what the instruction should depict.

\textbf{Environment Modeling} represents the current physical setting through the entities, states, and relationships evident in observations.
This representation forms the visual context available to the system and may include objects and materials, bodily configurations, and spatial arrangements.
For example, a view of a flower arrangement provides evidence of the vase and stems (entities), which stems are already inserted (states), and their relative positions and the gaps between them (relationships).
The representation is updated as the user moves or acts~\cite{cho2023realityreplay}, but remains partial because a view may omit or obscure parts of the environment.

\textbf{Action Grounding} connects an action specified in a task model to the user's current physical situation.
The task model describes actions that contribute to the user's goal, including the objects involved and preconditions for performing them.
For example, an action in a flower-arranging task may be to insert a flower, involving a selected stem and a vase, with the precondition that the stem has been prepared for placement.
Grounding this action involves relating these requirements to the observed scene: which stem will be inserted, where it should go, and how the surrounding arrangement constrains its placement.
This action-specific context supports depicting the intended physical change, such as inserting the stem into a particular gap at a suitable height and angle.
\emph{Context readiness} concerns whether enough information is available to produce that depiction.
For instance, the stem may be prepared for insertion, but if the user's hand obscures the target gap, an additional observation may be needed before generation.

\textbf{Visual Instruction Generation} uses the action and its grounded context to produce a depiction of how to act or what the action should achieve.
Once sufficient context is available, a generation request can be issued.
For flower insertion, an image could show the intended placement, while a video could demonstrate the insertion direction and motion.
The resulting guidance must also relate to the user's situation in both space and time.
\emph{Environment alignment} concerns whether the user can connect the depiction to the physical scene~\cite{chenPaperToPlaceTransformingInstruction2023,shiCARINGAIAuthoringContextaware2025}.
Preserving the appearance and arrangement of the vase and existing stems, for example, can help the user locate the depicted target gap; presenting the instruction beside or within the scene can further support this correspondence.
\emph{Timeliness} concerns whether the instruction is available when the user needs it~\cite{ha2014wearable, chen2017latency, olguin2021delayed}.
The figure distinguishes the generation request, the returned result, and the time of need: generation latency separates the first two, while timely guidance requires the result to arrive by the third.
Generation must therefore be coordinated with both context availability and the unfolding task.

The user interprets the instruction and acts, changing the environment and potentially advancing the task.
Subsequent observations provide evidence of these changes and inform guidance for later actions, closing the loop.
Producing an instruction does not itself establish that the action has been completed; therefore, further visual evidence needs to be continuously collected to update the task state and inform subsequent visual guidance depiction.

\section{Formative Model Evaluation}
\label{sec:formative-evaluation}
The conceptual framework positions visual instruction contextualization as the process of turning an action and its environmental context into visual guidance.
Realizing this process requires understanding the ability of state-of-the-art visual generative models to depict contextualized action and their potential limits of use across tasks.
We therefore began by conducting a formative model evaluation analyzing generated images and videos across a curated corpus involving 15 task domains.
Specifically, we examine (1) whether generated images and videos correctly depict the requested change while keeping other task-relevant scene content intact, and (2) which failures recur across media, models, and task domains, and what they imply for where generated guidance is currently feasible.

\begin{figure*}[t]
    \centering
    \includegraphics[width=\textwidth]{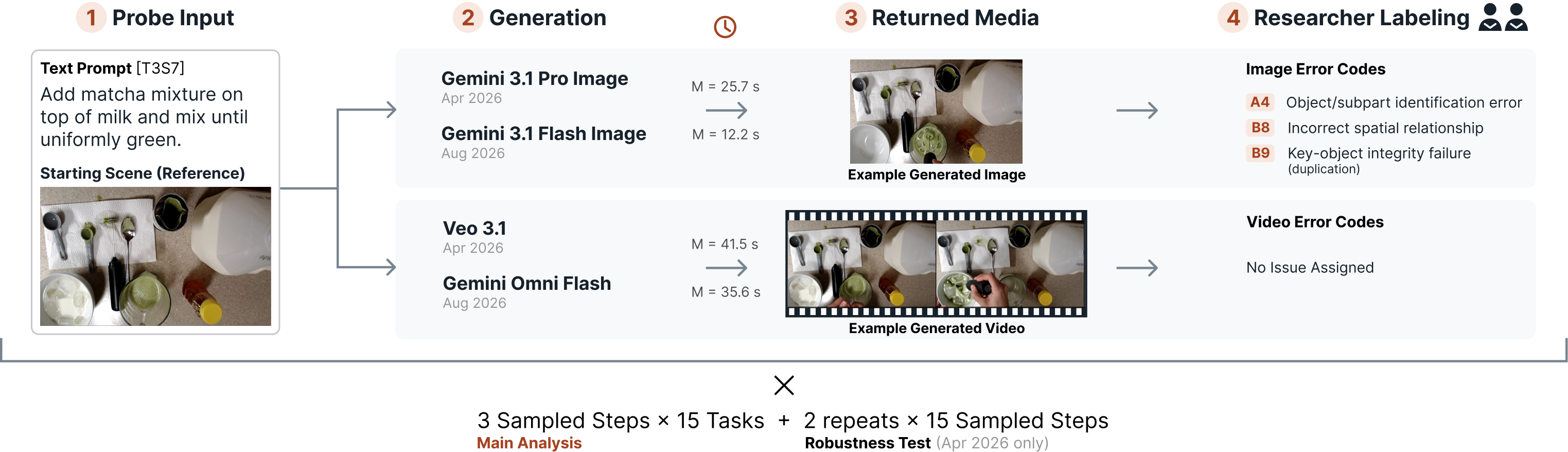}
    \caption{Formative model evaluation flow. A unit-action instruction and initial reference frame condition image and video generation across two model snapshots. Two researchers examine each returned artifact against both inputs and assign observable issue codes.
    In the illustrated example, the generated image uses the matcha glass rather than the milk bowl as the receiving vessel (A4: Object or subpart
identification error), placing the combined contents in the glass while leaving the milk bowl empty (B8: Incorrect spatial
relationship); an additional whisk appears alongside the original, indicating key-object duplication (B9: Key-object integrity
failure). The video receives no issue codes.}
    \Description{
    Composite overview of the formative model evaluation flow across four processing stages. Stage 1 shows the probe input, consisting of a unit-action text prompt for mixing matcha paired with an initial reference starting scene image. Stage 2 and Stage 3 show the generation and returned media phases, contrasting image models like Gemini 3.1 Pro and Flash against video models like Veo 3.1 and Gemini Omni Flash, complete with median generation latencies and representative output artifacts. Stage 4 shows the researcher labeling process, detailing specific visual error codes assigned to the generated image—such as object identification, spatial relationship, and duplication errors—while assigning no issues to the video, all scoped to an evaluation formula of sampled steps and tasks.
    }
    \label{fig:formative-evaluation-overview}
\end{figure*}

\subsection{Methodology}
\label{sec:formative-evaluation-procedure}

\subsubsection{Corpus}
We selected 15 everyday physical tasks for our formative exploration (see Appendix~\ref{sec:appendix-formative-dataset}).
While we did not intend this selection to be exhaustive, we purposefully included activities varying in motor involvement (\eg\ fine motor control: crochet, gross motor movement: yoga), temporal organization (\eg\ discrete: tennis, serial: LEGO assembly, continuous: jump rope), and object deformability (\eg\ rigid: LEGO pieces, deformable: origami paper, mixed: gift wrapping).
One author curated the corpus by selecting a YouTube tutorial for each task and decomposing its demonstrated procedure into unit actions.
Because the original videos varied in resolution, visual annotations, and camera perspective, the experimenter reenacted these actions in a local environment, recorded using a 1080p webcam, with a first-person view by default and a third-person view for physical exercise tasks to capture whole-body movement.
The corpus comprised 198 annotated steps, each with a unit-action instruction and a reference frame.
To balance task representation, we randomly sampled 3 unit actions per task, yielding 45 instances for the main analysis.

\subsubsection{Experiment Setup \& Model Selection}
For each sampled instance, we provided the initial reference frame, which establishes the physical context, and the unit-action instruction, which specifies the change to depict, as inputs to image and video generation.
This setup remained constant across models to examine recurring failure patterns.
In pilot testing, we experimented with additional setups such as prompt augmentation~\cite{hao2023optimizing}, set-of-mark prompting~\cite{yang2023set}, and semantic segmentation of the input frame~\cite{kirillov2023segment}, which show stochastic variation in output quality and behavior without establishing consistent benefits.
We therefore omitted these additional methods from the evaluation.

While pilot testing explored open-weight models (\eg\ FLUX.2 Klein, LTX-2.3, Wan-2.2) and other proprietary off-the-shelf models (\eg\ Seedance-2.0, Sora 2),
for the April 2026 evaluation, we selected \texttt{gemini-3-pro-image-preview} for images and \texttt{veo-3.1-\allowbreak{}generate-\allowbreak{}001} for videos based on their generation speed and perceived accuracy.
While our formative evaluation does not intend to comprehensively benchmark the visual generation model landscape,
in August 2026, we repeated the evaluation with the newer and lighter \texttt{gemini-3.1-\allowbreak{}flash-\allowbreak{}image} and \texttt{gemini-omni-\allowbreak{}flash-\allowbreak{}preview} models, featuring faster inference speeds for potentially more practical live use.

The main evaluation yielded 176 returned artifacts (86 images, 4 failed with no image output, and 90 videos). As a robustness test, we additionally generated two repeats for a subset of 15 instances (1 per task, both image and video).

\subsubsection{Analysis}
Two authors examined each artifact against its initial reference frame and unit-action instruction, assessing whether it depicted the requested change and preserved other task-relevant objects, states, and relationships.
The researchers iteratively developed a bottom-up codebook of 27 observable issues in four families:
(A) interpretation and grounding, (B) action and target-state accuracy, (C) contextual and physical coherence, and (D) presentation and temporal legibility (see Appendix~\ref{sec:appendix-formative-codebook}).
An artifact could receive multiple codes when issues co-occurred.
They further mapped the issues to 6 usefulness dimensions describing the aspect of instruction it could undermine:
(E1) whether the result makes the target state clear,
(E2) communicates the action sequence, (E3) preserves continuity with the source scene,
(E4) stays within the requested unit action,
(E5) exposes critical detail,
and (E6) provides useful intermediate states or trajectories.

Comparing August 2026 and April 2026 batches, we observed a decrease in generation time (from $M=25.7$\,s to $M=12.2$\,s for images and from $M=41.5$\,s to $M=35.6$\,s for videos;
however, we did not observe a corresponding reduction in assigned failure codes.
Across 45 instances, we assigned 49 issue codes to the original-model images and 69 to the updated-model images, as well as 88 to the original-model videos and 104 to the updated-model videos.
Robustness test results revealed persistent task-specific errors despite the variation across runs.
Figure~\ref{fig:formative-evaluation-overview} illustrates the overall workflow.

\subsection{Findings}
\label{sec:formative-probe-findings}

We report recurring failures in the requested action and its physical context, then consider their implications across media, models, and tasks.

\begin{figure*}[t]
    \centering
    \includegraphics[width=0.8\textwidth]{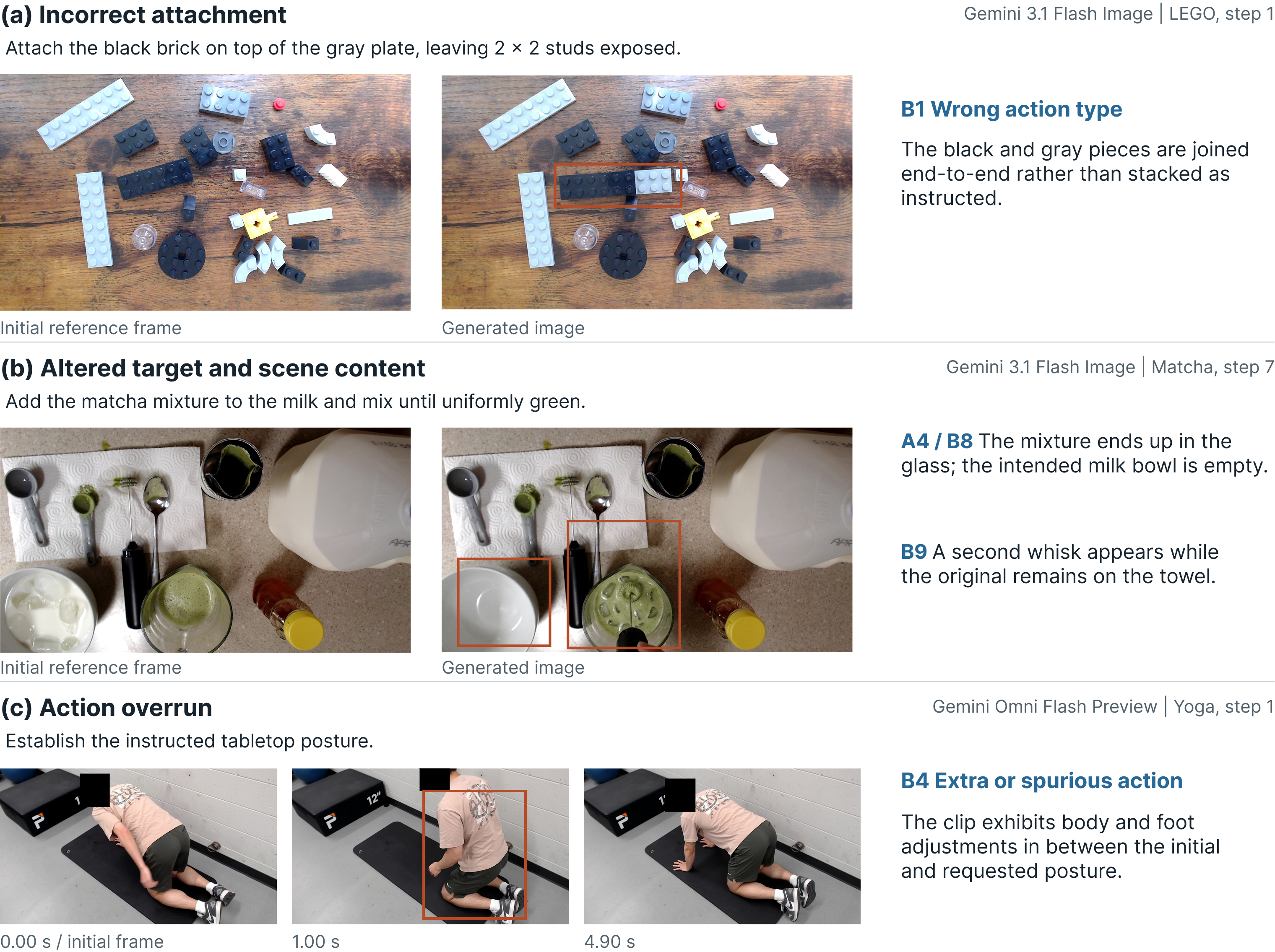}
    \caption{Representative failures in generated guidance. Each panel pairs a condensed instruction with its reference frame and generated output. (a) The black and gray LEGO pieces are joined end-to-end instead of stacked (B1). (b) The mixture appears in the wrong vessel (A4/B8), and a second whisk appears while the original remains (B9). (c) The yoga video adds body and foot adjustments beyond the requested posture (B4). Orange outlines highlight relevant image regions.}
    \Description{
    Composite overview of representative failures in generated visual guidance across different task models. Panel (a) shows an 'Incorrect attachment' failure in a LEGO assembly task, paired with reference and generated images demonstrating that pieces were joined end-to-end rather than stacked as instructed (B1). Panel (b) shows 'Altered target and scene content' during a matcha preparation task, highlighting that the mixture appeared in the wrong vessel (A4/B8) alongside an erroneously duplicated whisk (B9). Panel (c) shows an 'Action overrun' in a yoga video sequence, demonstrating extra body and foot adjustments that continue beyond the initially requested tabletop posture (B4).
    }
    \label{fig:formative-evaluation-examples}
\end{figure*}

\subsection{Task and Workspace Representation}
\label{sec:task-modeling}

\subsubsection{Action Fidelity and Context Preservation}
At least one observable issue occurred in 67 of 86 images and 82 of 90 videos.
A recognizable scene could still depict the wrong physical change.
In the LEGO example (Figure~\ref{fig:formative-evaluation-examples}a), the image preserves much of the surrounding arrangement but joins the pieces end-to-end rather than attaching one on top of the other (B1).
The distinction is small visually but detrimental to the instruction: the generated configuration cannot serve as an accurate target for the requested attachment.

We also observed failures that affected both the intended change and the source context.
In the matcha example (Figure~\ref{fig:formative-evaluation-examples}b), the model uses the wrong receiving vessel (A4: Object or subpart
identification error), places the combined contents in the glass while leaving the milk bowl empty (B8: Incorrect spatial
relationship), and duplicates the whisk (B9: Key-object integrity
failure).
We interpreted these co-occurring issues as threats to both target-state legibility and source-context continuity.
Together, the LEGO and matcha examples show why we assessed the requested transformation and the task-relevant properties that should remain unchanged.

For videos, we additionally examined whether the depicted motion remained within the requested action.
In the yoga example (Figure~\ref{fig:formative-evaluation-examples}c), we coded continued body and foot adjustments during the requested posture establishment as an extra or spurious action (B4: Extra or spurious action).
This failure highlights how unrequested movements can obscure where the instructed action ends and what the learner is expected to reproduce.

\begin{figure*}[tp]
    \centering
    \includegraphics[width=\textwidth]{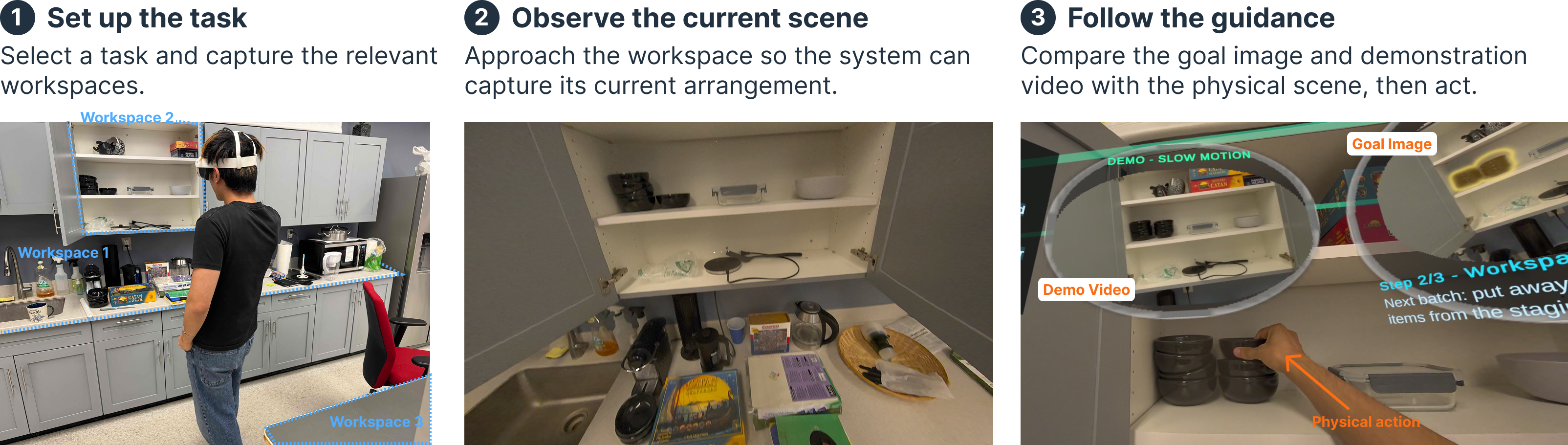}
    \caption{Prototype walkthrough illustrated through shelf organizing.
    (1) Set up the task: the user selects a task and captures the relevant workspaces.
    (2) Observe the current scene: as the user approaches the shelf, the headset captures its current arrangement to inform guidance generation.
    (3) Follow the guidance: the user compares the goal image and demonstration video with the physical scene and performs the action, here stacking dishes on the shelf.}
    \Description{
    Composite overview of a prototype walkthrough illustrated through a shelf organizing task. Panel (1) shows a user setting up the task by selecting relevant workspaces in a kitchen environment, paired with bounding box overlays capturing the spatial layout. Panel (2) shows the system observing the current scene from a first-person perspective as the user approaches the shelf to capture its initial physical arrangement. Panel (3) shows the user following the guidance to physically stack dishes on the shelf, guided by floating, system-generated goal image and demonstration video overlays situated within the physical workspace.
    }
    \label{fig:system-walkthrough}
\end{figure*}

\subsubsection{Recurring Failures and Feasibility Across Media, Models, and Tasks}
Across tasks, we found that the implications of recurring failures depended on the physical details the guidance needed to convey.
Beyond the incorrect LEGO attachment in Figure~\ref{fig:formative-evaluation-examples}a, our codebook examples include a crochet image whose grip and finger placement did not match the requested hold-and-pinch technique (B5: Imprecise action
execution), a gift-wrapping image whose paper flap formed a physically incoherent triangular configuration (C7: Impossible physical
state or transformation), and a furniture-assembly image whose fastening tool became malformed at the bolt (C5: Impossible or
malformed tool).
We interpret these examples as limitations in depicting fine manipulation, deformable-material states, and tool-mediated actions.

We also found that different actions within a domain could require different validations.
The yoga example in Figure~\ref{fig:formative-evaluation-examples}c illustrates unrequested movement beyond a posture (B4: Extra or spurious action), while the jump-rope sizing example in Appendix~\ref{sec:appendix-formative-codebook} shows raised handles without the required check against armpit or mid-chest height (B3: Omitted subaction).
Thus, whole-body guidance could fail through either extra movement or an omitted check, even when the general activity remained recognizable.
We therefore interpret feasibility in relation to the requirements of individual actions; these examples do not establish that one domain is more reliable than another.

Examples also reveal limits on whether a depicted action could be inspected.
In a \texttt{gemini-omni-flash-preview} circuit-wiring clip, the fingers hid the final LED insertion, preventing us from checking the specified rows (D3: Key content outside the frame or time range); in a crochet clip from the same model, the yarn manipulation was too abrupt to follow reliably (D5: Abrupt or excessive motion).
They suggest a need to expose critical contact points and make motion inspectable, although changing the presentation can also compromise continuity.
We consequently regard clearer framing and pacing as design directions requiring evaluation, rather than demonstrated remedies for these failures.
Our sample of three actions per task supports these formative implications rather than estimates of domain-wide reliability.

\subsection{Implications for System Design}
\label{sec:formative-probe-implications}

Drawing on the formative evaluation, we identify three implications for our system design:
\par\begingroup
\setlength{\parindent}{0pt}
\setlength{\parskip}{0.2\baselineskip}
\hangindent=2em \hangafter=1
\makebox[2em][l]{\textbf{I1}}\textbf{Support actions with verifiable outcomes.} We prioritize low-consequence, reversible actions whose outcomes can be visually checked for our initial exploration.\par
\hangindent=2em \hangafter=1
\makebox[2em][l]{\textbf{I2}}\textbf{Validate the change and its context.} We treat validation against the requested action and initial scene as a design requirement, including object relationships and action scope.\par
\hangindent=2em \hangafter=1
\makebox[2em][l]{\textbf{I3}}\textbf{Make target states and action details inspectable.}
We present goal images for inspecting the intended outcome and demonstration videos for inspecting motion, with framing and playback adjustment that let users examine task-relevant details.\par
\endgroup

\section{Prototype System}
\label{sec:system}

Drawing on the formative findings, we instantiated Generative Tutorial as an AR prototype system that proactively generates contextualized visual instructions based on the user's situated environment and task progress.

\subsection{Walkthrough}
\label{sec:system-walkthrough}

A user begins by selecting a task and confirming camera views of the relevant workspace.
In the shelf-organizing example in Figure~\ref{fig:system-walkthrough}, these views capture items on a staging surface and the shelf where they will be placed, providing visual context about the available objects and their arrangement (\emph{environment modeling}).
The system relates this context to actions in the task model, such as which dishes to stack and where to place them (\emph{action grounding}).
Once the required context is available (\emph{context readiness}), it requests a goal image depicting the intended arrangement and a video demonstrating the action (\emph{visual instruction generation}).
Guidance for upcoming actions can be prepared using predicted outcomes of preceding actions, allowing generation to advance ahead of the user (\emph{timeliness}).
The active goal image appears beside the physical workspace, followed by the demonstration video when available.
This placement allows the user to compare the depicted outcome and motion with the actual shelf (\emph{environment alignment}).
The user follows the guidance and explicitly confirms completion, advancing task progress and bringing the next action’s prepared guidance into view.
Fresh observations provide updated physical context when required, continuing the interaction loop.

The system contextualizes a user's high-level goal using observations of the relevant workspace.
Each workspace is represented as a 3D bounding box defined by its location and dimensions recorded before runtime.
As the system continuously samples RGB frames, it associates them with workspaces using camera-view intersection and estimated visibility.
A VLM then combines a workspace image with the user's goal, retained workspace context, and any prior task model to identify relevant objects, their observed states, and the actions needed to achieve the goal \textbf{(I1)}.

Following prior work~\cite{ye2026regular,huhVid2CoachTransformingHowTo2025}, the resulting task model consists of atomic actions, each representing the smallest independently instructable unit.
Each action specifies its input objects, intended output states, and dependencies on other actions.
Later observations can extend the model with additional actions, while previously published actions and object identities remain fixed.

\begin{figure*}[tp]
    \centering
    \includegraphics[width=.8\textwidth]{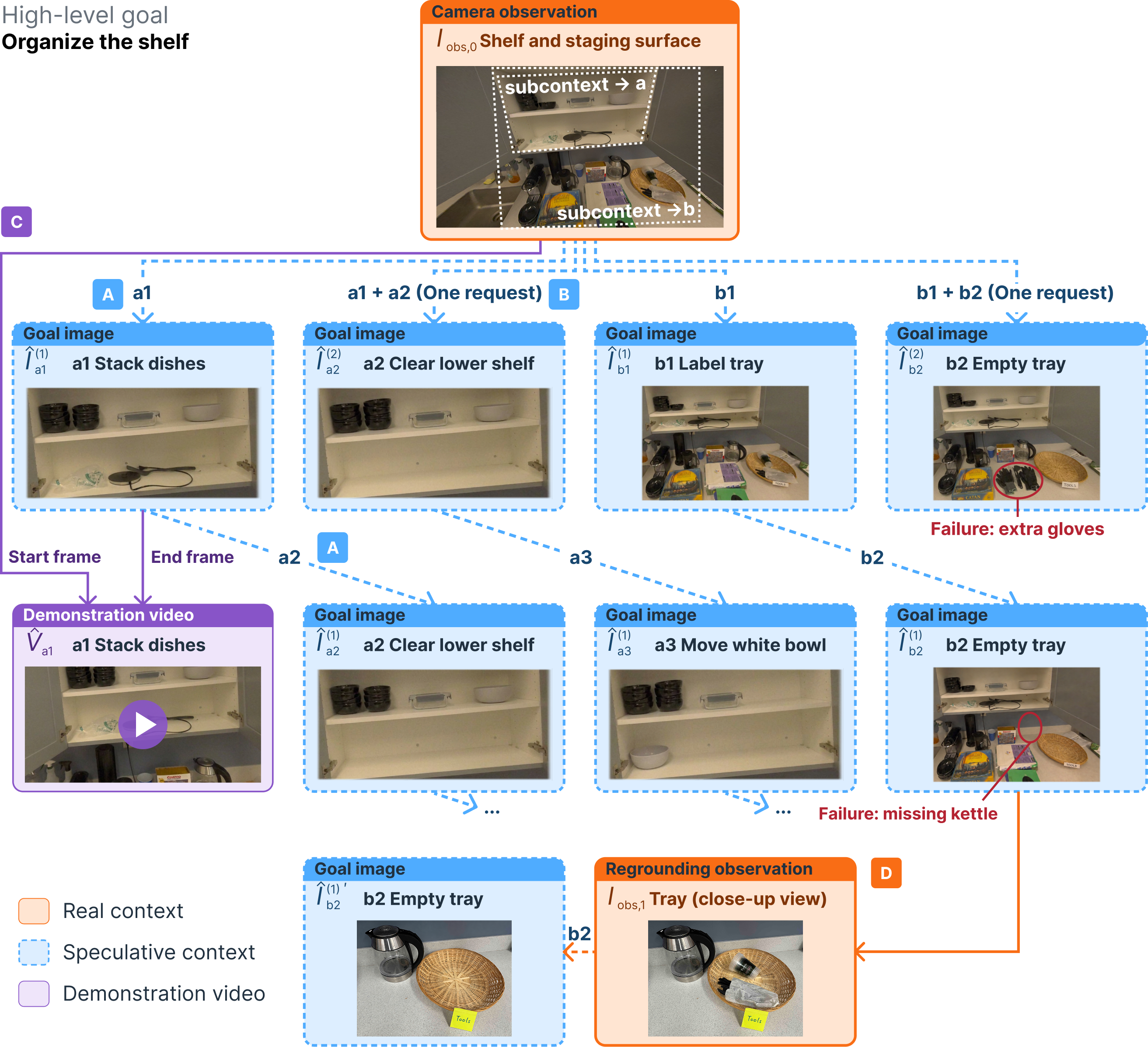}
    \caption{
    Visual state propagation illustrated through shelf organizing.
    Orange denotes camera observations (real context), blue predicted outcomes (speculative context), and purple demonstration video.
    The initial observation contains two workspace-specific subcontexts: the shelf ($a$) and the staging surface containing the tray ($b$).
    (A) Sequential prediction uses each predicted outcome as the next action's source.
    (B) Two-action prediction generates the combined outcome in one request, providing an alternative route to the same intended state.
    (C) Video generation uses source and predicted outcome images as start and end frames.
    (D) Regrounding replaces speculative context with a fresh observation. Red annotations mark observed errors, not automatic detection.}
    \Description{
    Composite overview of the context-driven generation pipeline for goal images and demonstration videos. Flow (A) shows the generation of a workspace-grounded goal image for stacking dishes derived directly from real subcontext camera observations. Flow (B) shows speculative generation chains for future tasks like clearing a shelf or emptying a tray, paired with annotations highlighting hallucination failures such as extra gloves or a missing kettle. Flow (C) shows the generation of a demonstration video utilizing the initial observation as a start frame and the generated goal image as an end frame. Flow (D) shows a regrounding observation where the system captures the updated real context of the workspace to correct prior speculative errors and accurately generate the final goal image.
    }
    \label{fig:visual-state-propagation}
\end{figure*}

\subsection{Proactive Generation through Visual State Propagation}
\label{sec:proactive-generation}

Generating images and videos takes time;
if generation begins only when the user reaches an action, the guidance may arrive after it is needed.
The prototype therefore prepares guidance for upcoming actions while the user performs earlier ones.
The backend compiles the task model into a graph of object-state and action nodes, then propagates predicted visual outcomes along its dependencies.
Because the physical outcomes of earlier actions may not yet be observable, camera observations provide \emph{real context} and generated outcomes provide \emph{speculative context} for subsequent requests.

For each action, the image generator combines an available observation or predicted goal image with the action instruction to produce a new goal image, while preserving the viewpoint and unrelated objects.
The result can guide that action and be propagated to a dependent prediction.
Figure~\ref{fig:visual-state-propagation} begins with $I_{\mathrm{obs},0}$, which contains the shelf subcontext ($a$) and the staging-surface-and-tray subcontext ($b$).
From this observation, the system prepares predictions for stacking dishes ($a_1$), the combined shelf actions ($a_1,a_2$), labeling the tray ($b_1$), and the combined tray actions ($b_1,b_2$).
The $a_1$ prediction ($\hat{I}_{a1}^{(1)}$) then supports the prediction for clearing the lower shelf ($a_2$) while the $b_1$ prediction ($\hat{I}_{b1}^{(1)}$) supports the sequential empty-tray prediction ($b_2$).

The \emph{prediction span} specifies how many actions a single request anticipates.
A one-action request predicts the next outcome (Figure~\ref{fig:visual-state-propagation}.A); a two-action request concatenates two dependency-consistent instructions in one prompt to predict their combined result without first generating an intermediate image (Figure~\ref{fig:visual-state-propagation}.B).
For the shelf context, the two routes to the outcome of $a_2$ are
\[
\hat{I}_{a_2}^{(2)} = G(I_{\mathrm{obs},0},(a_1,a_2)),
\qquad
\hat{I}_{a_2}^{(1)} = G(\hat{I}_{a_1}^{(1)},a_2).
\]
Increasing prediction span prepares guidance farther ahead, but extends reliance on speculative context, increasing the chance of aggregated errors.

Demonstration videos remain specific to individual actions, even when goal images anticipate two actions (Figure~\ref{fig:visual-state-propagation}.C).
For $a_1$, the video generator produces $\hat{V}_{a_1}$ using $\hat{I}_{\mathrm{obs},0}$ as its start frame, $\hat{I}_{a_1}^{(1)}$ as its end frame, and the stacking instruction.
We use the final goal image as the video’s end frame to improve consistency between the generated demonstration and the intended final state.

Speculative context can also carry errors: the illustrated two-action tray prediction $\hat{I}_{b_1}^{(1)}$ introduces extra gloves, while the one-action prediction $\hat{I}_{b_2}^{(1)}$ omits the kettle.
Regrounding then needs to replace the speculative source with a new observation of the current workspace \textbf{(I2)}.
In Figure~\ref{fig:visual-state-propagation}.D, $I_{\mathrm{obs},1}$ shows the kettle beside the labeled tray with its contents still inside; the renewed request is $\hat{I}_{b_2}^{(1)\prime}=G(I_{\mathrm{obs},1},b_2)$.

Requests proceed when their dependencies and suitable source context images (real or speculative) are available.
When no single speculative context history covers those prerequisites, the system obtains real context.
After prerequisite confirmation, a VLM checks camera frames for workspace identity, prerequisite availability, and visibility.
This check establishes readiness to generate.
Initial goals may be displayed and propagated without a mandatory correctness check; when a replacement goal image becomes available (triggered via alternative paths), a VLM-based comparison between the existing and replacement images can retain the more useful one.
Neither this comparison nor the illustrative failure annotations establish reliable error detection or guarantee correct downstream images and video.

\subsection{Situated Media Presentation}
\label{sec:spatial-placement}

Situated presentation aims to make the correspondence between generated guidance and its physical referents clear while minimizing perceptual load during manipulation~\cite{lindlbauer2019contextaware, cheng2021semanticadapt}.
Through pilot testing, we designed a side-by-side layout (Figure~\ref{fig:system-walkthrough}.3) that keeps the goal image and demonstration video available together, above and beside the workspace, using its geometry and the user's pose at reference capture, allowing users to compare the intended outcome, the demonstrated motion, and their own actions \textbf{(I3)}. We discuss alternative immersive visualization designs in Section~\ref{sec:discussion-representation}.

The presentation develops as media become available: \eg\ the goal image initially occupies the central view; when the demonstration is ready, the image and video occupy the view adjacently.
Inspired by WorldScribe's approach to aggregating and presenting lively information~\cite{changWorldScribeContextAwareLive2024}, we do not pause to present available guidance while additional generation proceeds.
Initial goal images appear without waiting for correctness checks; as replacements arrive, the system retains the image judged more useful.
This supports uninterrupted presentation but does not guarantee the accuracy of displayed guidance.
The interface further supplements text instructions and text-to-speech audio narration.
Based on pilot testing insights, the prototype also explores optional presentation features:
segmentation-based highlighting draws attention to relevant objects (see Figure~\ref{fig:system-walkthrough}.3),
slower video playback supports inspection of motion,
and alternative perspectives (\eg\ close-up) for additional views of an action.

\subsection{Implementation}
\label{sec:implementation}
The prototype uses a Meta Quest 3 client built in Unity 6 and a Python/Flask backend hosted on a Windows laptop.
Meta XR SDK, Mixed Reality Utility Kit, and OpenXR support passthrough capture, room registration, tracking, and input.
The client handles capture, interaction, and spatial presentation; the backend coordinates model requests and serves structured JSON and media URLs over HTTP on a shared local wireless network.
The backend pipeline defaults to \texttt{gemini-\allowbreak{}3.5-\allowbreak{}flash-\allowbreak{}lite} for optional task-model contextualization, \texttt{gemini-\allowbreak{}3.5-\allowbreak{}flash} for physical-context readiness checks and same-action replacement validation comparisons, \texttt{gemini-\allowbreak{}3.1-\allowbreak{}flash-\allowbreak{}image} for goal images, \texttt{gemini-\allowbreak{}omni-\allowbreak{}flash-\allowbreak{}preview} for demonstration videos, and \texttt{gemini-\allowbreak{}3.1-\allowbreak{}flash-\allowbreak{}tts-\allowbreak{}preview} for complementary narration.
Full prompts for generations and validation appear in Appendix~\ref{sec:appendix-prompts}.

\section{Evaluation User Study}
\label{sec:evaluation-user-study}

We conduct a lab study with 24 participants who complete four physical tasks using our prototype system and a baseline system featuring pre-authored guidance.
We address two research questions:

\par\medskip
\begingroup
\setlength{\parindent}{0pt}
\hangindent=3em \hangafter=1
\makebox[3em][l]{\textbf{RQ1}}How effectively does our prototype system support users in performing physical tasks?
\par
\hangindent=3em \hangafter=1
\makebox[3em][l]{\textbf{RQ2}}What do participants' experiences with the prototype suggest about the opportunities and pitfalls of the broader Generative Tutorial concept for physical-task guidance?\par
\endgroup
\medskip

\begin{figure*}[t]
    \centering
    \includegraphics[width=0.92\textwidth]{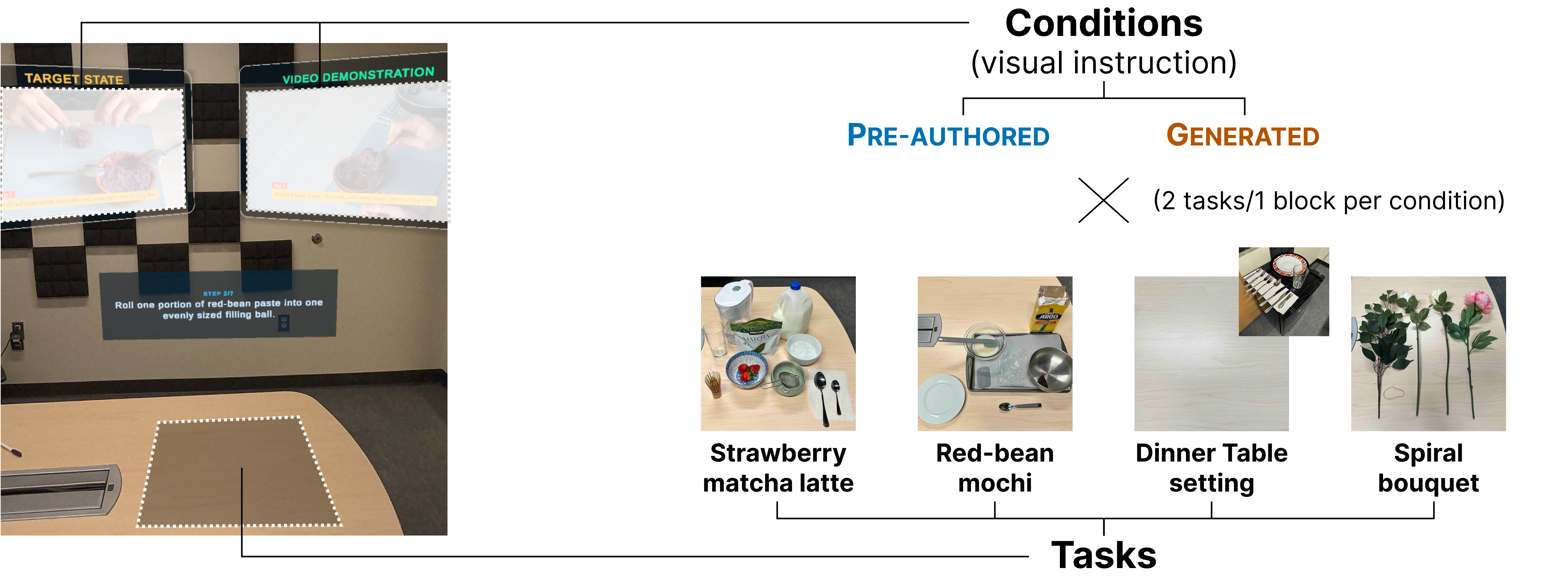}
    \caption{User-study apparatus, conditions, and tasks. Both conditions present guidance beside the physical workspace using the same layout; they differ in whether visual guidance is authored in advance or generated during task performance. Each participant completes all four tasks once, with two tasks per condition. Task order and task--condition assignments are counterbalanced across participants.}
    \Description{
    Composite overview of the user-study apparatus, experimental conditions, and physical tasks. The left section shows the mixed-reality interface from a participant's perspective, demonstrating how both conditions present goal images, demonstration videos, and text instructions in an identical spatial layout beside the physical workspace. The top right section shows the two core visual instruction conditions—comparing manually pre-authored guidance against system-generated guidance. The bottom right section shows the physical material setups for the four evaluated tasks: Strawberry matcha latte, Red-bean mochi, Dinner Table setting, and Spiral bouquet, which are counterbalanced across the two conditions for each participant.
    }
    \label{fig:study-apparatus}
\end{figure*}

\subsection{Participants}
\label{sec:evaluation-participants}

We recruited 24 participants through university mailing lists. Participants were required to be at least 18 years old, able to perform light
two-handed tabletop tasks, able to follow English instructions, and have normal
or corrected-to-normal vision.
The study was approved by our institution's IRB, and participants received a
\$30 Amazon gift card as compensation.

Participants were aged 20--32 years (\statsum{25.33}{3.19}{\mbox{years}}); eight were women and sixteen were men. Eight reported normal vision and sixteen reported corrected-to-normal vision. The background questionnaire asked about prior VR/AR headset use, AI image/video generation or editing, use of AR or AI guidance for physical tasks, and current alertness~\cite{shahid2012stanford}. Median headset use was ``a few times ever,'' and median AI image/video generation or editing was ``a few times a month.'' Median alertness was ``functioning at high levels, but not fully awake.'' Fourteen participants reported prior use of AI tools for physical-task guidance; ten reported none.

\subsection{Study Design}
\label{sec:evaluation-design}

We use a counterbalanced repeated-measures design in which each participant completes four trials across four different tasks.
Trials are grouped into two blocks of two: one with \condPreauthored guidance and one with \condGenerated guidance.
Each task is performed only once per participant, avoiding direct repetition and associated task-specific learning effects.
We cross four Williams task orders with two condition-block orders.
The eight possible combinations were assigned equally across the 24 participants (three participants per order).

\subsubsection{Conditions}
\label{sec:evaluation-conditions}

The comparison examines how producing visual guidance from the user's evolving workspace affects task performance and experience relative to guidance authored in advance.
To compare the two forms of guidance under equivalent task requirements, we define a common step-by-step procedure for each task. We adapt reference tutorial procedures to the materials, ingredients, and tools prepared for the study, keeping the required actions and step granularity consistent across conditions (see Appendix~\ref{app:user-study-tasks}).

\textbf{\condPreauthored.}
We select popular YouTube tutorials whose procedures correspond to the study tasks and extract video clips and goal-state frames for the prescribed steps.
Aside from procedural correspondence, the source media differ from the study setup at various steps in terms of tools, materials, object appearance, workspace arrangement, and details of execution.
This mirrors the common real-world scenario where users must translate context-agnostic instructions to fit their localized workspaces.
Segmenting the source media into video clips and goal-state frames allows us to control for presentation format across conditions, while also enabling participants to navigate directly between steps to access the relevant guidance.

\textbf{\condGenerated.}
We disable runtime task contextualization to keep the prescribed procedure constant across conditions.
We alternatively encode the fixed procedure in the prototype's task model as unit actions with input objects, intended output states, and dependencies between actions.
This representation supports proactive generation by identifying which upcoming actions can use observed or predicted visual context.
Participants receive goal images and demonstration videos generated during the trial from camera observations of the workspace or speculative states propagated from earlier generated outcomes.
Media are presented as they become available, exposing participants to actual generation latency and output variation.

Both conditions use the same workspace-relative UI layout, step instructions (text and audio narration), and controls.
Participants advance to the next step with an explicit confirmation thumbs-up gesture;
we omit any automatic completion detection mechanism to capture natural task behaviors and personal judgments of step completion across both conditions.

\subsubsection{Tasks and Apparatus}
\label{sec:evaluation-tasks}

Participants (A) prepare a strawberry matcha latte, (B) make red-bean mochi, (C) set a Western-style dinner table, and (D) arrange a spiral bouquet with artificial stems.
Informed by the capabilities observed in our formative evaluation, we select these activities as feasible contexts for exploring generated guidance across ingredient preparation, material shaping, object placement, and stem manipulation. 
We held the physical environment constant across studies, using the same quiet room and meeting table (see Figure~\ref{fig:study-apparatus}).

\subsubsection{Session Procedure}
\label{sec:evaluation-procedure}

Sessions lasted approximately 75 minutes. After consent and a background questionnaire, participants were fitted with the headset, the workspace was registered, and they completed a brief Rubik's Cube practice task to learn the navigation controls.

Participants then completed four trials in their assigned order. Before each trial, the researcher reset the materials and introduced the task goal. Participants were instructed to work at a normal pace while considering both speed and quality, with no imposed time limit. Trials were captured through system logs, headset recordings, and researcher observations.

After each trial, participants completed a post-task questionnaire. They completed the System Usability Scale (SUS) after each two-trial condition block (Trials 2 and 4). After all four trials, an exit survey collected condition preferences and comparisons between the two guidance styles, followed by a retrospective semi-structured interview. Survey and interview prompts are provided in Appendix~\ref{sec:appendix-study-questionnaires}.

\subsection{Measures and Analysis}
\label{sec:evaluation-measures}
\paragraph{Task performance and behavior.}
We measure task quality and execution timing from synchronized system logs and video recordings.
Task quality is rated using the step-specific checklists (0 or 1) in Appendix~\ref{app:study-rubrics}, as the percentage of observable criteria satisfied within each step.
Two authors rated the checklist questions for all steps independently, blinded to the condition, before resolving conflicted ratings through discussion.
Agreement between the initial ratings was 94.5\% across 982 paired binary rubric judgments (Cohen’s $\kappa = .740$).
Task time sums non-setup step durations, excluding between-step gaps.
Within each step, annotations further separate pre-action (from step start to the participant's first physical action), physical actions, and post-action confirmation (from end of the last physical action to step-completion confirmation) intervals to characterize behavioral phases.
\paragraph{Generated Guidance timing and quality.}
For \condGenerated condition,
we record its generation latency and whether each artifact is available at step start and, otherwise, the delay from step onset.
We evaluate the quality of generated media primarily through participants' free-response accounts of any ``incorrect, impossible, outdated, or mismatched'' guidance (Appendix~\ref{sec:appendix-study-questionnaires} Post-Task Survey Q18 and Q19).
This differentiates the approach in formative model evaluation (\ie\ codebook) and aims to capture perceived quality in the context of use.
\paragraph{Self-reported measures.}
Post-task and block questionnaires measure workload (NASA--TLX), system usability (SUS), and subjective perceptions (instruction-workspace correspondence, clarity, confidence, reliance, availability, co-visibility, and distraction), helpfulness of each guidance modality (images, video, narration, text), perceived repeatability, future use, and any noticed guidance errors.
The exit questionnaire measures overall condition preferences and comparative perceptions of speed, mistakes, trust, and effort.
Retrospective semi-structured interview elicits participants' accounts of following, double-checking, or ignoring the generated guidance; comparisons among the presentation modalities; desired system changes; and anticipated use cases.

\paragraph{Quantitative analysis.}
We analyze task-time and questionnaire outcomes using linear mixed-effects models that account for task, trial order, and repeated observations within participants.
Step-level models additionally account for step and repeated observations within trials, while SUS is analyzed at the condition-block level.
Duration measures are log-transformed, using $\log(1+t)$ for pre-action intervals.
We report condition effects with 95\% confidence intervals and apply Holm corrections within the performance, workload, experience, and helpfulness outcome families.
Overall preference is evaluated with an exact binomial test.
We additionally examine task-specific effects and sensitivity to modeling choices.
\paragraph{Qualitative analysis.}
We conduct affinity analysis on corrected interview transcripts from all 24 participants (P1--P24), grouping interpretation notes and supporting quotes into affinities and higher-level concepts.
Open-ended responses are interpreted alongside task outcomes and media-availability records to contextualize the quantitative findings and further address our two research questions.

\section{Findings}
\label{sec:evaluation-findings}

We organize the findings around the prototype system's support for physical tasks (RQ1) and the opportunities and pitfalls that participants' experiences suggest for the concept of \projectName\ (RQ2).

Throughout the findings, $N$ denotes the number of participants (out of 24) who raised a theme in the post-task interview, and $n$ denotes the number of usable questionnaire responses for a given item.
Interview counts reflect spontaneous mentions rather than agreement with a prompt, so they index salience rather than prevalence.

\subsection{Prototype System Support for Physical Tasks (RQ1)}
\label{sec:findings-prototype}

\begin{figure*}[t]
    \centering
    \includegraphics[width=0.95\textwidth]{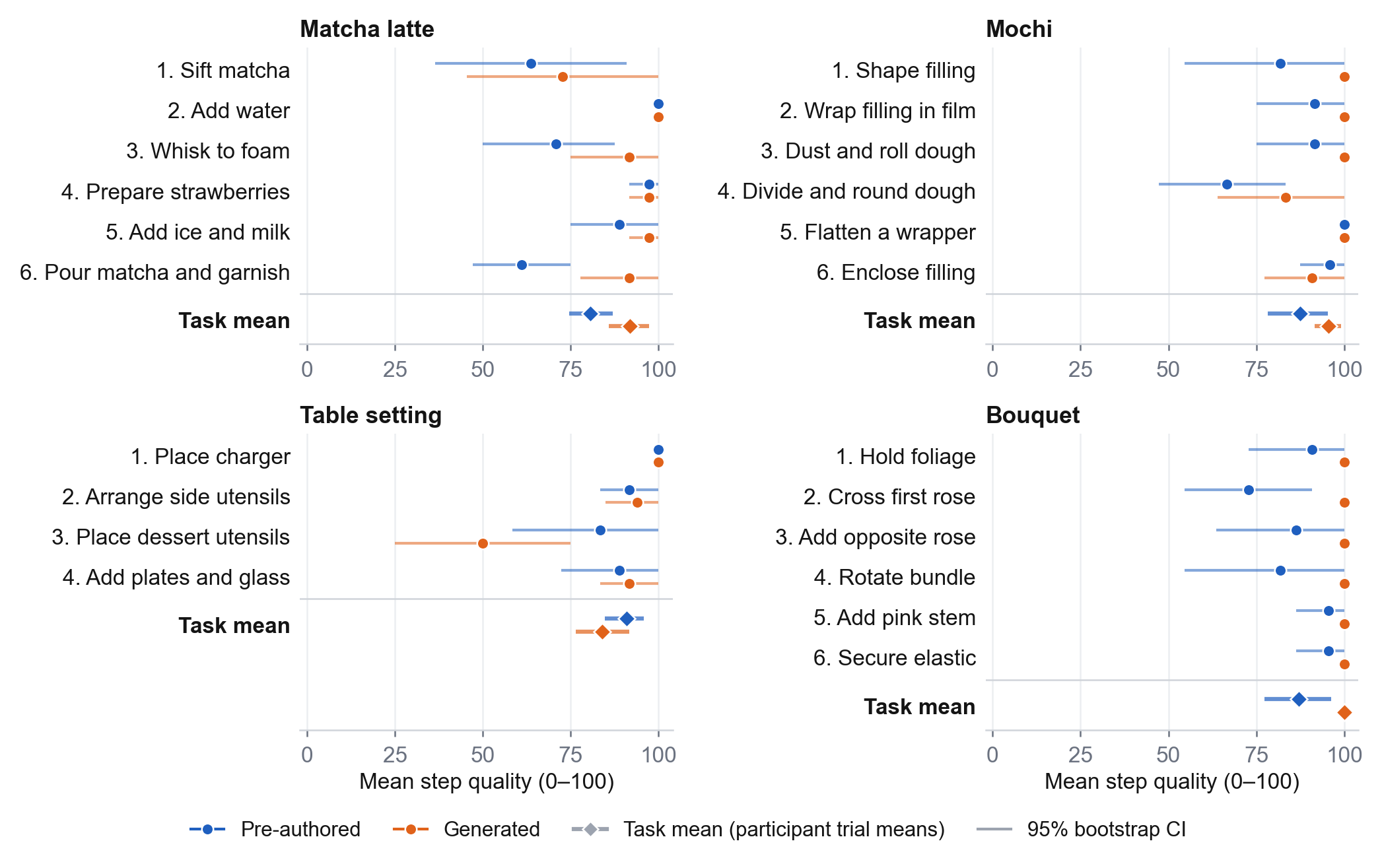}
    \caption{Task quality by task, step, and condition. Points show mean scores.}
    \Description{
    Composite overview of task quality by task, step, and condition, detailing mean scores and 95\% bootstrap confidence intervals. The top section presents dot-and-whisker plots for the Matcha latte and Mochi tasks, illustrating generally high step quality scores across both instruction types, with generated guidance frequently approaching or hitting maximum scores. The bottom section contrasts the Table setting and Bouquet tasks; the Table setting panel highlights a notable variance and quality dip for generated instructions specifically on the 'Place dessert utensils' step, while the Bouquet panel shows generated guidance achieving consistently near-perfect scores that outperform the pre-authored baseline.
    }
    \label{fig:findings-quality}
\end{figure*}

\subsubsection{Generated Visual Guidance Supported Higher Task Quality}
\label{sec:findings-quality}

Task-quality scores were significantly higher with \condGenerated\ guidance.
Mean step quality was $92.84/100$ with \condGenerated\ and $86.62/100$ with \condPreauthored, an adjusted difference of $7.02$ points (CI $[2.50,11.54]$, $p_{\mathrm{adj}}=.041$).

This advantage was not uniform across tasks (Figure~\ref{fig:findings-quality}).
Descriptive task means favored \condGenerated\ for bouquet ($100.00$ versus $87.12$), matcha ($91.76$ versus $80.51$), and mochi ($95.56$ versus $87.50$), but not for table setting ($84.03$ versus $90.97$).
The study was powered for the overall comparison rather than for task-level contrasts, and no task-level contrast was significant after correction.
We therefore report the following patterns as exploratory observations that motivate closer inspection, not as evidence of task-specific effects.

Two steps illustrate where \condGenerated\ was most favored.
In matcha, the final pouring and garnishing step scored $91.67$ versus $61.11$; in mochi, the rounded-balls criterion was met by 9 of 12 participants with \condGenerated\ versus 4 of 12 with \condPreauthored, while equal-size portions were met by 10 of 12 in both conditions.
Both steps involve specific target states and tool--object interactions, and the mochi contrast was confined to the shape criterion rather than portioning.
This is consistent with \condGenerated\ helping participants relate intended results to their own materials more directly than the pre-authored demonstrations could.

\begin{figure}[h]
    \centering
    \includegraphics[width=\linewidth]{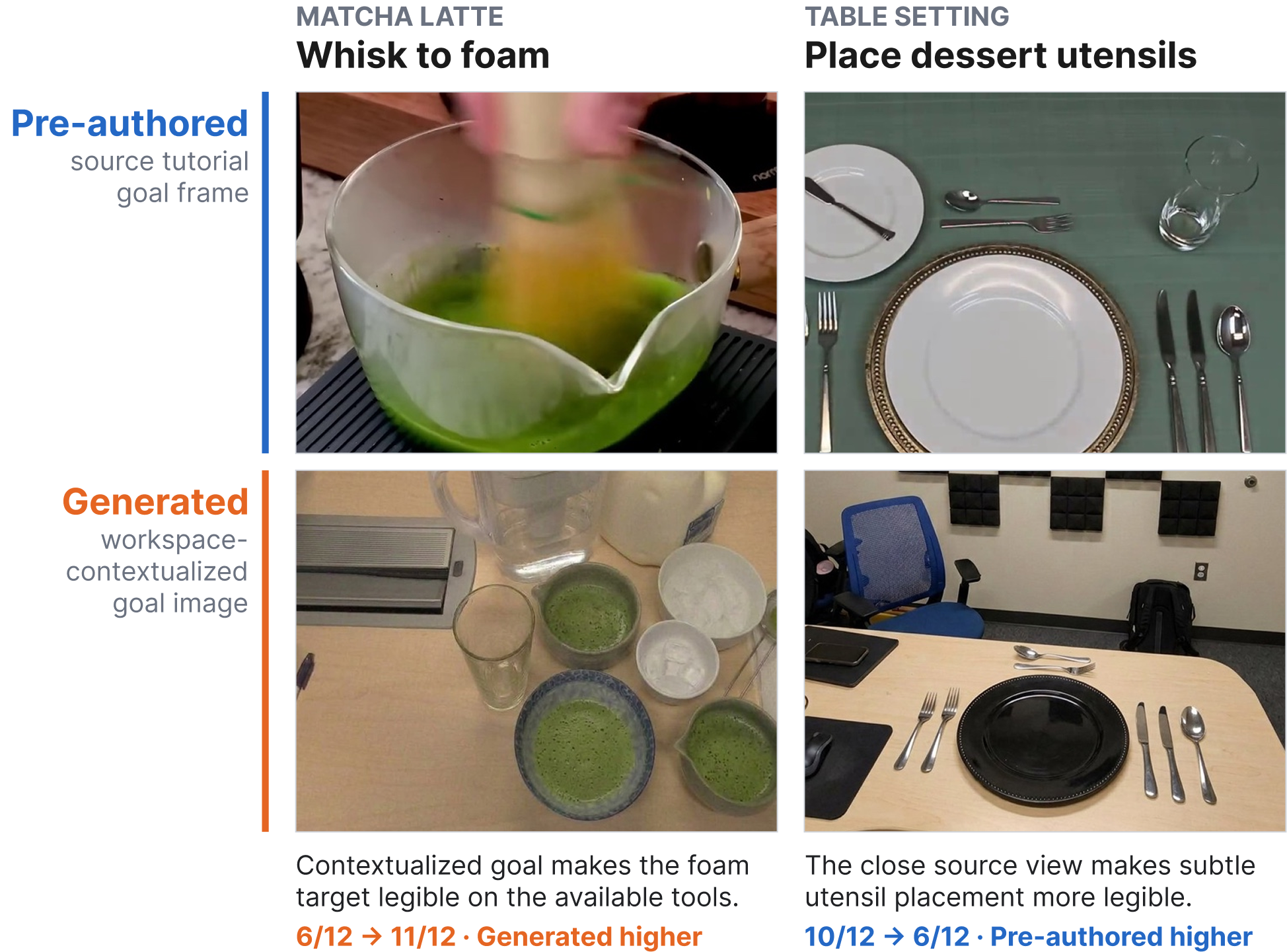}
    \caption{Two paired examples of visual guidance at steps with large descriptive condition differences.
    Generated guidance was associated with higher criterion completion for foam (6/12 to 11/12); the close pre-authored table-setting frame was associated with higher spoon-and-fork placement completion (10/12 to 6/12). 
    }
    \Description{Composite overview of condition-specific goal-image comparisons across tasks. A two-row, three-column grid contrasts pre-authored source-tutorial goal frames with generated workspace-contextualized goal images. The Matcha foam column shows higher criterion completion under the generated condition, whereas the Table-setting utensil-placement column shows higher completion under the pre-authored condition.}
    \label{fig:findings-condition-contrasts}
\end{figure}

Table setting was the one task where the descriptive difference ran the other way, and it points to a possible boundary of this account.
The gap was concentrated in placing dessert utensils (Step~3), which averaged $50.00$ versus $83.33$; the spoon and fork placement criteria were each met by 6 of 12 participants with \condGenerated\ versus 10 of 12 with \condPreauthored (see Figure~\ref{fig:findings-condition-contrasts}).
Unlike the matcha and mochi steps above, these criteria depend on a conventional arrangement---correct relative position \emph{and} orientation---rather than on a target state that varies with the participant's own materials.
Where the intended result is generic, the pre-authored demonstration may convey it more reliably, and scene-conditioned generation offers less to gain while introducing opportunities for the depicted arrangement to deviate from the convention.

\begin{figure*}[t]
    \centering
    \includegraphics[width=0.9\textwidth]{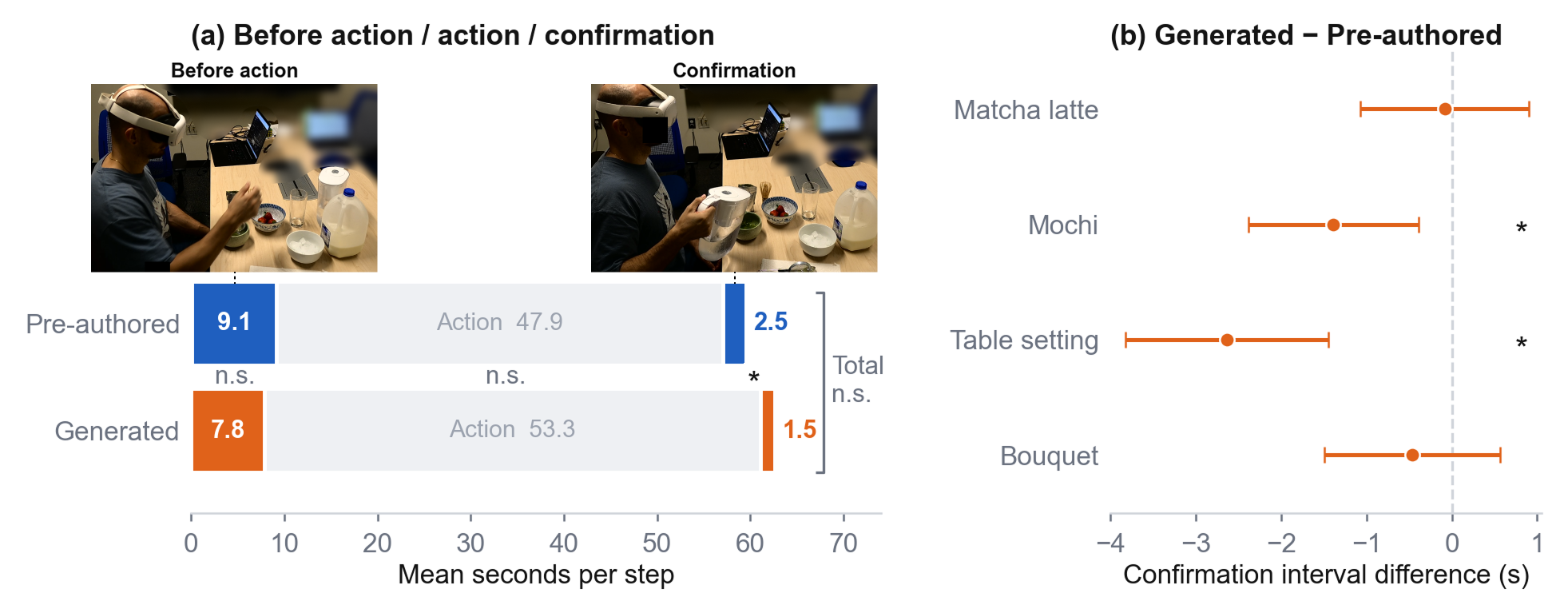}
    \caption{Action timing by condition.
    (a) Mean durations before action, during action, and before confirmation,
    illustrated with Step 2 of the Matcha latte task: before action, the participant identifies the water pitcher; at confirmation, the participant check final state.
    (b) Adjusted confirmation interval differences (Generated minus Pre-authored), with 95\% confidence intervals. * $p < .05$.}
    \Description{
    Composite overview of action timing by condition. Panel (a) shows mean durations for the before-action, action, and confirmation phases. Inset photographs illustrate Step 2 of the Matcha latte task: before acting, the participant identifies the water pitcher, and at confirmation, the participant verifies the final state. Initial planning and action-execution times do not differ significantly between conditions, whereas the confirmation phase is significantly faster by 1.06 seconds with generated instructions. Panel (b) shows adjusted generated-minus-pre-authored confirmation-interval differences, with 95\% confidence intervals, for Matcha latte, Mochi, Table setting, and Bouquet. The reductions are statistically significant for Mochi and Table setting.
    }
    \label{fig:findings-timing}
\end{figure*}
\subsubsection{Generated Visual Guidance Showed No Clear Overall Time Savings but Shortened Step Confirmation}
\label{sec:findings-timing}

We measured overall task time as the sum of step durations, using it as an indicator of task efficiency.
Across 92 trials with complete timing, there was no clear condition difference: the adjusted ratio of \condGenerated\ to \condPreauthored\ task time was $0.99$ (CI $[0.91,1.08]$).

To examine where participants spent time, we further divided each step into three intervals.
The \emph{pre-action interval}, from step start to the first physical action, serves as a proxy for orienting to the guidance and preparing to act.
The \emph{action interval}, from the start of the first physical action to the end of the last, captures the period of task execution, including any intervening pauses.
Pre-action intervals were descriptively shorter and action intervals longer with \condGenerated, but neither difference survived correction.

The clearest difference occurred in the \emph{confirmation interval}, from the end of physical action to the participant's step-confirmation input.
We use this as a proxy for completion assessment and the decision to proceed.
Across 508 steps, the confirmation interval averaged $1.50$\,s with \condGenerated\ and $2.49$\,s with \condPreauthored.
The model-adjusted reduction was $1.06$\,s (CI [$0.37,1.74$], $p_{\mathrm{adj}}=.041$).
The clearest reductions happened in table setting and mochi (Figure~\ref{fig:findings-timing}).
Participants described using target images to judge completion or correctness ($N=4$): P10 used the goal image as a benchmark, while P23 explained, \textit{``I can compare my environment with the visual to check whether I am right.''}
These accounts suggest that goal images could support quicker (but not necessarily more accurate) completion judgments.

\begin{figure*}[t]
    \centering
    \includegraphics[width=0.9\textwidth]{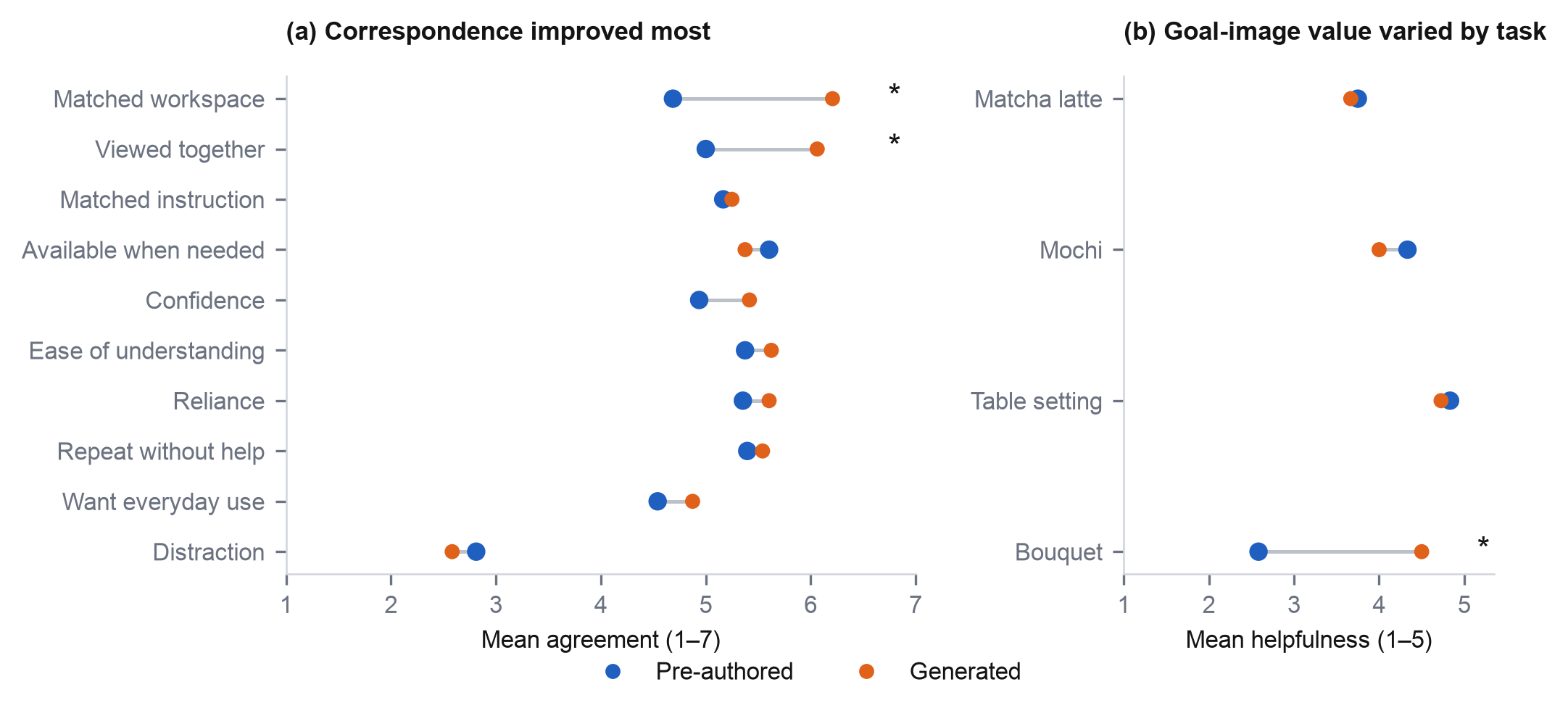}
    \caption{Participant ratings by condition.
    Points show mean (a) agreement with questionnaire statements and (b) goal-image helpfulness by task.
    * p < .05.}
    \Description{
    Composite overview of participant ratings by condition across questionnaire statements and task-specific helpfulness. Panel (a) shows mean agreement with various subjective metrics on a 1–7 scale, illustrating that the generated condition significantly improved perceived correspondence—specifically how well the guidance matched the workspace and was viewed together with the physical task—compared to the pre-authored baseline, while maintaining similar levels of confidence and ease of understanding. Panel (b) shows goal-image helpfulness on a 1–5 scale across four activities, demonstrating that while helpfulness ratings were relatively uniform between conditions for the matcha, mochi, and table setting tasks, the generated goal images provided a statistically significant benefit for the visually complex spiral bouquet task.
    }
    \label{fig:findings-correspondence}
\end{figure*}

\subsubsection{Generated Visual Guidance Made Instructions Easier to Relate to the Workspace}
\label{sec:findings-outcomes}
Participants rated the guidance's correspondence with their workspace higher with \condGenerated\ than \condPreauthored\ guidance ($M=6.21$ versus $4.69$ on a seven-point scale, $p_{\mathrm{adj}}=.024$).
They also reported greater ease viewing the guidance and workspace together without interference (Q12; $M=6.06$ versus $5.00$,
$p_{\mathrm{adj}}=.017$).
Participants described less work identifying objects and translating between settings when the guidance matched their workspace ($N=10$).
P6 no longer needed to \textit{``translate whatever shown in a video to that shown in front of me''};
P19 explained, \textit{``Generated guidance gave me the exact items I had, so it was less confusing.''}
Conversely, P2 found the pre-authored flower demonstration difficult to apply when its number and mix of stems differed from the available materials.
In the exit comparison, 17 participants agreed that generated guidance required less effort, five were neutral, and two disagreed ($p_{\mathrm{adj}}<.001$).

Broader evaluations were less conclusive (see Figure~\ref{fig:findings-correspondence}). Workload means were close ($26.02$ versus $25.78$), as were SUS scores ($75.31$ versus $74.48$). Overall preference was also mixed: 15 of 24 participants chose \condGenerated, without a significant preference imbalance.

\subsubsection{Generated Visual Guidance Could Match the Workspace and Still Contain Errors}
\label{sec:findings-guidance}

Improved workspace correspondence did not establish more correct guidance.
Ratings that the visuals matched the task instruction were similar between \condGenerated and \condPreauthored ($M=5.25$ versus $5.17/7$), with no clear condition difference.
Participants reported guidance that seemed incorrect, impossible, outdated, or mismatched in 26 of 48 generated trials and 21 of 48 pre-authored trials; the difference was not significant ($p=.102$).
Reports about \condGenerated\ guidance included both contextual mismatches and procedural inconsistencies:
For contextual mismatches, P12 reported that the dessert spoon and fork looked different in the generated images, while P5 described an unexpected bowl appearing during the matcha task.
Procedural inconsistencies concerned what participants were expected to do: P10 reported that the mochi video showed five dough balls although the text and narration specified four.

\subsubsection{Under Proactive Generation, Goal Images Were Usually Ready at Step Start; Videos Often Were Not}
\label{sec:findings-use}

Ratings of visual guidance being available when needed (Q14) showed no clear condition difference ($M=5.38$ for \condGenerated\ versus $5.60$ for
\condPreauthored\ on a seven-point scale), despite generated guidance being
produced during the trial.
Interview accounts of missing or delayed visual guidance ($N=9$) point to differences in readiness across media.
P19 explained, \textit{``The second (\condGenerated) condition helped me identify the items, but the video was not necessarily there, so I wondered what to do.''}

Across 213 image and 159 video readiness events logged before step end, most media were already ready when the step began: 201 of 213 images ($94.4\%$) and 114 of 159 videos ($71.7\%$).
Among the remaining arrivals, the median delay from step start was $15.6$\,s for images ($n=12$) and $22.6$\,s for videos ($n=45$).
Goal images were thus more often ready at step start, while videos more often became ready during the step.

\begin{figure}[h]
    \centering
    \includegraphics[width=0.8\linewidth]{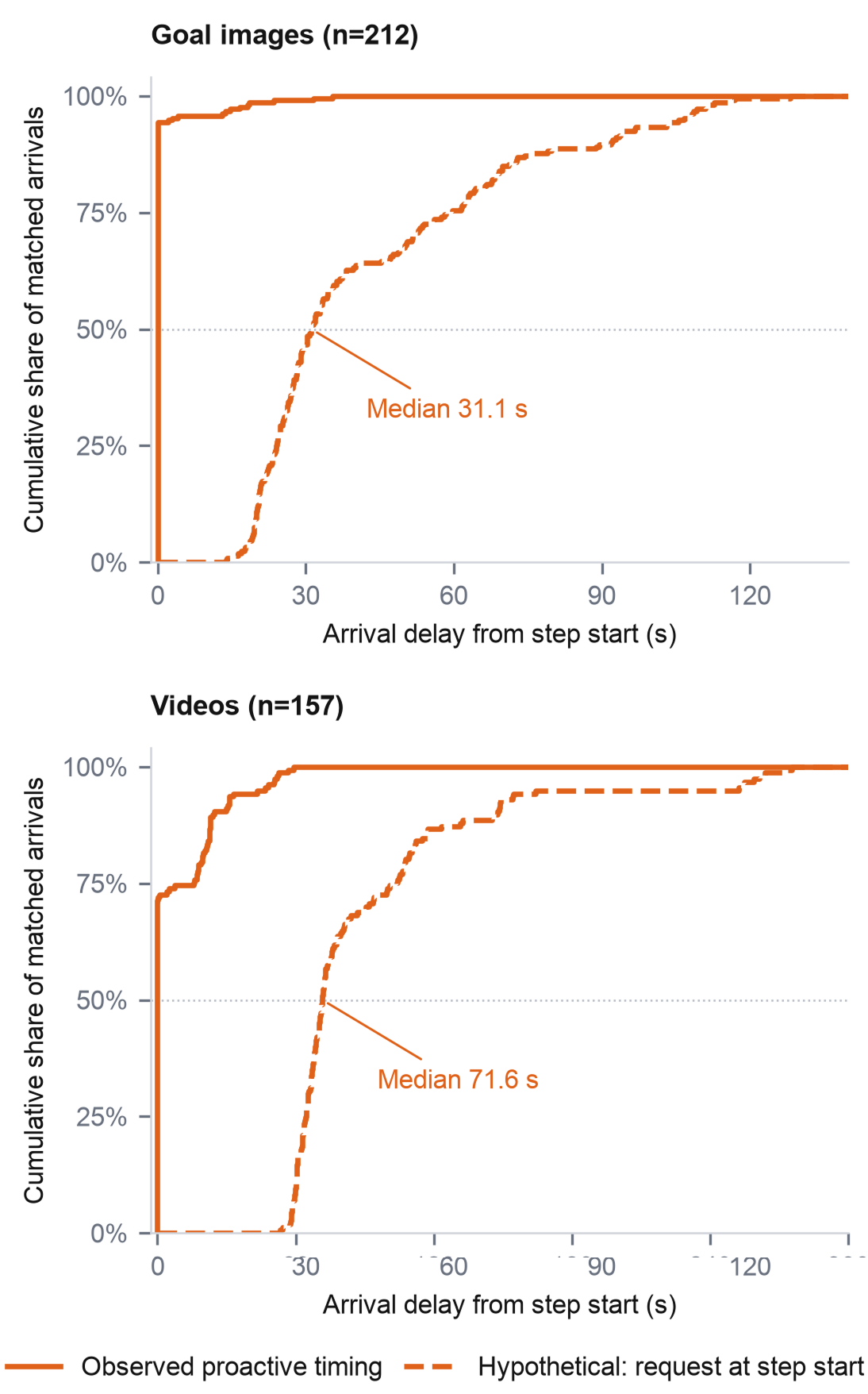}
    \caption{Generated Media readiness relative to step start.
    Curves show cumulative shares of
    matched
    image and video arrivals under observed proactive timing and hypothetical requests
    at step start, holding recorded latencies fixed.
    }
    \Description{
    Composite overview of Generated Media readiness relative to step start, contrasting proactive and reactive generation models. The left panel plots the cumulative share of goal image arrivals, demonstrating that observed proactive timing ensures nearly all images are available at the exact moment a step begins (0s delay), whereas a hypothetical system requesting generation at the step start would incur a median latency of 31.1 seconds. The right panel plots video arrivals, illustrating that proactive timing delivers the majority of videos instantly at step start, effectively eliminating what would otherwise be a median delay of 71.6 seconds under a reactive generation baseline.
    }
    \label{fig:findings-readiness-counterfactual}
\end{figure}

Figure~\ref{fig:findings-readiness-counterfactual} contrasts the observed proactive timing (solid curves) with a hypothetical  replay (cf.~\cite{gmeiner2026counterfactual}) in which each step’s first request begins at step start (dashed curves).
Holding the observed request-to-readiness intervals and subsequent request offsets fixed, the dashed curves reach their medians at $31.1$\,s for images and $71.6$\,s for videos, compared with $0$ for observed proactive timing.

\begin{figure*}[t]
    \centering
    \includegraphics[width=0.9\textwidth]{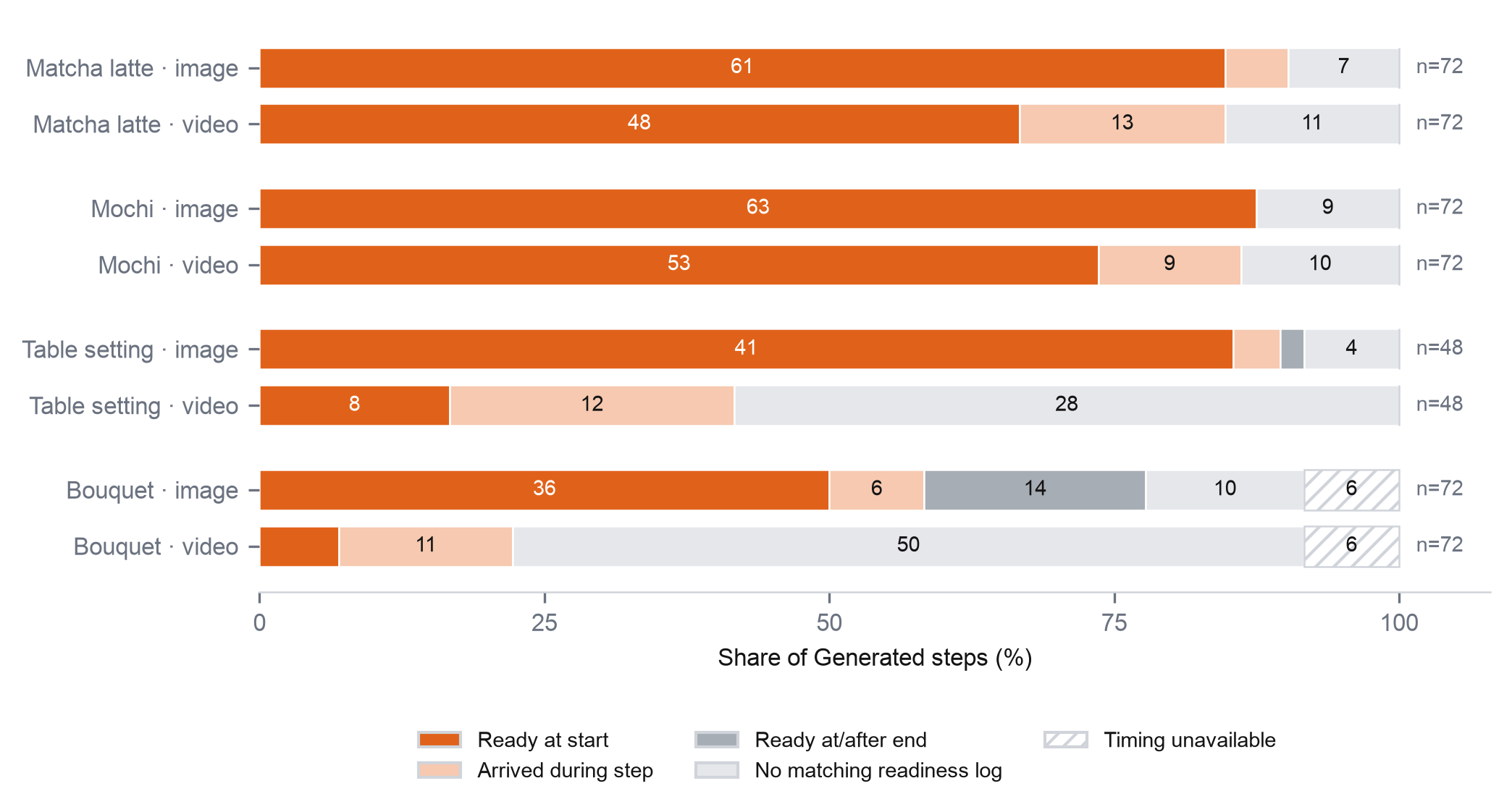}
    \caption{Generated-media readiness by task and modality.
    Bars show the share of planned steps in each readiness category; segment labels give counts, and n gives the total planned steps.}
    \Description{
    Composite overview of generated-media readiness broken down by task and modality. The top section covers the Matcha latte and Mochi tasks, illustrating robust proactive system performance where the vast majority of both generated images and videos were successfully 'Ready at start.' The bottom section covers the Table setting and Bouquet tasks, highlighting a significant reliability divergence between modalities; while goal images remained predominantly ready at the start of a step, demonstration videos frequently failed to process, represented by massive portions of the data falling into the 'No matching readiness log' category.
    }
    \label{fig:findings-availability}
\end{figure*}

The task-level chart complements this timing distribution by showing coverage across all planned steps (Figure~\ref{fig:findings-availability}).
Videos were ready by step start in five of 72 bouquet steps, eight of 48 table-setting steps, 48 of 72 matcha steps, and 53 of 72 mochi steps.
In addition to differences in media-generation time across tasks, this variation was partly due to differences in step duration: bouquet steps were generally completed more quickly, leaving less time for videos to be generated before the next step began.

Together, these results suggest that proactive generation can preserve access to contextual visual guidance with reduced latency as tasks unfold by shifting generation ahead of the moment of need.
This benefit was strongest for goal images whereas video readiness remained sensitive to longer generation times and rapid task progression.

\subsection{Opportunities and Pitfalls of Generative Tutorials (RQ2)}
\label{sec:findings-concept}

\subsubsection{Some Trusted Familiar Settings; Others Trusted Real Demonstrations}
\label{sec:findings-dependability}
Trust is critical for contextualized visual guidance as local resemblance can make a generated depiction seem immediately applicable without independently establishing that the depicted action is correct.
Participants differed in the cues they relied on to judge an instruction's credibility.
Local scene correspondence acted as a credibility cue ($N=3$).
P20 trusted generated guidance more because it followed a scan of the local environment and objects.
P23 similarly treated the use of their own materials as a reason to accept the depicted action: \textit{``That makes me think this must be how it should be done.''} For these participants, local correspondence could provide reassurance as well as practical assistance.

Others valued evidence that a real person had performed the procedure ($N=3$).
P14 trusted pre-authored demonstrations because they showed \textit{``steps that someone has actually taken.''}
P2 trusted a real demonstration's physical plausibility even when it was harder to apply locally, while P10 accepted a different reference setting to avoid generation uncertainty.
P21 explicitly separated usefulness from trust, acknowledging the generated guidance's effectiveness in the study tasks while remaining concerned about hallucinations.
The exit trust ratings reflect these contrasting accounts: nine participants trusted generated guidance more, nine disagreed, and six were neutral.

Trust also had to be negotiated when generated images and videos disagreed.
P18 described a fallback sequence: \textit{``Usually I look at video first. If it seems wrong, I look at the image; if that also seems wrong, I read the text.''}
P22 followed the video when its placements conflicted with the goal image, whereas P14 favored narration over a conflicting video.
Participants thus supplied their own judgments about which source to follow.
Multiple representations offered opportunities to cross-check, but their presence did not establish a shared or consistently authoritative account of the next action.

\subsubsection{Contextual Correspondence Made Visual Anomalies Harder to Interpret}
\label{sec:findings-adaptation}

Contextualized visuals could make the physical scene directly comparable with the instruction and encourage close matching.
For example, P6 described avoiding the need to translate from another setting, and P23 used goal images to check both the arrangement and whether all objects were present.
P10 reported, \textit{``I would just do it again. And wait till I matched exactly the goal image.''}

Yet the correspondence also made unexpected changes harder to dismiss.
Changes in generated quantities, objects, or action transitions could disrupt the sequence or prompt doubts about one's own actions ($N=5$).
P24 double-checked earlier actions after a rubber band appeared unexpectedly in the flower demonstration, and P16 reported in the post-task questionnaire that sudden changes in the video \textit{``made me unsure if what I did was correct.''}
The uncertainty concerned whether a depicted difference required an action, revealed a personal mistake, or was a generation artifact.

For P10, whose account spans both effects, resolving that uncertainty was what mattered.
Having treated the goal image as an exact standard, they described mismatches as a source of anxiety: \textit{``when they mismatch, you just feel very nervous. You don't feel that you're doing everything correctly.''}
The anxiety eased once they attributed a mismatch to generation rather than to personal failure.
Other participants made this attribution readily, disregarding an artifact while retaining an understandable action ($N=6$).
P21 noted, \textit{``A cup might grow two handles, but that did not affect understanding or doing the task,''} and P18 rejected an extra generated cup because there was only one real instance.

Attribution was harder when the depicted goal was plausible but the real materials behaved differently.
P15 found the depicted ball larger and more perfect than their available material seemed able to produce, and P16 did not know how to proceed when matcha failed to sieve as smoothly as the demonstration.
In these cases nothing in the image marked itself as an artifact, so the discrepancy remained an open question about the participant's own execution.

These accounts suggest a shift in the work of interpretation: local resemblance reduced the need to map between different scenes, but users still had to decide which details were meaningful and achievable.
Where that distinction was unclear, apparent correspondence could invite unnecessary checking or attempts to reproduce an unreliable target~\cite{liu2026overreliance}.

\subsubsection{Images Showed the Goal; Videos Explained the Action}
\label{sec:findings-media-roles}

Modality helpfulness ratings showed no overall corrected condition differences.
The exception was bouquet arrangement, where generated goal images were rated more helpful than pre-authored ones ($M=4.50$ versus $2.58$ on a five-point scale, $p_{\mathrm{adj}}<.001$; Figure~\ref{fig:findings-correspondence}).
Bouquet is the task in our set whose outcome is most fully specified by its final arrangement and least governed by a conventional target, so a goal image carries proportionally more of the guidance there; we read the result in that light rather than as a general modality effect.

Participants' accounts point to a division of labor between the two representations.
Images were valued when the uncertainty concerned the intended result: P1 wanted an image when \textit{``you just need the final state, the goal state of each step,''} and P19 used a visible goal to plan their own action sequence.
Videos were needed when the uncertainty concerned how to act ($N=8$): P16 needed video to understand how to roll dough, and P17 found a photograph insufficient for tying a bouquet.
Which representation was sufficient therefore varied with the kind of uncertainty a step presented, and the same task could require both.

\subsubsection{Guidance Had to Fit Participants' Pace and Physical Actions}
\label{sec:findings-media-control}

Participants often sought alternative modalities to complement visual instructions, and the most useful form of guidance depended on how they allocated their attention.
P19 could act while listening to narration, whereas P6 found listening difficult while concentrating on hand movements.
Language comfort did not imply a single preferred channel either: P18 favored readable text, while P6 found reading instructions in a non-native language difficult.
No single modality was therefore preferable across participants.

Persistent text mattered for a different reason than modality preference: it gave participants a way to revisit information, check instructions, or control the pace ($N=6$).
P9 relied on text when narration could not be repeated, and P13 used it to control the pace.
Choosing text therefore sometimes reflected control or necessity rather than a general preference for text over generated visuals.

Presentation could also interfere with the physical work itself.
The headset or overlays sometimes obstructed participants' view of physical materials or surroundings ($N=6$): P19 could not monitor spilled cornstarch because the instructions blocked the view.
Pacing and controls interfered in other ways.
P13 was still carrying out an earlier action when narration had reached its third sentence, and P8 found the right-thumb confirmation gesture difficult while holding flowers.
These accounts qualify the favorable co-viewability ratings: usable assistance depended on whether the information, pacing, and controls fit the participant's current action, as well as whether the depicted scene matched the workspace.

\subsubsection{Further Adapting Contextualized Visual Instructions}
\label{sec:findings-value}

Participants proposed extending contextualization to available resources and intermediate task states.
Requests included substitutions or quantities suited to available materials ($N=3$): P8 wanted to ``achieve the task with whatever is in hand,'' and P12 raised missing equipment.
Participants also proposed support for on-demand generation ($N=5$).
P8 wanted specific corrections such as thinner dough or less filling, and P23 wanted the system to ``generate a path from that intermediate state to the goal.''
These proposals concern adapting the instruction itself to the user's situation, extending the study's focus on generating contextualized visuals for fixed procedures.

Other suggestions concerned how users could access and assess contextualized visuals.
P21 wanted to replay part of a clip, P8 requested visible loading status, and P24 wanted to switch between generated and pre-authored guidance.
P23 suggested combining an established recipe-like procedure with demonstrations generated for local objects, while P22 proposed cross-checking instructions against evidence for the specific model being used.
Such controls could help users revisit an action or consult another source when a generated visual was insufficient.

Participants expected these adaptations to make contextualized visual instructions more useful when local circumstances differed from available demonstrations.
P19 described bicycle repairs involving unfamiliar parts, whereas P21 anticipated less benefit for standardized furniture with fixed parts and procedures.
They also identified constraints: P16 noted that stove cooking would continue while guidance was delayed and P18 anticipated difficulty distinguishing personal mistakes from system errors in complex tasks.
These anticipated uses and limitations suggest that further contextualization should account for changing task states and the practical costs of obtaining and acting on updated guidance.

\section{Discussion and Future Work}
Our findings show that contextualized live visual generation can help people relate instructions to their environmental context and assess intended outcomes as tasks unfold, while revealing opportunities and limitations in how users build trust, interpret generated errors, and adjust their actions in response beyond current model capabilities.
In this section, we discuss what this combination establishes about the \projectName{} concept.

\subsection{Prototype Effectiveness and Broader Implications}
\label{sec:discussion-effectiveness}
The study provides evidence of the benefits of our prototype system in supporting higher task-quality scores, greater correspondence with the workspace, and shorter interval between an action and confirming step.
These results do not establish the prototype as a broadly superior alternative to pre-authored guidance: benefits varied across tasks and measures, with no clear overall time savings, similar workload and SUS scores, and mixed preferences.
Instead, they suggest that contextualized live visual generation can support specific aspects of physical-task performance, in which variations in the appearance of local materials and difficulty assessing intended outcomes contribute to task complexity.

While our system leverages state-of-the-art visual generative models to prototype the experience of live contextualized visual instruction, our current exploration is also constrained by their capabilities.
We reckon that with newer models, the specific instantiation of the system components can be invalidated or unjustified: 
faster generation could reduce the need for speculative pre-generation,
while more reliable preservation of objects, spatial relationships, and action scope could reduce the need for corrective processing and candidate comparison.
Beyond this technical instantiation, our work conceptualizes live contextualized visual instruction and provides preliminary evidence of its value and an empirical account of how users experience it.
Recent machine learning advances have enabled increasingly coherent instructional sequences; for example, ShowHowTo~\cite{souvcek2025showhowto} generates step-by-step image sequences conditioned on an initial scene and textual instructions.
Complementing these advances, our findings highlight the need to address context mismatches as they arise and accommodate users’ evolving task progress, situated needs, and perceptual abilities.

\subsection{Task Scope and Cognitive Support}
We selected four low-consequence tasks with visually assessable outcomes to explore contextualized visual instruction within the capabilities identified by the formative evaluation.
Although participants envisioned applications beyond these settings, this scope limits the generalizability of our findings to tasks requiring greater precision, involving less observable states, or carrying more serious consequences (\eg\ surgical tasks).

Our findings also motivate closer examination of the finer-grained cognitive activities involved in physical task performance.
Although we observed no clear overall time savings, shorter confirmation intervals suggest that contextualized goal images may facilitate assessment of whether a step has been completed successfully.
Future designs could target specific activities or psychomotor phases~\cite{hendersonAugmentedRealityPsychomotor2011, klatzky2023wca}---identifying objects, interpreting spatial relationships, judging completion, or deciding how to recover---and evaluate whether generated visual instruction supports that activity.

Our focus is immediate guidance rather than learning.
Participants reported relatively high perceived ability to perform the task again without guidance, with no clear condition difference ($M=5.54/7$ for generated versus $M=5.40/7$ for pre-authored guidance).
However, this item measures perceived capability, and the study did not assess retention or transfer.
Guidance for immediate performance may externalize targets and procedural information;
future work on contextualized visual instruction designs that aim to develop independent skill should therefore investigate questions on supporting understanding of why actions are appropriate~\cite{zhangFollowingUnderstandingInvestigating2025}, acquiring the knowledge needed to make decisions~\cite{koedinger2012knowledge}, and developing metacognitive strategies for deciding when to seek, question, or proceed without assistance.

\subsection{Adaptive Visual Representations}
\label{sec:discussion-representation}

Our work explores photorealistic image and video generation to depict instructional changes in the user’s environmental context.
This initial exploration covers only part of the design space for using visual generative models in physical-task guidance.
While photorealism may help users with direct mapping of materials, more abstract representations---such as sketches, diagrams, or motion traces---could emphasize task-relevant relationships and reduce incidental details.
Future work should examine which representations and levels of detail best support the perceptual demands of different tasks.

We partially explored ``customization'' through playback adjustment and alternative perspectives for inspecting action details.
A more systematic exploration could examine adaptations to users’ experience and abilities.
For example, for a left-handed person, a corresponding left-handed demonstration can preserve the intended relationships among tools, materials, and actions rather than simply mirror an asymmetric scene.

In the context of mixed reality, presenting 2D instructional media still requires users to translate depicted spatial relationships into three-dimensional action~\cite{fidalgo2025handaug}.
Our prototype adopted side-by-side image and video presentation to make both representations available for comparison while limiting interference with manipulation.
Future work could investigate more spatially integrated presentations as well as probing the feasibility of generating 3D visual representations (\eg\ Gaussian splats~\cite{kari2025reality,vachha2025dreamcrafter}) for scaffolding effective physical task performance.

\subsection{Limitations}
Other limitations of this work include the coverage of the formative evaluation and the controlled scope of the user study.
The formative evaluation sampled 45 actions from a curated corpus using selected model snapshots; its findings characterize recurring issues rather than establish reliability across domains.
Future work should examine how different errors affect users' interpretations and actions.

The user study involved 24 participants in a prepared laboratory workspace with fixed procedures.
Participants were recruited from a university population and were relatively young (20--32 years) and familiar with AI tools: 14 of 24 reported prior use of AI assistance for physical tasks.
This sample is plausibly more receptive to generated guidance than a general population, so the observed advantages may not transfer to users with less exposure to such tools.
The study does not evaluate automatic task planning or adaptation to deviations during execution.
The comparison also evaluates contextualized generation and proactive delivery together, without isolating their contributions.
Future studies with more diverse users and environments (\eg\ deployment), alongside component-level comparisons, could clarify where the observed benefits generalize.

As a research prototype, our system triggers proactive generation to prioritize media readiness without optimizing for cost-efficiency which is impractical for real-world deployment.
We do not evaluate visual state propagation as a technical mechanism.
In particular, we do not measure the accuracy of predicted outcomes, nor compare one-action and two-action prediction spans on cost and fidelity; the span used in the study was fixed after pilot testing.
Our claims concern how participants performed and experienced tasks with the resulting guidance, not the correctness of the propagation itself.
Relatedly, the hypothetical timing replay (Section~\ref{sec:findings-use}) holds observed latencies and inputs fixed, so it quantifies how much lead time proactive generation supplied rather than establishing its causal benefit.

Our measurement of completion assessment depends on an explicit confirmation gesture.
The gesture takes time to perform, so the absolute duration of the confirmation interval likely overstates the cost of assessing completion alone; P8's difficulty confirming while holding flowers indicates that this cost also varied with what participants were holding.
Because the same mechanism was used in both conditions, this offset does not account for the observed condition difference, but the absolute values should be read as an upper bound rather than as a direct measure of judgment time.
More generally, logged readiness and behavioral intervals do not directly measure viewing, waiting, or cognitive processes that more advanced sensing hardware may capture.

The side-by-side layout was settled through pilot testing rather than systematic comparison against alternative presentations.
Participants' reports of occluded materials indicate that this choice shaped the experience we observed, and the presentation design should be treated as one point in a larger space (Section~\ref{sec:discussion-representation}) rather than as a validated arrangement.

\section{Conclusion}
In this paper, we introduced \projectName{}, a conceptual framework for live visual instruction that situates intended outcomes and actions within a user's physical environment and task flow.
Based on a formative evaluation, we built an AR prototype that proactively generates goal images and demonstration videos using observed and speculative context.
An evaluation with 24 participants across four tasks showed that \projectName{} improved performance quality, increased perceived workspace correspondence, and reduced step-confirmation intervals.
Qualitative insights further revealed how contextual resemblance shapes trust, how generation errors impact interpretation, and how guidance must flexibly accommodate user needs.
Ultimately, these findings define interaction requirements for integrating generative visual models into live physical tasks.
As generative AI advances, we hope this work inspires future systems that coordinate contextual relevance, timely delivery, and human judgment to support users in their own environments.

\begin{acks}
This research was funded, in part, by the U.S. Government under ARPA-H contract 1AY2AX000062 and by the National Science Foundation under Award IIS-2406218. The views and conclusions contained in this document are those of the authors and should not be interpreted as representing the official policies, either expressed or implied, of the U.S. Government.
\end{acks}

\bibliographystyle{ACM-Reference-Format}
\bibliography{references}

\onecolumn
\appendix
\raggedbottom
\setlength{\LTpre}{6pt}
\setlength{\LTpost}{6pt}
\section{Formative Model Evaluation}
\label{sec:appendix-formative}
\subsection{Dataset}
\label{sec:appendix-formative-dataset}

\begingroup
\footnotesize
\setlength{\tabcolsep}{3pt}
\renewcommand{\arraystretch}{1.08}
\begin{longtable}{@{}p{\dimexpr0.045\linewidth-\tabcolsep\relax}p{\dimexpr0.135\linewidth-2\tabcolsep\relax}p{\dimexpr0.22\linewidth-2\tabcolsep\relax}p{\dimexpr0.06\linewidth-2\tabcolsep\relax}p{\dimexpr0.1\linewidth-2\tabcolsep\relax}p{\dimexpr0.14\linewidth-2\tabcolsep\relax}p{\dimexpr0.14\linewidth-2\tabcolsep\relax}p{\dimexpr0.16\linewidth-\tabcolsep\relax}@{}}
\caption{Task metadata.  The four-dimensional descriptive annotations are inspired by the literature on motor skills~\cite {muratori2013motor} and material distinctions~\cite{arriolaRios2020deformable}.}\label{tab:appendix-formative-dataset} \\
\toprule
\raggedright \textbf{ID} & \raggedright \textbf{Task / source} & \raggedright \textbf{Procedure scope} & \raggedright \textbf{Steps} & \raggedright \textbf{Motor involvement} & \raggedright \textbf{Temporal organization} & \raggedright \textbf{Object deformability} & \raggedright \textbf{Environmental predictability} \tabularnewline
\midrule
\endfirsthead
\caption[]{Task metadata (continued).} \\
\toprule
\raggedright \textbf{ID} & \raggedright \textbf{Task / source} & \raggedright \textbf{Procedure scope} & \raggedright \textbf{Steps} & \raggedright \textbf{Motor involvement} & \raggedright \textbf{Temporal organization} & \raggedright \textbf{Object deformability} & \raggedright \textbf{Environmental predictability} \tabularnewline
\midrule
\endhead
\midrule
\multicolumn{8}{r}{\footnotesize\itshape Continued on next page} \\
\endfoot
\bottomrule
\endlastfoot
\raggedright T1 & \raggedright \href{https://youtu.be/Q2gClH258iU}{LEGO} & \raggedright Assemble a camera through stud alignment and attachment & \raggedright 25 & \raggedright Fine & \raggedright Serial & \raggedright Rigid & \raggedright Closed \tabularnewline
\addlinespace[1pt]
\raggedright T2 & \raggedright \href{https://youtu.be/aAxGTnVNJiE}{Crochet} & \raggedright Manipulate yarn loops into a slip knot and chain stitches & \raggedright 10 & \raggedright Fine & \raggedright Serial & \raggedright Deformable & \raggedright Closed \tabularnewline
\addlinespace[1pt]
\raggedright T3 & \raggedright \href{https://youtu.be/JaYPSCEGWfQ}{Matcha latte} & \raggedright Measure, transfer, and mix drink ingredients & \raggedright 7 & \raggedright Mixed & \raggedright Serial; continuous actions & \raggedright Mixed & \raggedright Closed \tabularnewline
\addlinespace[1pt]
\raggedright T4 & \raggedright \href{https://youtu.be/GC_Szxdqh2Y}{Origami} & \raggedright Crease and fold layered paper into a crane & \raggedright 45 & \raggedright Fine & \raggedright Serial & \raggedright Deformable & \raggedright Closed \tabularnewline
\addlinespace[1pt]
\raggedright T5 & \raggedright \href{https://youtu.be/VJc3BxrRdFE}{Furniture assembly} & \raggedright Assemble a chair using fasteners and mating frame joints & \raggedright 28 & \raggedright Mixed & \raggedright Serial & \raggedright Mixed & \raggedright Closed \tabularnewline
\addlinespace[1pt]
\raggedright T6 & \raggedright \href{https://youtu.be/l_pp-1qu9Ig}{Gift wrapping} & \raggedright Measure, fold, and secure paper around a rigid box & \raggedright 24 & \raggedright Mixed & \raggedright Serial & \raggedright Mixed & \raggedright Closed \tabularnewline
\addlinespace[1pt]
\raggedright T7 & \raggedright \href{https://youtu.be/dCGS067s0zo}{Cutting vegetables} & \raggedright Coordinate knife cuts and supporting grip to dice an onion & \raggedright 5 & \raggedright Mixed & \raggedright Serial & \raggedright Deformable & \raggedright Closed \tabularnewline
\addlinespace[1pt]
\raggedright T8 & \raggedright \href{https://youtu.be/jcQvBnM370I}{Packing a box} & \raggedright Arrange and cushion contents, fill voids, and seal & \raggedright 7 & \raggedright Mixed & \raggedright Serial & \raggedright Mixed & \raggedright Closed \tabularnewline
\addlinespace[1pt]
\raggedright T9 & \raggedright \href{https://youtu.be/M8DNQvyGnf0}{Tying shoelaces} & \raggedright Cross, loop, and tighten flexible laces into a knot & \raggedright 4 & \raggedright Fine & \raggedright Serial & \raggedright Deformable & \raggedright Closed \tabularnewline
\addlinespace[1pt]
\raggedright T10 & \raggedright \href{https://youtu.be/KdojWs2nldY}{Folding a T-shirt} & \raggedright Fold unsupported fabric using crossed-arm pinches and a flip & \raggedright 7 & \raggedright Mixed & \raggedright Serial & \raggedright Deformable & \raggedright Closed \tabularnewline
\addlinespace[1pt]
\raggedright T11 & \raggedright \href{https://youtu.be/s8LAtR8pmDg}{Circuit wiring} & \raggedright Connect and power an LED circuit through pin and polarity alignment & \raggedright 6 & \raggedright Fine & \raggedright Serial & \raggedright Mixed & \raggedright Closed \tabularnewline
\addlinespace[1pt]
\raggedright T12 & \raggedright \href{https://youtu.be/I8_9JE_PuyA}{Sandwich assembly} & \raggedright Cut, spread, and layer food to cover a sandwich roll & \raggedright 11 & \raggedright Mixed & \raggedright Serial; continuous actions & \raggedright Deformable & \raggedright Closed \tabularnewline
\addlinespace[1pt]
\raggedright T13 & \raggedright \href{https://youtu.be/kDOGb9C5kp0}{Jump rope} & \raggedright Size a rope and coordinate repeated hops, rotation, and posture & \raggedright 9 & \raggedright Gross & \raggedright Continuous; discrete setup & \raggedright Mixed & \raggedright Closed \tabularnewline
\addlinespace[1pt]
\raggedright T14 & \raggedright \href{https://youtu.be/Kp2PlwLrlic}{Yoga flow} & \raggedright Link floor-based poses through coordinated body transitions & \raggedright 6 & \raggedright Gross & \raggedright Serial & \raggedright N/A & \raggedright Closed \tabularnewline
\addlinespace[1pt]
\raggedright T15 & \raggedright \href{https://youtu.be/yyQ-v4V3NU8}{Tennis swing} & \raggedright Establish stance, turn, and execute a bounded forehand swing & \raggedright 4 & \raggedright Gross & \raggedright Discrete; serial preparation & \raggedright Rigid & \raggedright Closed \tabularnewline
\addlinespace[1pt]
\midrule
\multicolumn{3}{@{}l}{\textbf{Total (15 tasks)}} & \textbf{198} & & & & \\
\end{longtable}
\endgroup
\subsection{Codebook and Visual Examples}
\label{sec:appendix-codebook}
\label{sec:appendix-formative-examples}
\label{sec:appendix-formative-codebook}
\label{sec:appendix-formative-task-examples}

Table~\ref{tab:appendix-codebook-issues} combines the 27 retained issue codes with visual examples and operational explanations.
Image pairs show the source frame and generated output; video sequences show sampled frames with timestamps.
Examples cover all 15 tasks.
E1--E6 denote mapped usefulness threats, not positive artifact scores, and are defined at the end of the table.

\begingroup
\newcommand{\probePair}[4]{%
\begin{tabular}[t]{@{}p{.485\linewidth}@{\hspace{.03\linewidth}}p{.485\linewidth}@{}}%
\centering\scriptsize #3 & \centering\scriptsize #4\tabularnewline[-2pt]
\includegraphics[width=\linewidth]{#1} & \includegraphics[width=\linewidth]{#2}
\end{tabular}}
\newcommand{\probeTriple}[6]{%
\begin{tabular}[t]{@{}p{.32\linewidth}@{\hspace{.02\linewidth}}p{.32\linewidth}@{\hspace{.02\linewidth}}p{.32\linewidth}@{}}%
\centering\scriptsize #4 & \centering\scriptsize #5 & \centering\scriptsize #6\tabularnewline[-2pt]
\includegraphics[width=\linewidth]{#1} & \includegraphics[width=\linewidth]{#2} & \includegraphics[width=\linewidth]{#3}
\end{tabular}}
\newcommand{\probeQuad}[8]{%
\begin{tabular}[t]{@{}p{.235\linewidth}@{\hspace{.02\linewidth}}p{.235\linewidth}@{\hspace{.02\linewidth}}p{.235\linewidth}@{\hspace{.02\linewidth}}p{.235\linewidth}@{}}%
\centering\scriptsize #5 & \centering\scriptsize #6 & \centering\scriptsize #7 & \centering\scriptsize #8\tabularnewline[-2pt]
\includegraphics[width=\linewidth]{#1} & \includegraphics[width=\linewidth]{#2} & \includegraphics[width=\linewidth]{#3} & \includegraphics[width=\linewidth]{#4}
\end{tabular}}

\begingroup
\footnotesize
\setlength{\tabcolsep}{4pt}
\renewcommand{\arraystretch}{1.08}
\begin{longtable}{@{}p{\dimexpr.19\linewidth-\tabcolsep\relax}p{\dimexpr.46\linewidth-2\tabcolsep\relax}p{\dimexpr.35\linewidth-\tabcolsep\relax}@{}}
\caption{Formative evaluation codebook with source--output image pairs and sampled video frames.}
\label{tab:appendix-codebook-issues}
\label{tab:appendix-codebook-e}
\label{tab:appendix-formative-examples}
\label{tab:appendix-formative-examples-a}
\label{tab:appendix-formative-examples-b}
\label{tab:appendix-formative-examples-c} \\
\toprule
\textbf{Codebook} & \textbf{Explanation} & \textbf{Visual example} \\
\midrule
\endfirsthead
\caption[]{Formative evaluation codebook with source--output image pairs and sampled video frames (continued).} \\
\toprule
\textbf{Codebook} & \textbf{Explanation} & \textbf{Visual example} \\
\midrule
\endhead
\midrule
\multicolumn{3}{r}{\footnotesize\itshape Continued on next page} \\
\endfoot
\bottomrule
\endlastfoot
\multicolumn{3}{@{}l}{\textbf{A. Interpretation and grounding}} \\*
\raggedright \textbf{A1} Semantic overreach or intrusion\newline\textit{E1, E3} & \raggedright An unsupported object, person, part, or semantic element appears.\par\smallskip\textit{Example:} LEGO, step 4 (Gemini 3.1 Flash Image). The clip brick appears on an additional black assembly rather than only modifying the existing one. & \raggedright \probePair{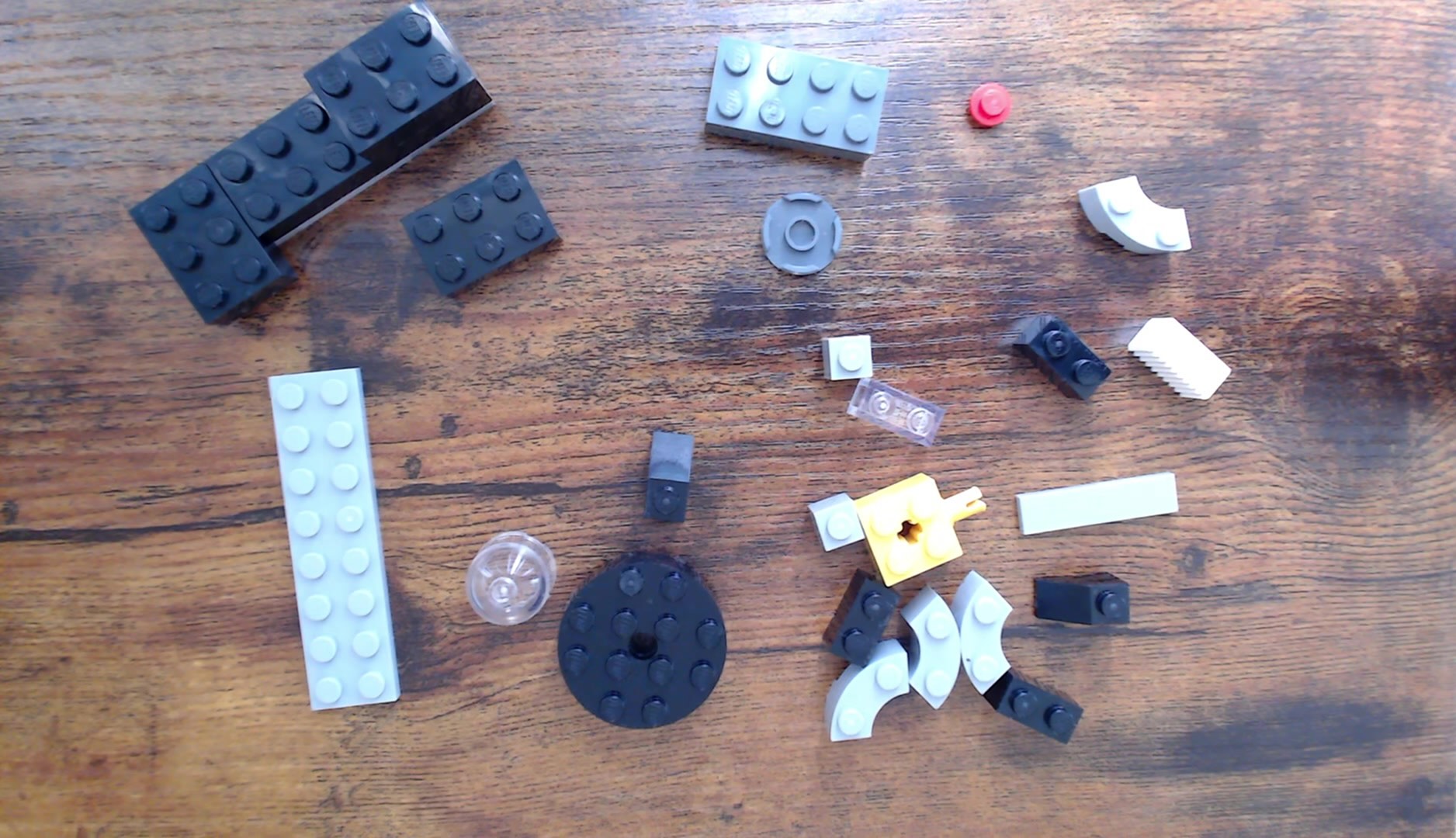}{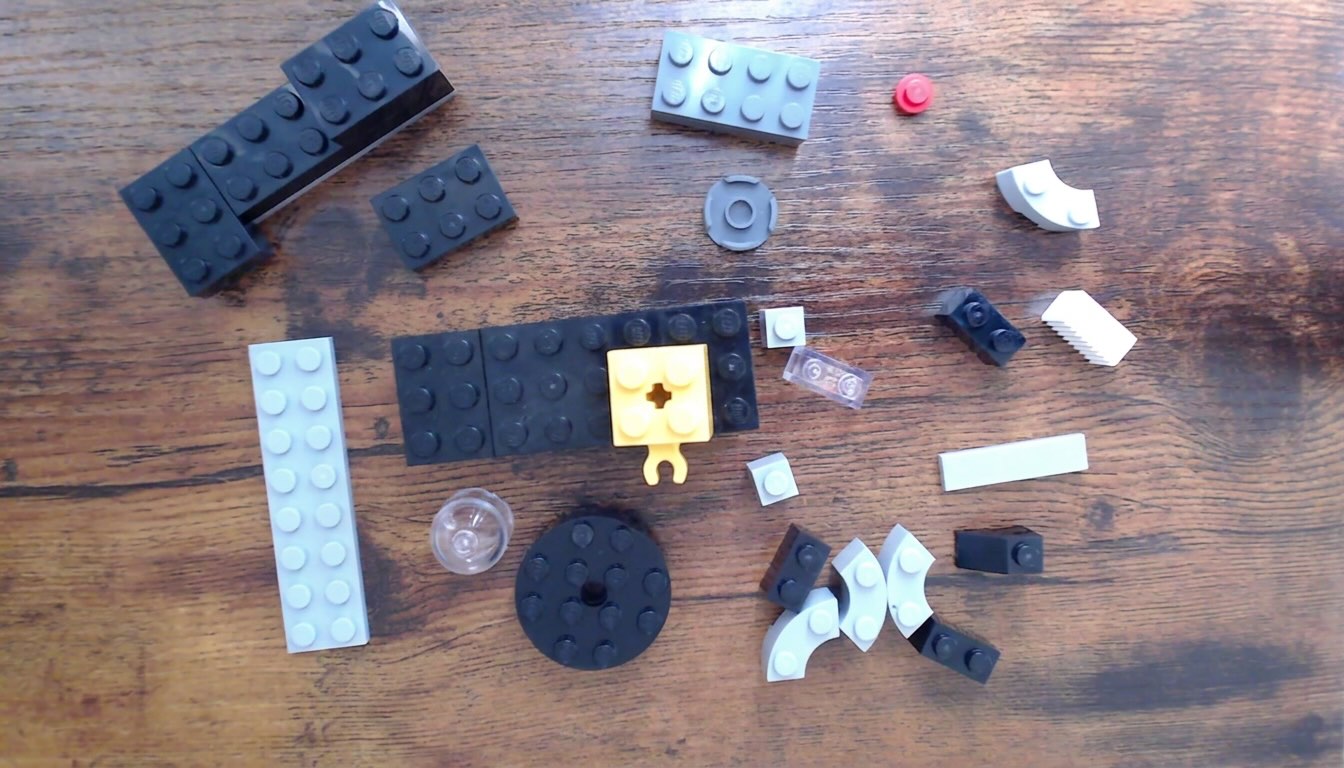}{Source}{Generated} \tabularnewline
\addlinespace[4pt]
\raggedright \textbf{A2} Ambiguity-resolution failure\newline\textit{E1, E2} & \raggedright The artifact adopts an unsupported resolution of ambiguity in the instruction or scene.\par\smallskip\textit{Example:} Packing a box, step 7 (Gemini 3.1 Flash Image). The tape placement resolves an underspecified sealing instruction in a way coded as unsupported. & \raggedright \probePair{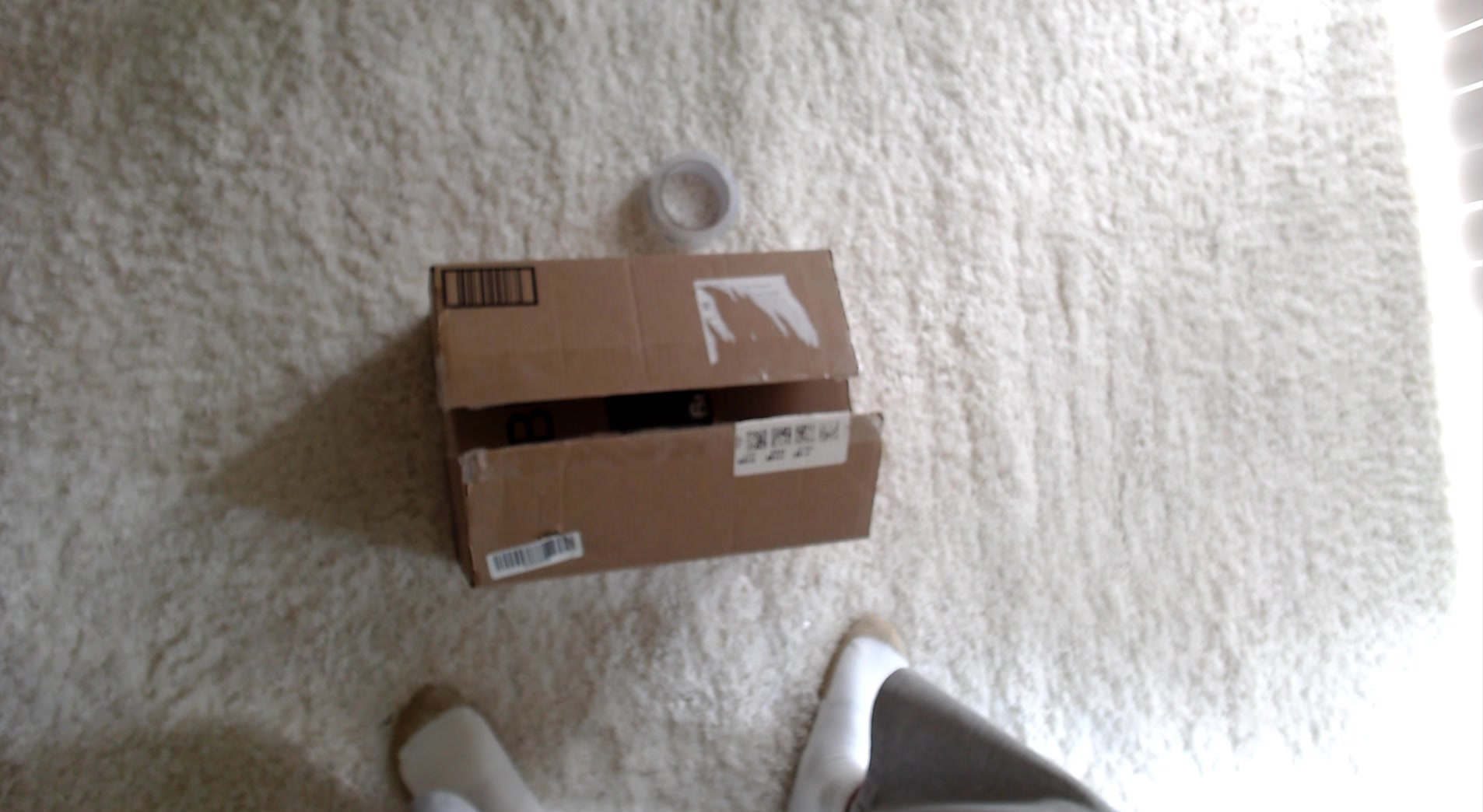}{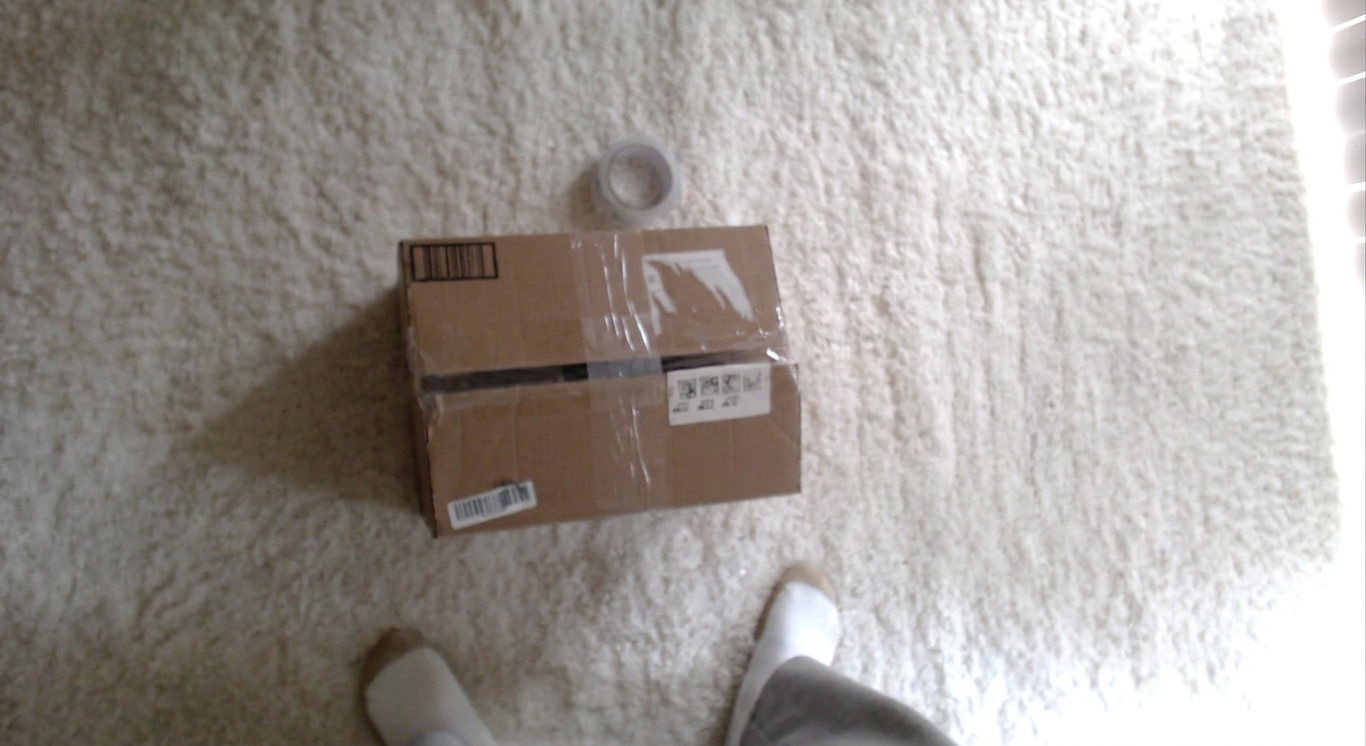}{Source}{Generated} \tabularnewline
\addlinespace[4pt]
\raggedright \textbf{A3} Hidden-state inference error\newline\textit{E1, E3} & \raggedright An occluded or latent property, such as orientation, connectivity, or a hidden surface, is inferred incorrectly.\par\smallskip\textit{Example:} Furniture assembly, step 6 (Gemini 3.1 Flash Image). The hidden fastening arrangement is inferred incorrectly while the bolt is being secured. & \raggedright \probePair{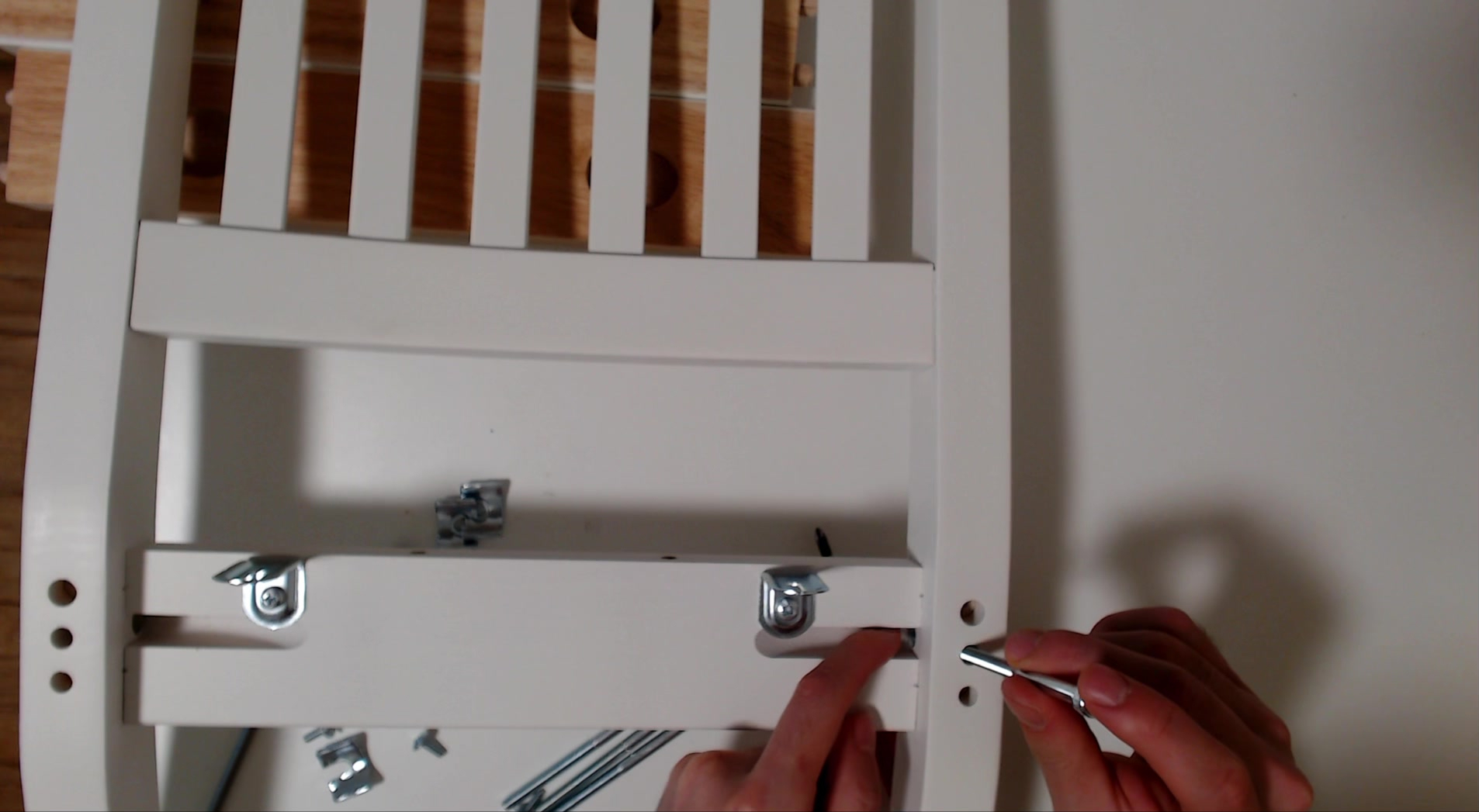}{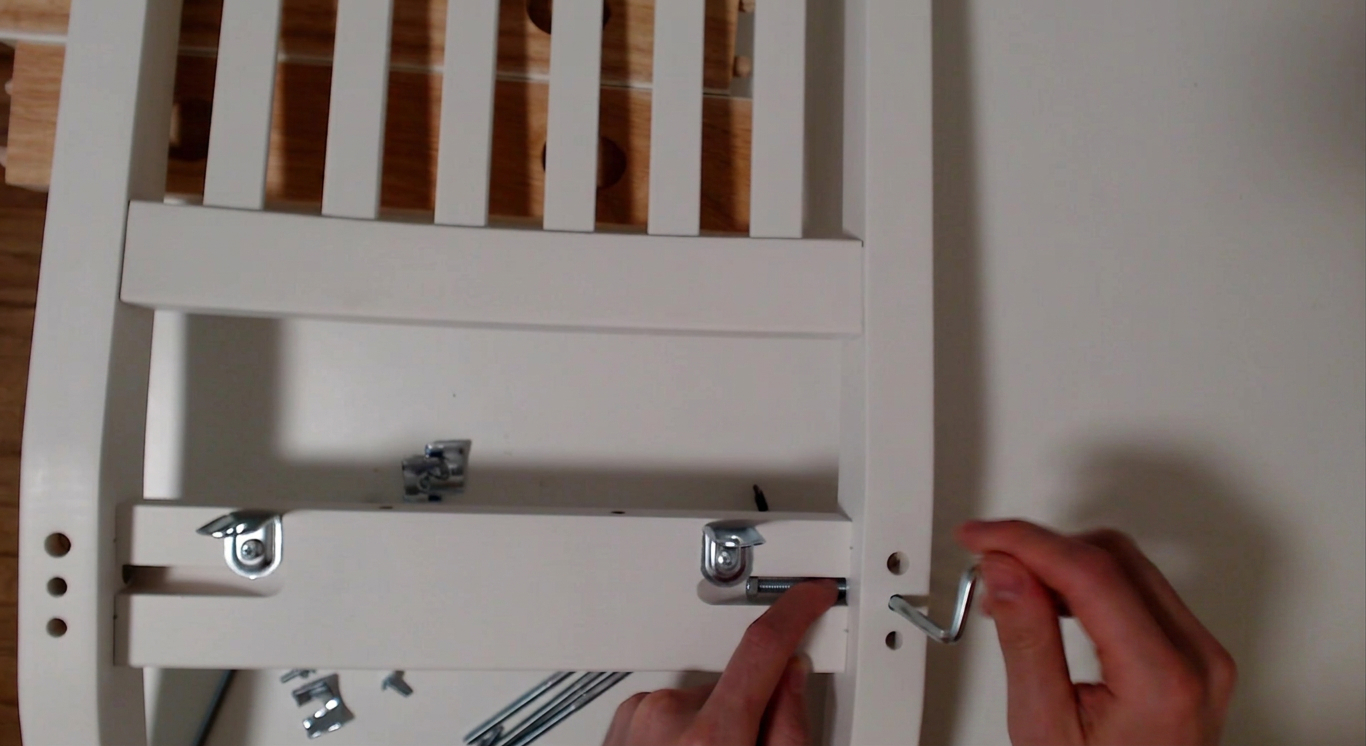}{Source}{Generated} \tabularnewline
\addlinespace[4pt]
\raggedright \textbf{A4} Object or subpart identification error\newline\textit{E1, E2, E3} & \raggedright The artifact acts on, replaces, or emphasizes the wrong existing object or subpart.\par\smallskip\textit{Example:} Matcha latte, step 7 (Gemini 3.1 Flash Image). The green mixture and ice appear in the glass while the intended milk bowl is emptied. & \raggedright \probePair{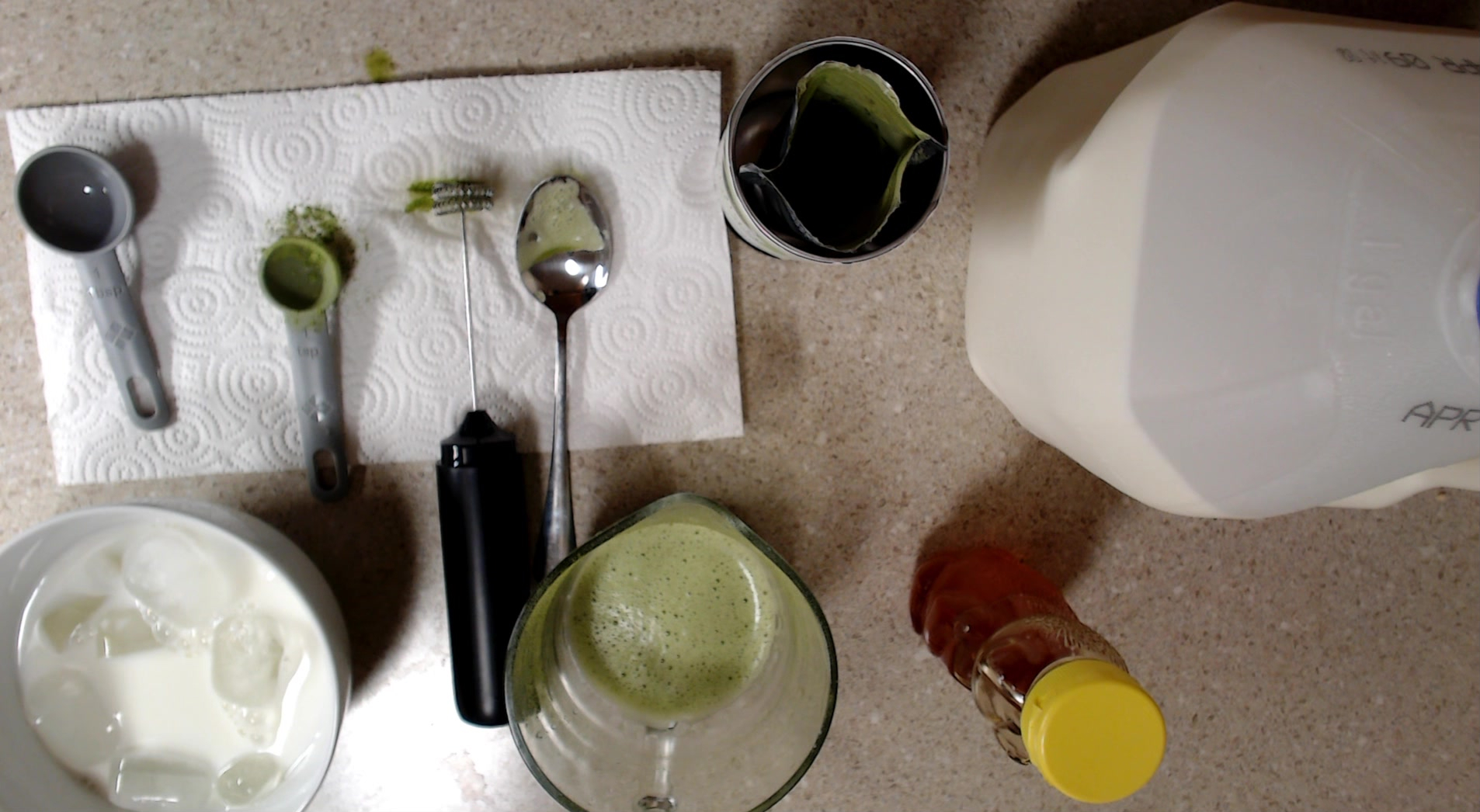}{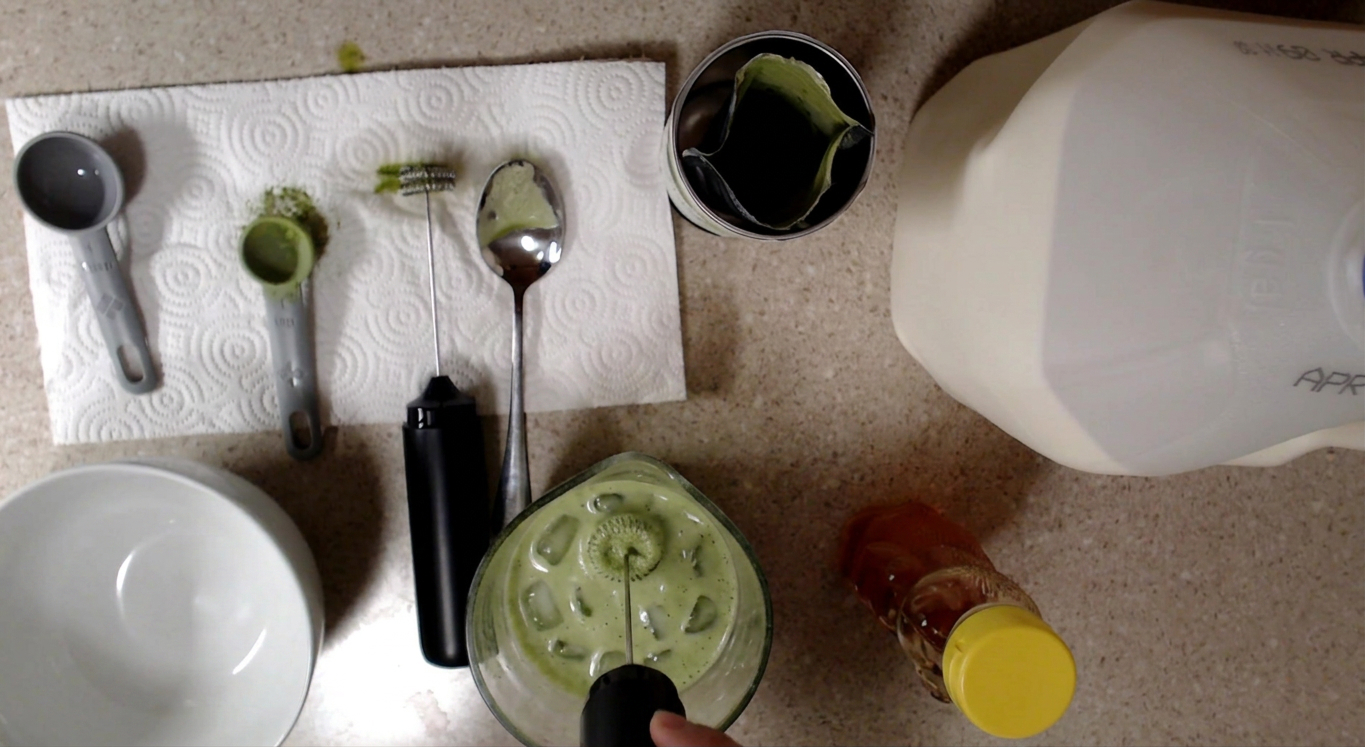}{Source}{Generated} \tabularnewline
\addlinespace[4pt]
\midrule
\multicolumn{3}{@{}l}{\textbf{B. Action and target-state accuracy}} \\*
\raggedright \textbf{B1} Wrong action type\newline\textit{E1, E2} & \raggedright The depicted operation belongs to a different action category from the instruction.\par\smallskip\textit{Example:} LEGO, step 1 (Gemini 3.1 Flash Image). The black brick is joined end-to-end with the gray piece rather than attached on top. & \raggedright \probePair{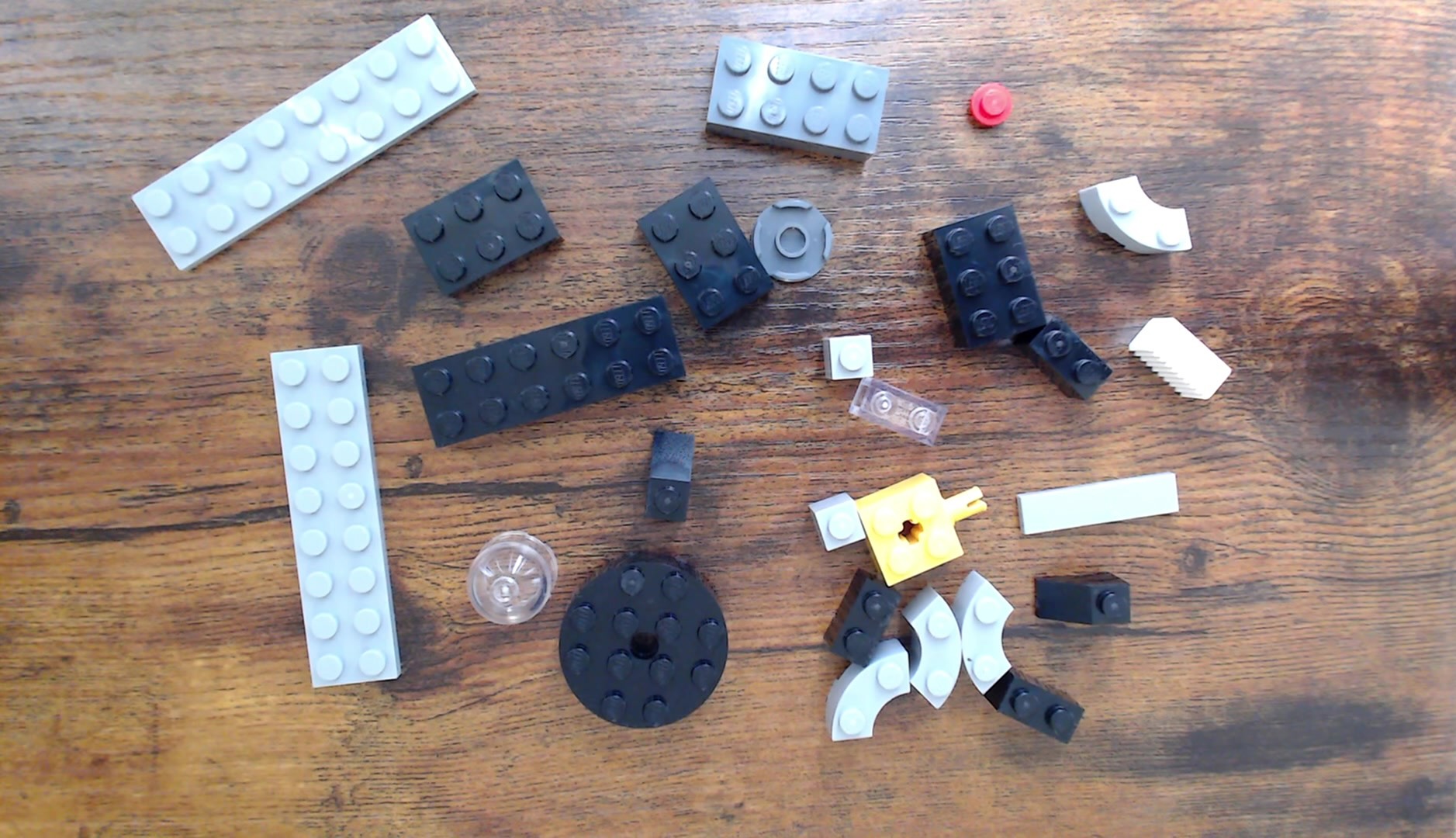}{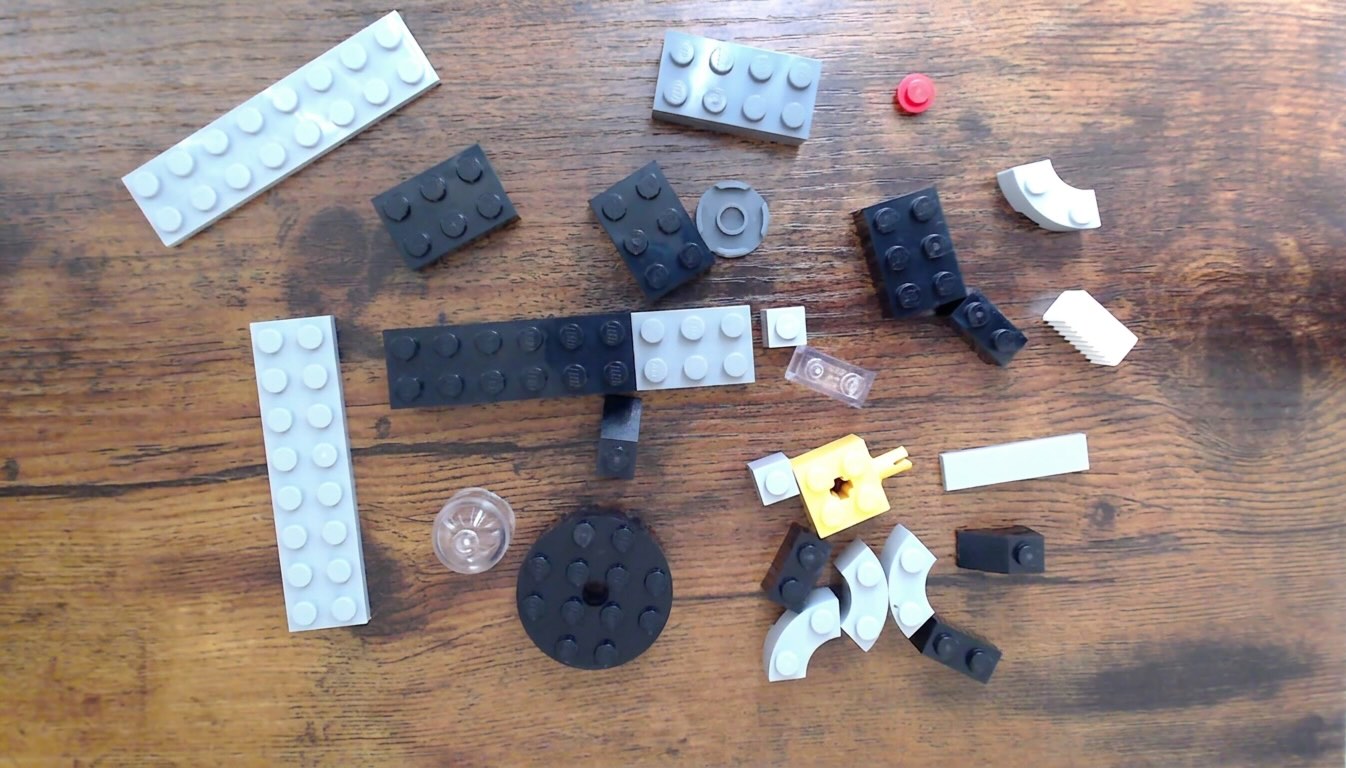}{Source}{Generated} \tabularnewline
\addlinespace[4pt]
\raggedright \textbf{B2} Incomplete action extent\newline\textit{E1, E2} & \raggedright The correct action is partly present but does not reach the required extent or endpoint.\par\smallskip\textit{Example:} Gift wrapping, step 9 (Gemini 3.1 Flash Image). The box is repositioned, but the requested centered placement with equal paper margins is not reached. & \raggedright \probePair{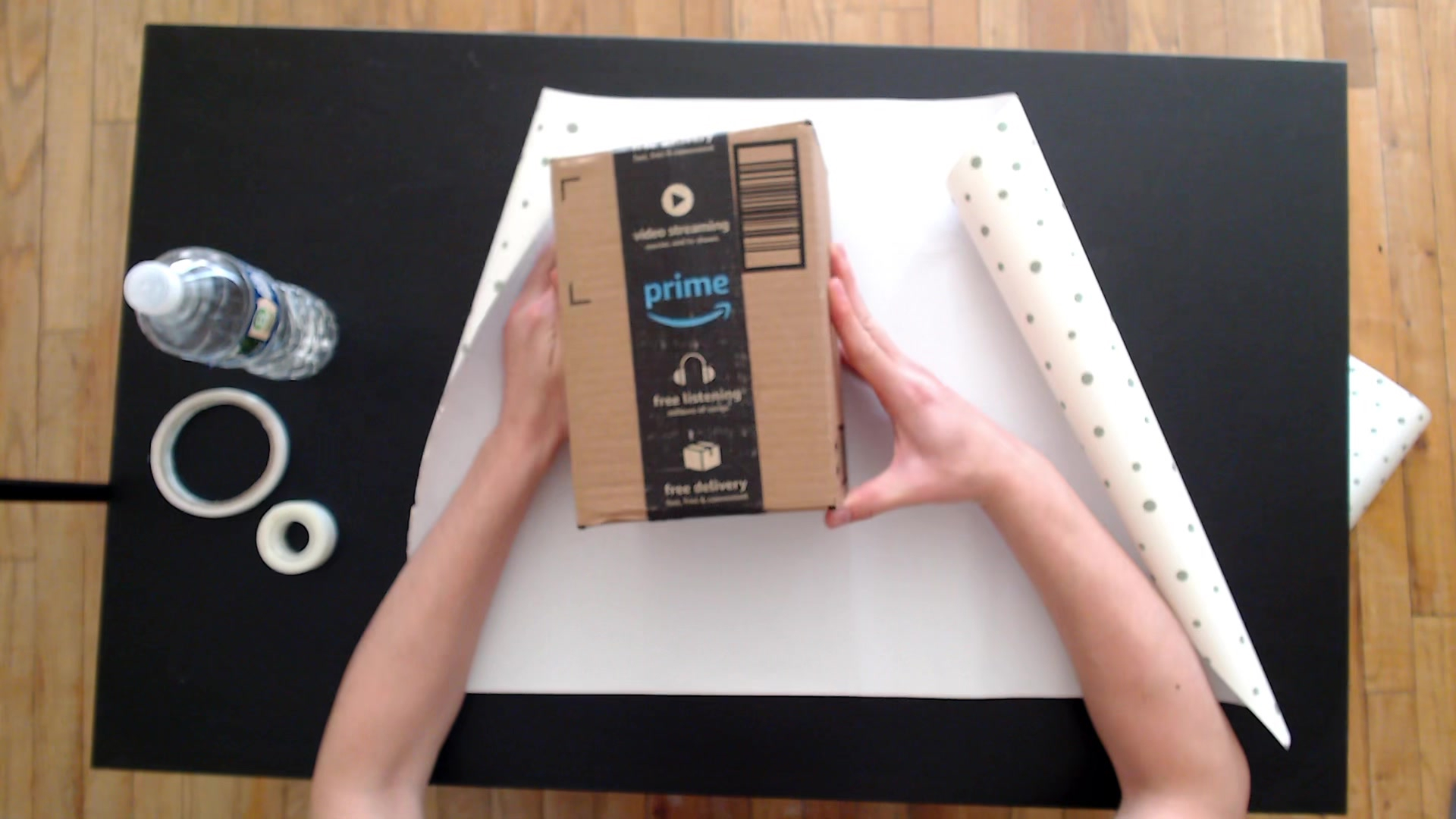}{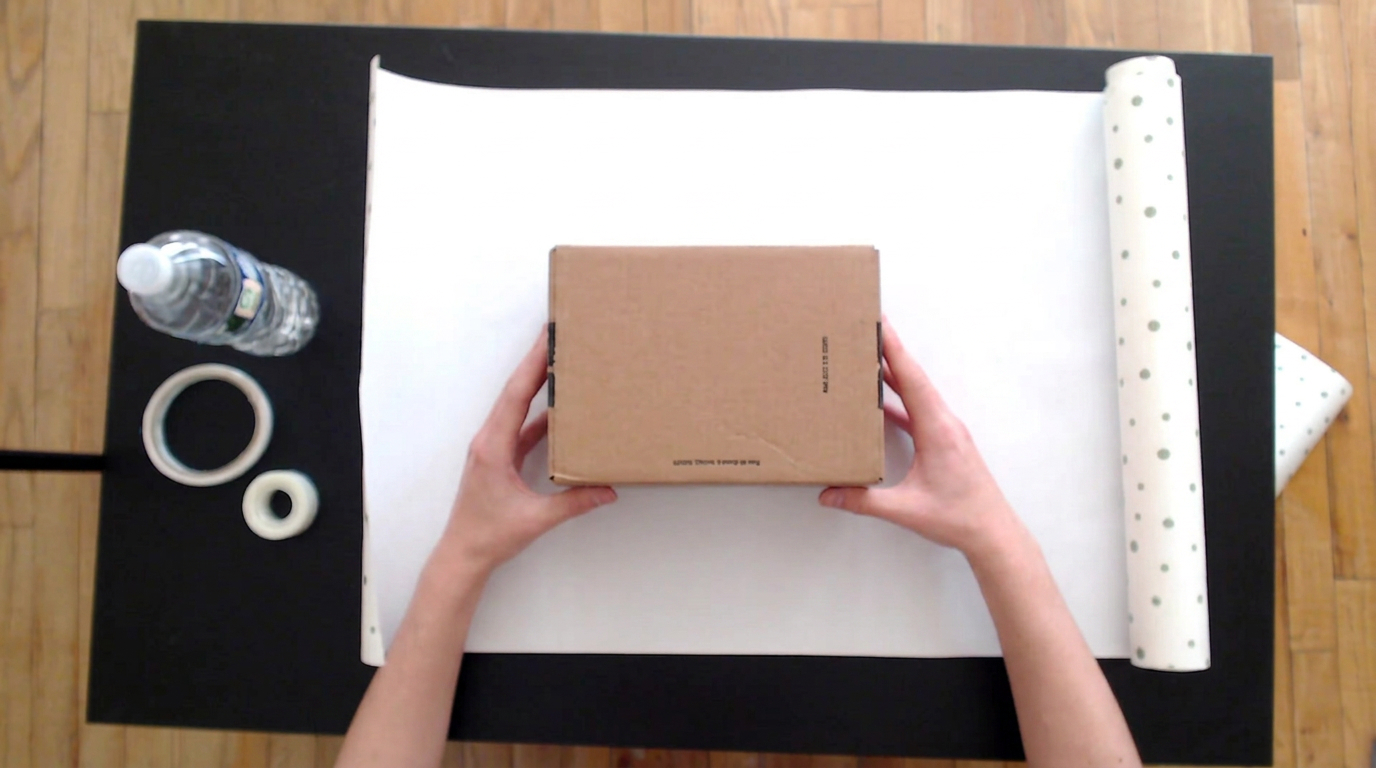}{Source}{Generated} \tabularnewline
\addlinespace[4pt]
\raggedright \textbf{B3} Omitted subaction\newline\textit{E1, E2} & \raggedright A distinct required component of a compound instruction is absent.\par\smallskip\textit{Example:} Jump rope, step 2 (Gemini 3.1 Flash Image). The handles are raised, but the required check against armpit or mid-chest height is omitted. & \raggedright \probePair{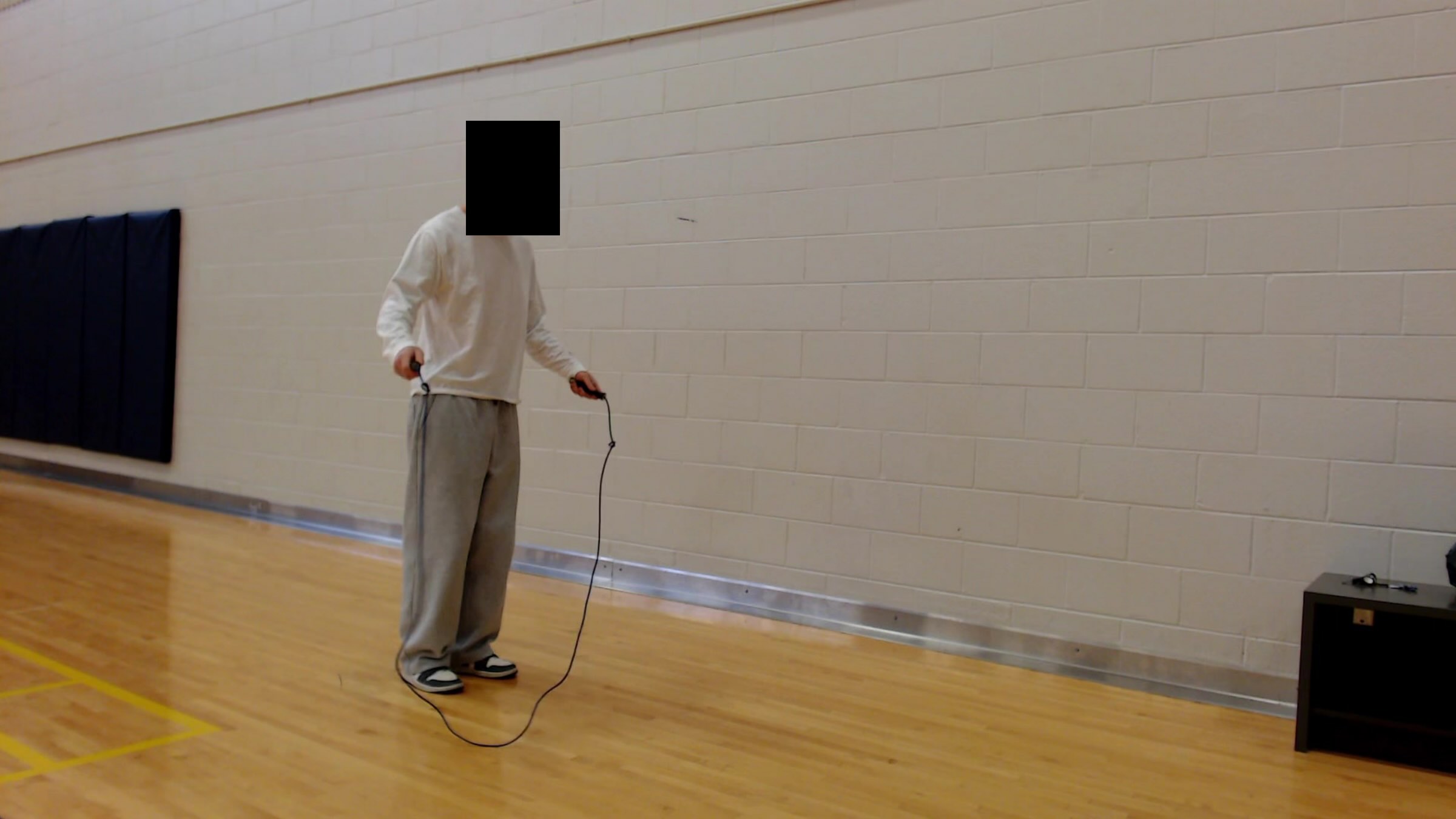}{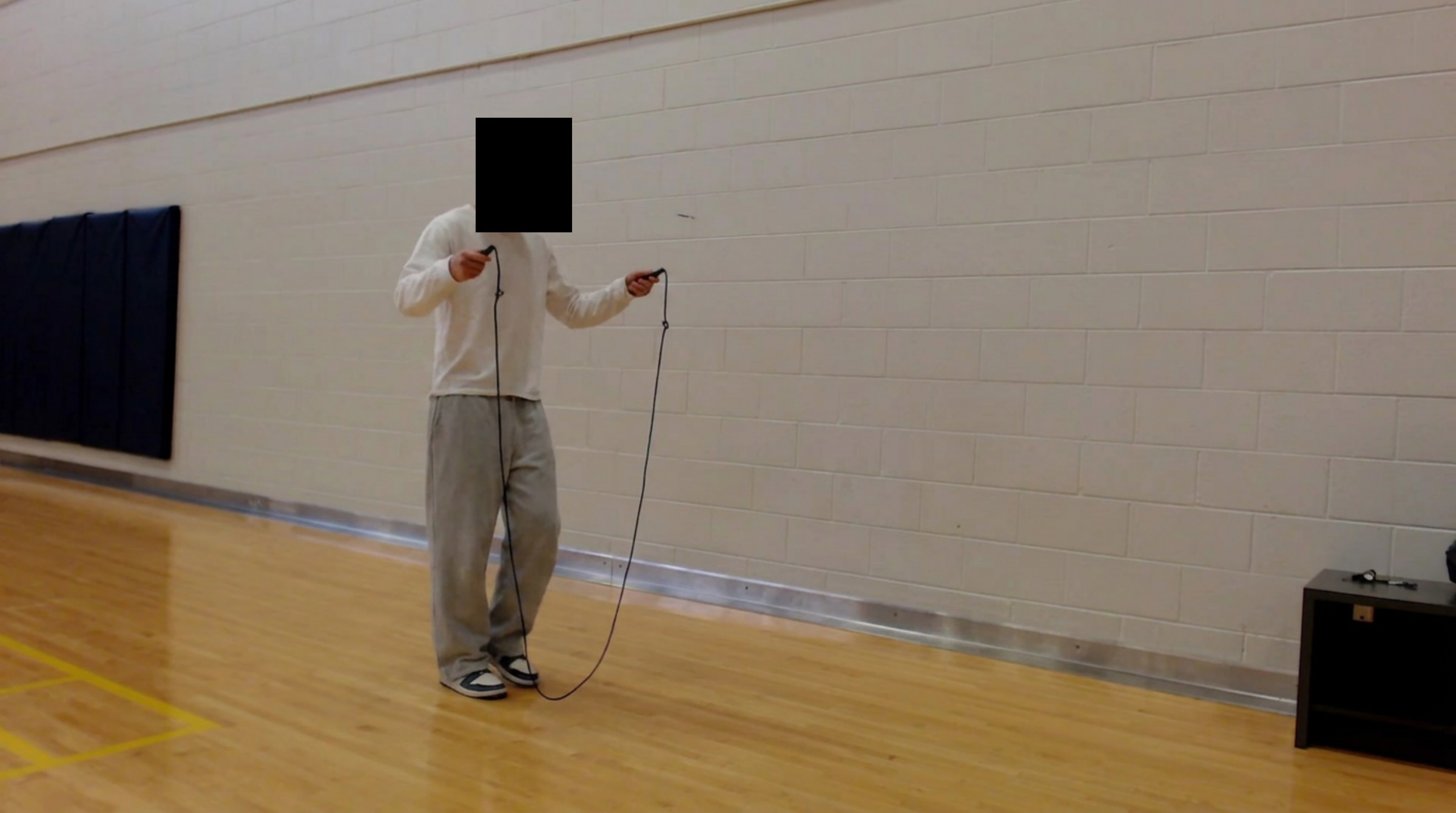}{Source}{Generated} \tabularnewline
\addlinespace[4pt]
\raggedright \textbf{B4} Extra or spurious action\newline\textit{E4} & \raggedright An unrequested action is performed, or the artifact continues into a later step.\par\smallskip\textit{Example:} Yoga flow, step 1 (Gemini Omni Flash Preview). The clip adds further body and foot adjustments beyond establishing the requested tabletop posture. & \raggedright \probeTriple{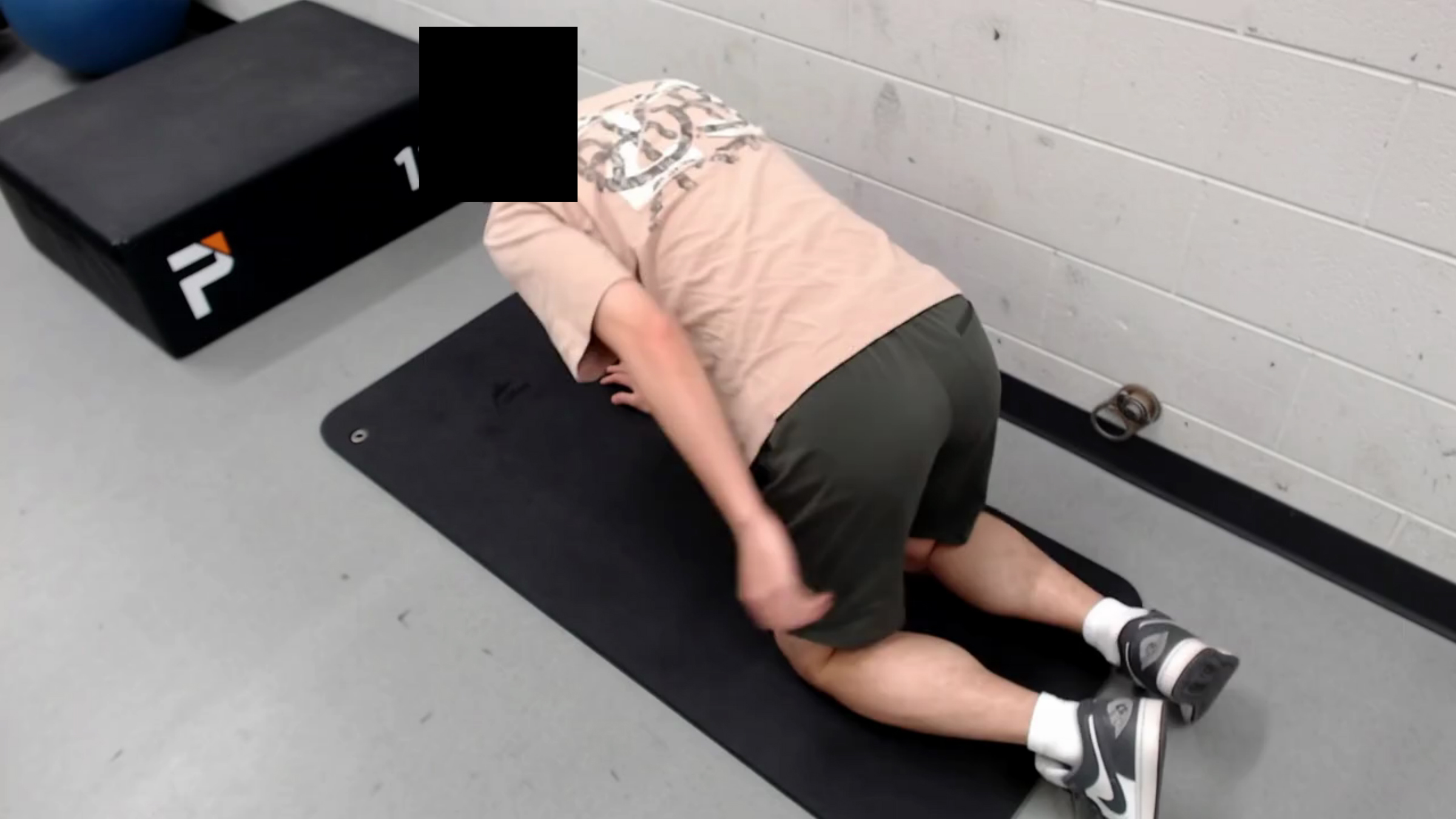}{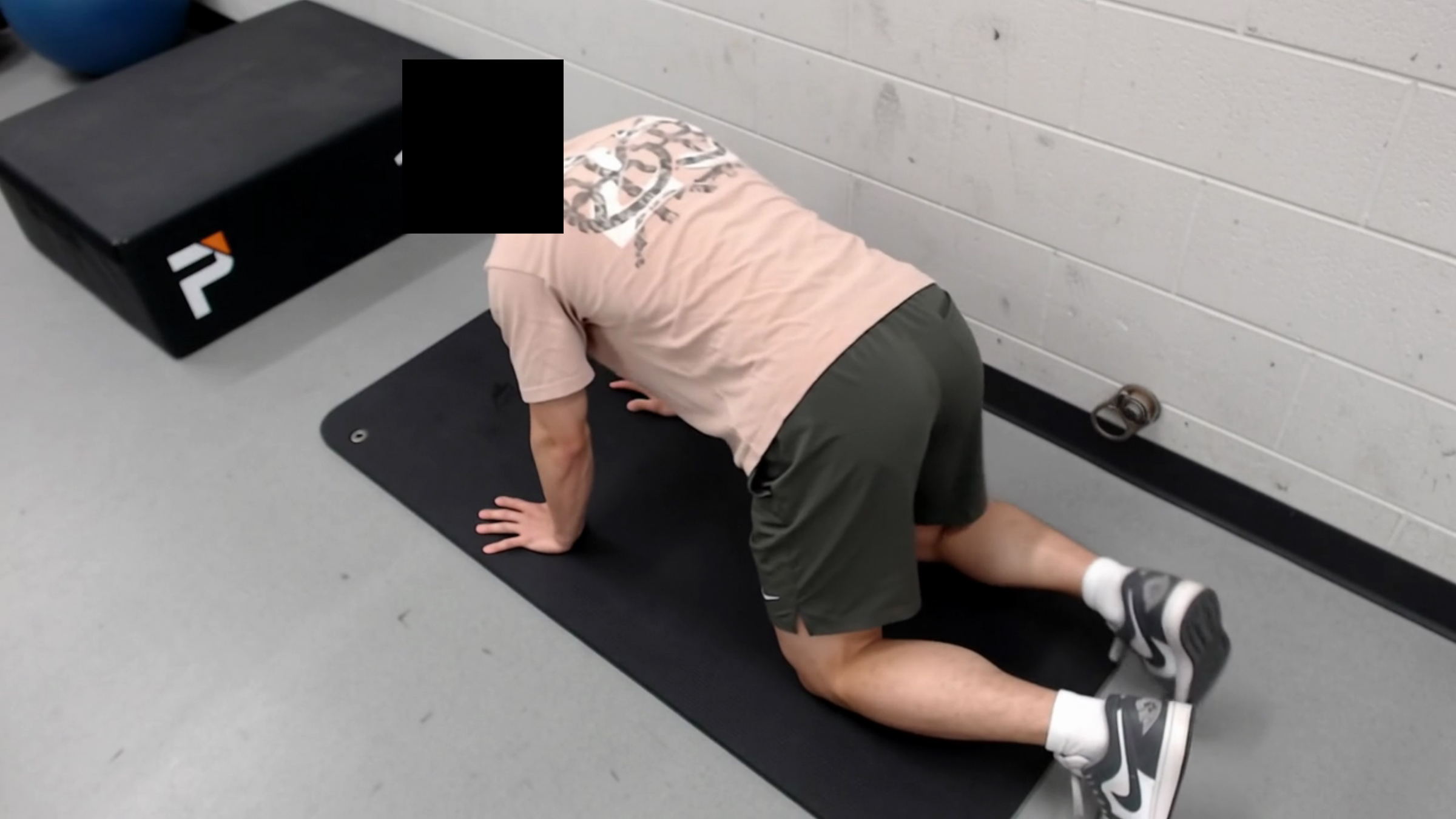}{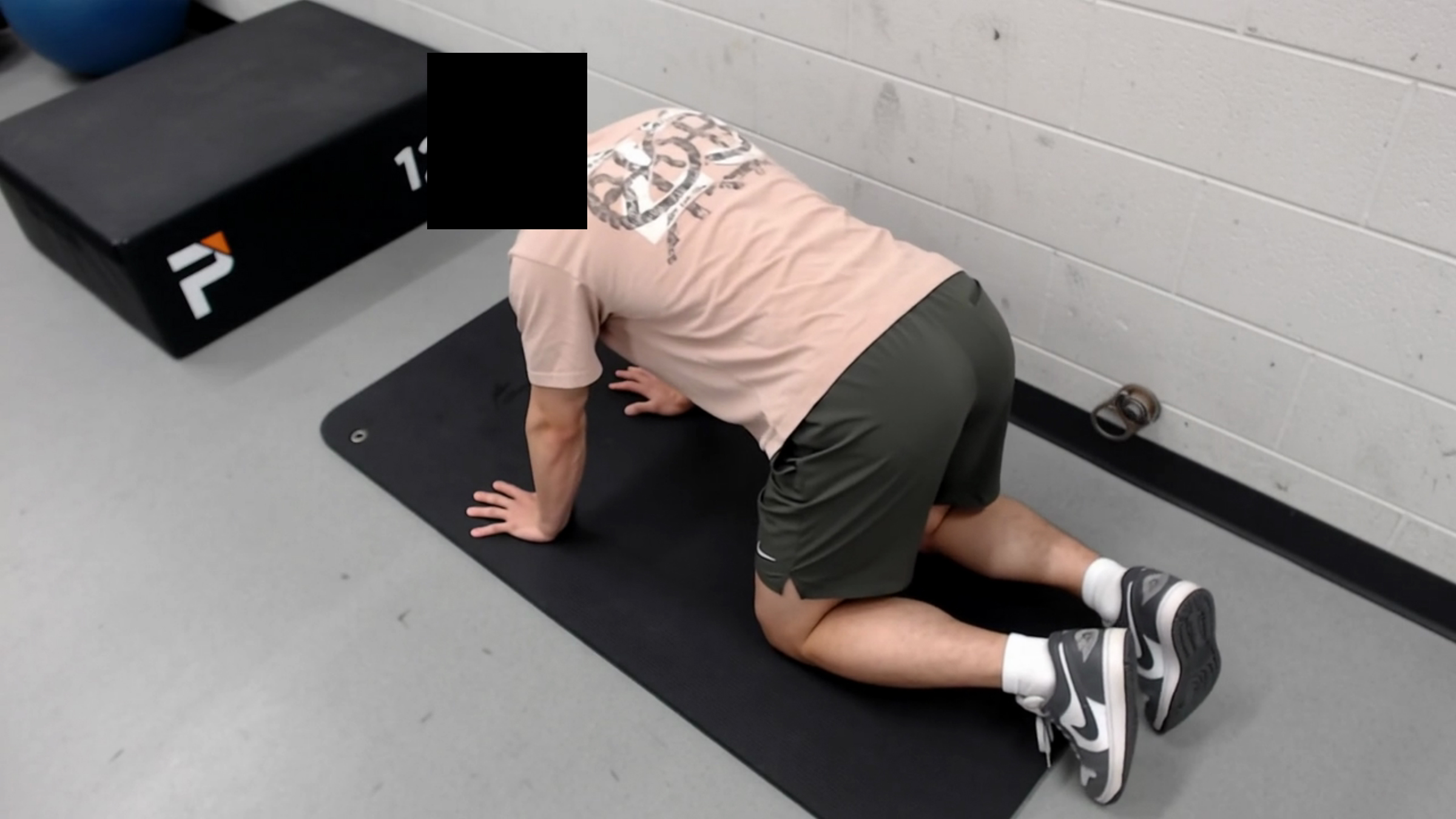}{0.00s/source}{2.50 s}{4.90 s} \tabularnewline
\addlinespace[4pt]
\raggedright \textbf{B5} Imprecise action execution\newline\textit{E1, E2} & \raggedright The action type is correct, but its angle, direction, posture, force, timing, or technique is inaccurate.\par\smallskip\textit{Example:} Crochet, step 7 (Gemini 3.1 Flash Image). The grip and finger placement do not match the specified hold-and-pinch technique. & \raggedright \probePair{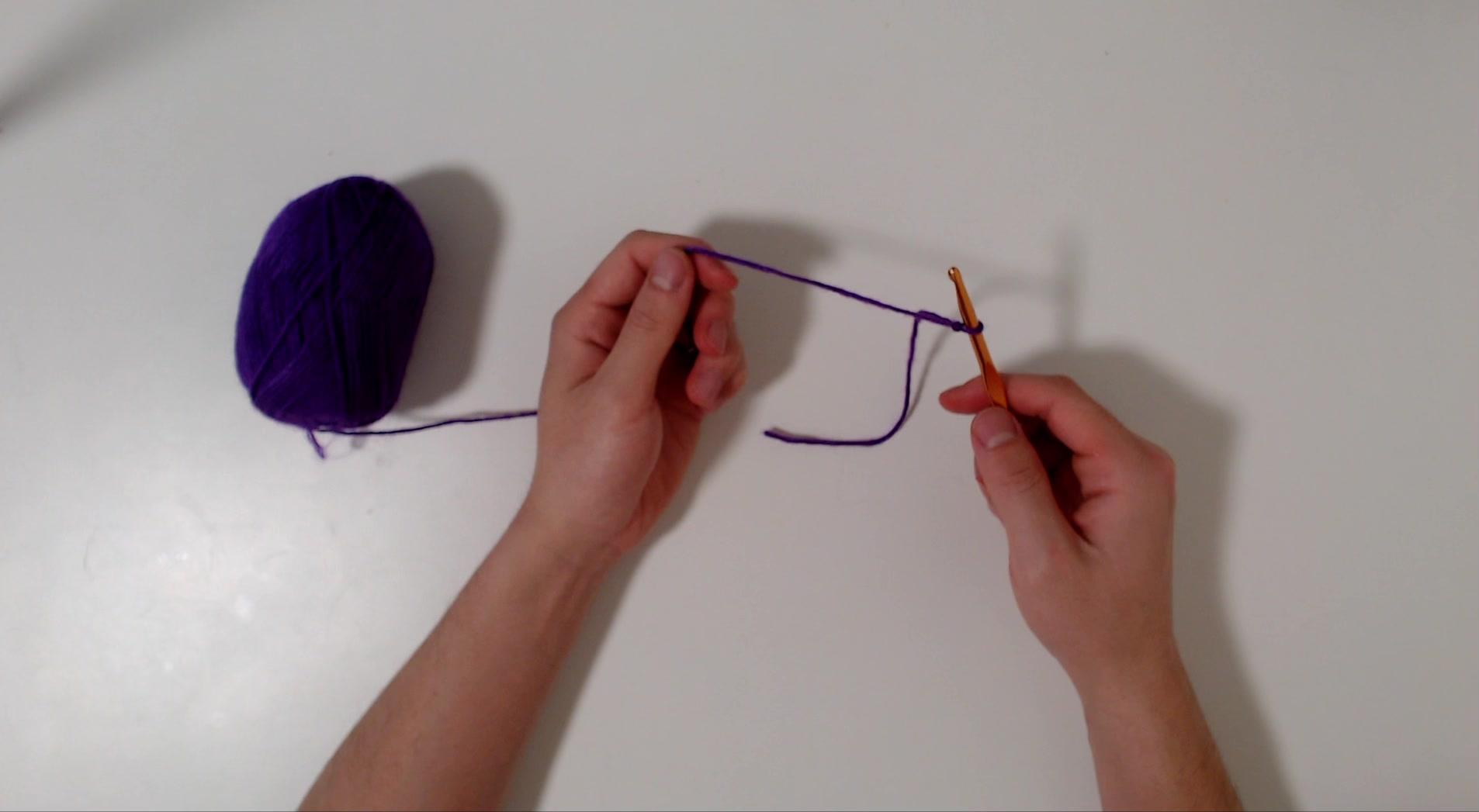}{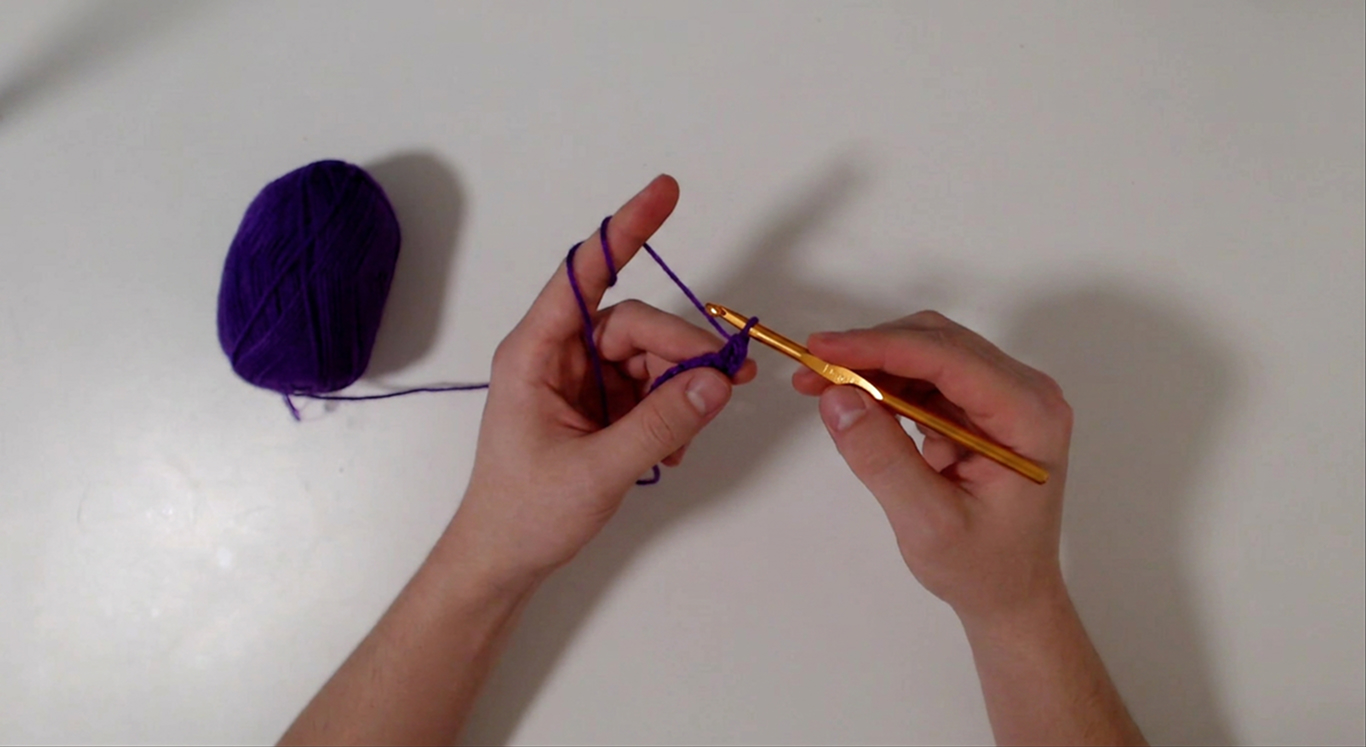}{Source}{Generated} \tabularnewline
\addlinespace[4pt]
\raggedright \textbf{B6} No instructional transformation\newline\textit{E1, E2} & \raggedright The requested transformation is not meaningfully shown.\par\smallskip\textit{Example:} Origami, step 30 (Gemini 3 Pro). The top-layer inward fold is not meaningfully shown; the narrow shape remains essentially unchanged. & \raggedright \probePair{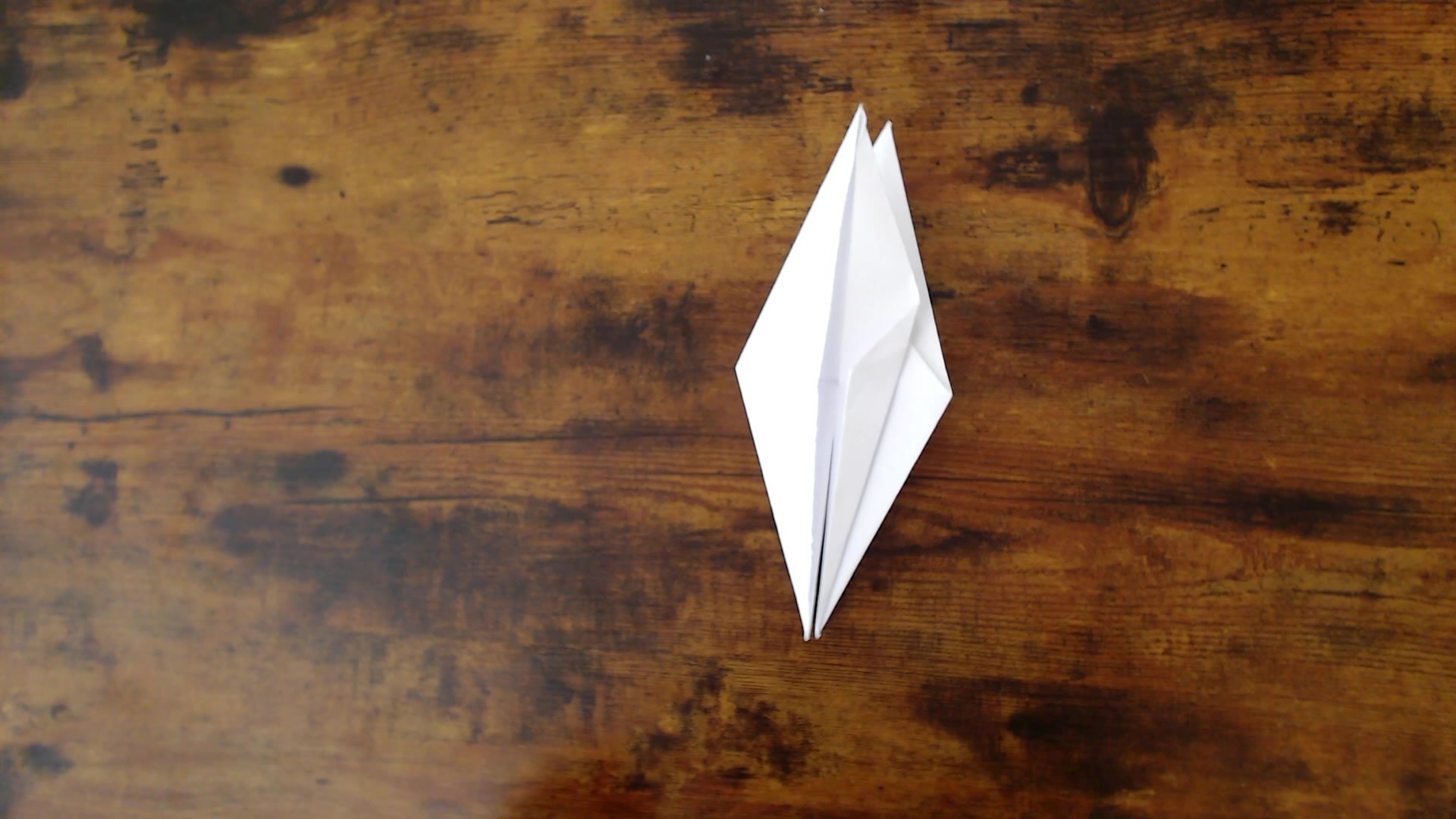}{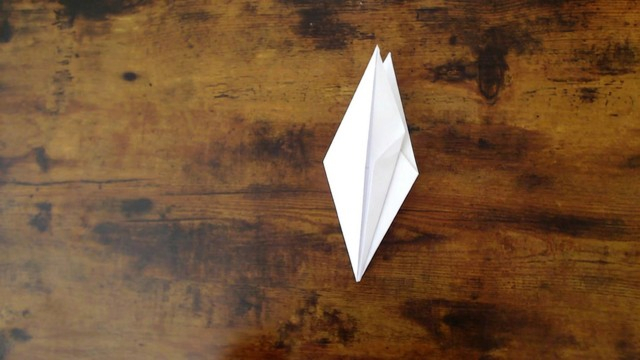}{Source}{Generated} \tabularnewline
\addlinespace[4pt]
\raggedright \textbf{B7} Incorrect target shape or configuration\newline\textit{E1} & \raggedright The resulting geometry, posture, fold, assembly, or overall configuration is incorrect.\par\smallskip\textit{Example:} Folding a T-shirt, step 7 (Gemini 3.1 Flash Image). Instead of one folded shirt, the output shows a long flat garment beside a separate folded form. & \raggedright \probePair{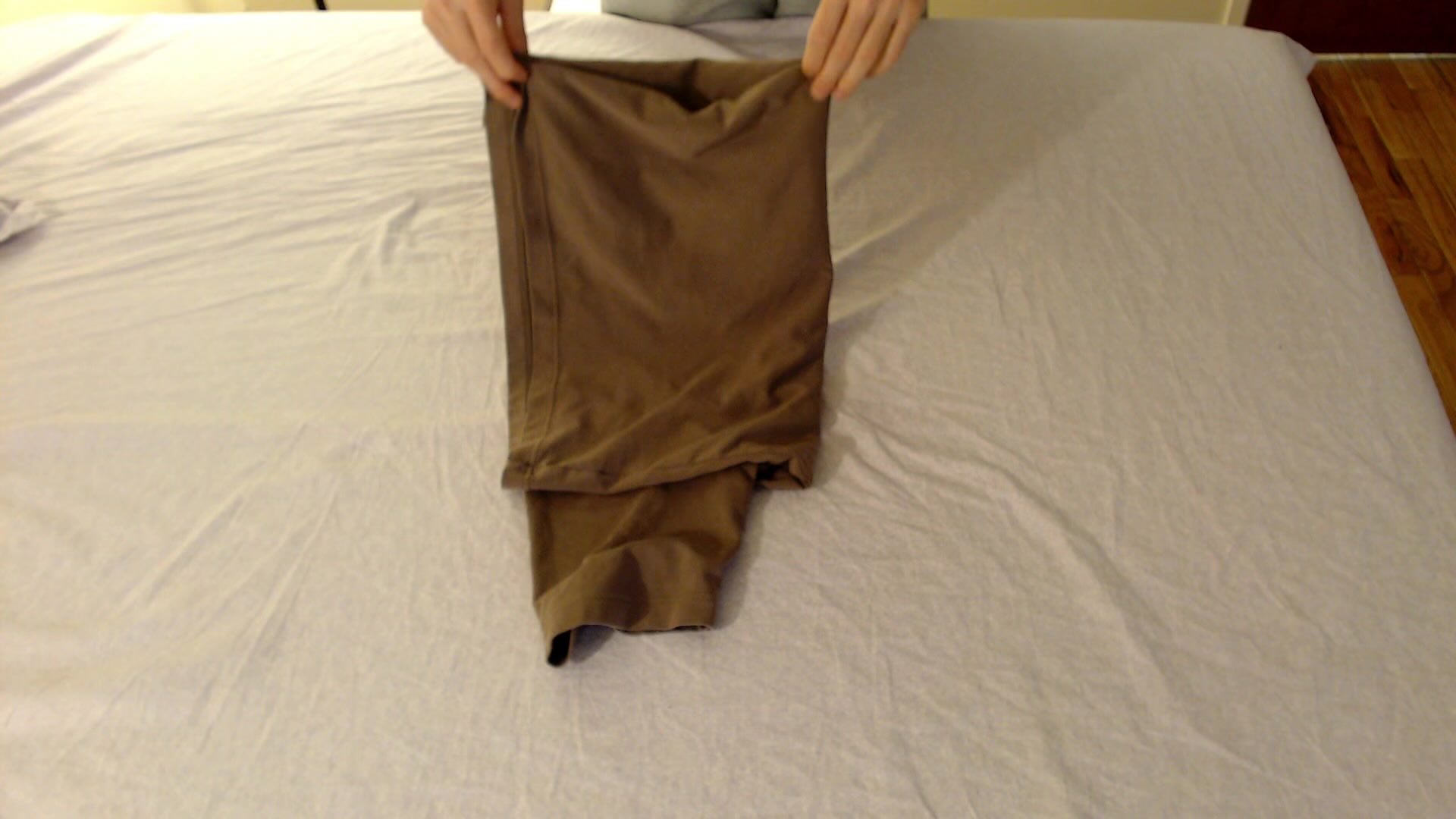}{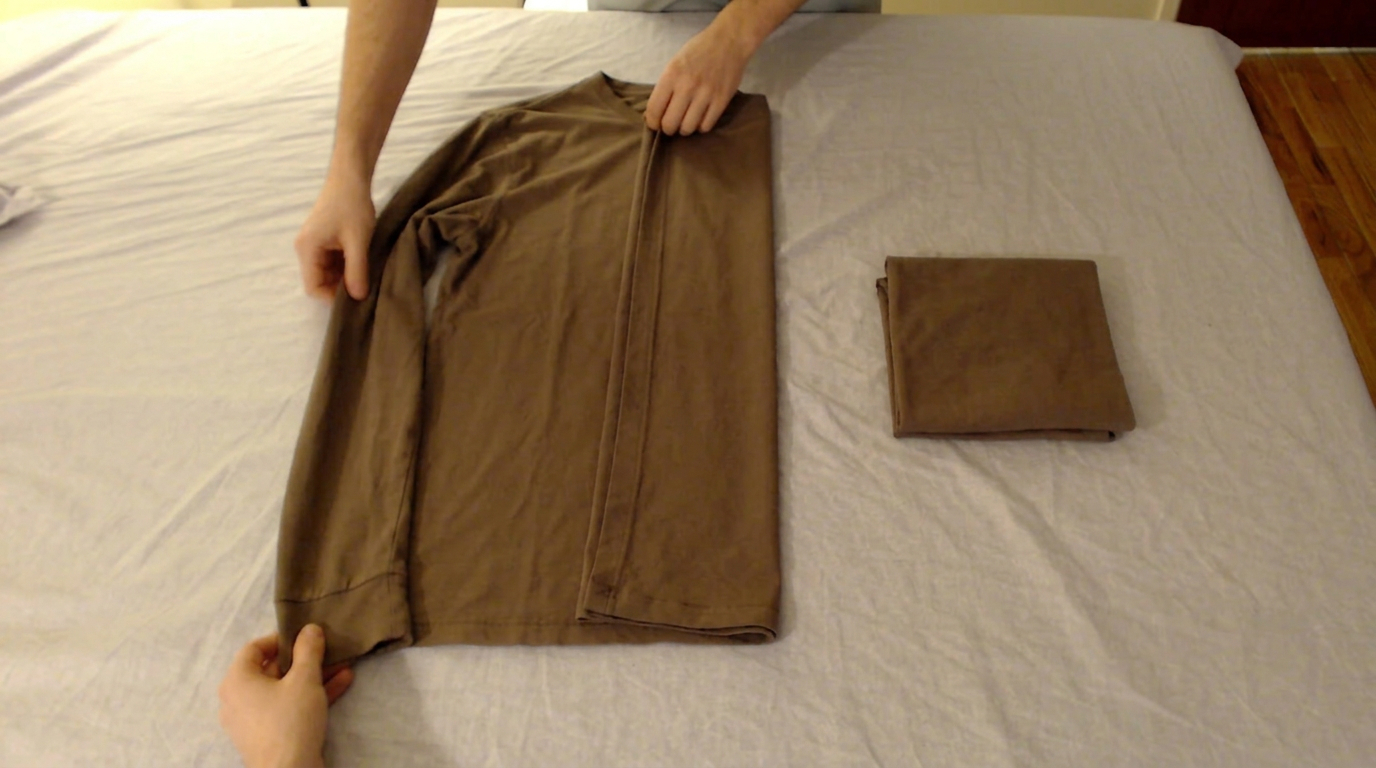}{Source}{Generated} \tabularnewline
\addlinespace[4pt]
\raggedright \textbf{B8} Incorrect spatial relationship\newline\textit{E1, E3} & \raggedright A task entity has the wrong absolute position or the wrong relative, contact, attachment, or connective relation.\par\smallskip\textit{Example:} Cutting vegetables, step 2 (Gemini 3.1 Flash Image). The knife and guiding fingers do not have the instructed curled-knuckle relationship. & \raggedright \probePair{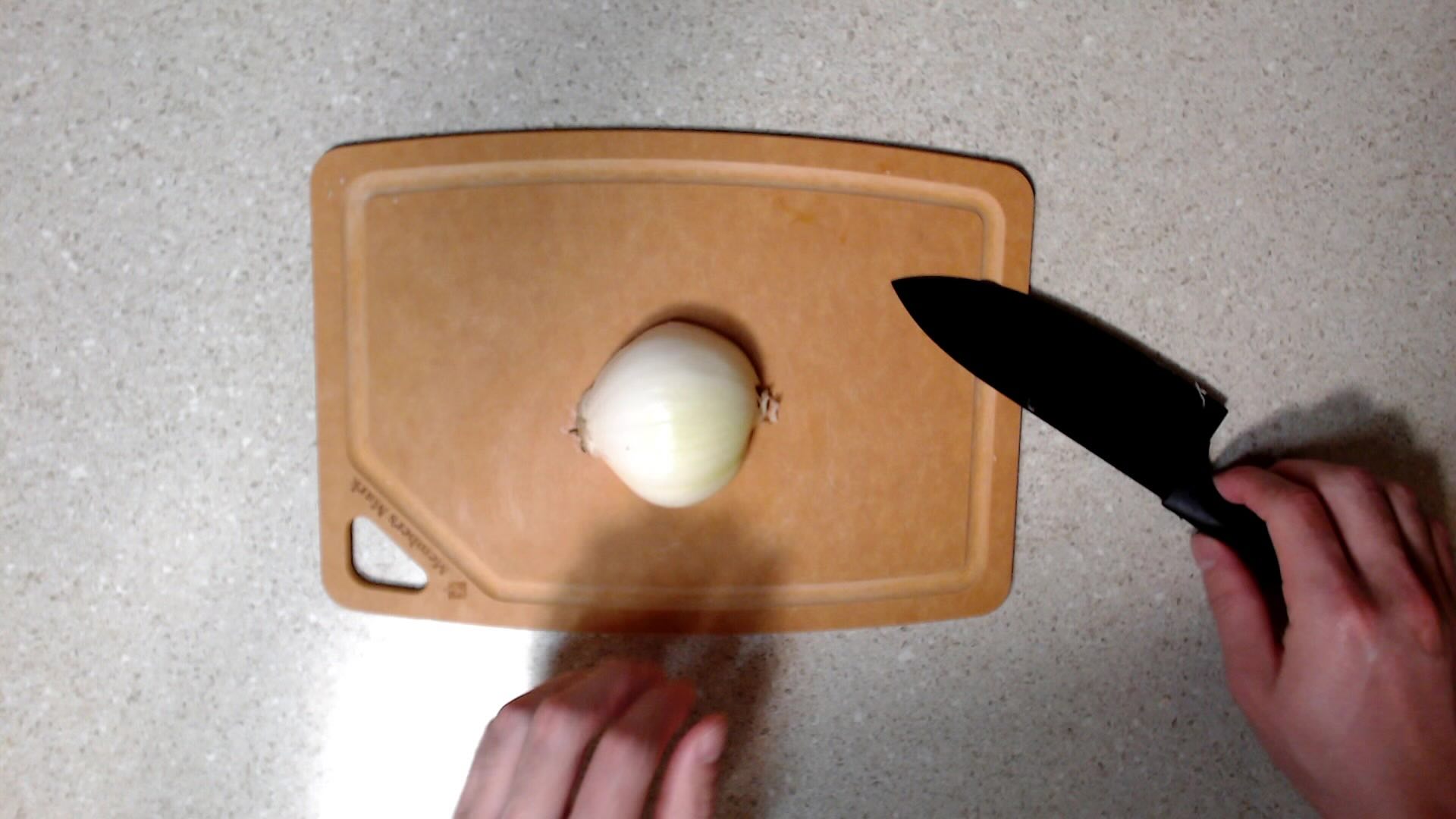}{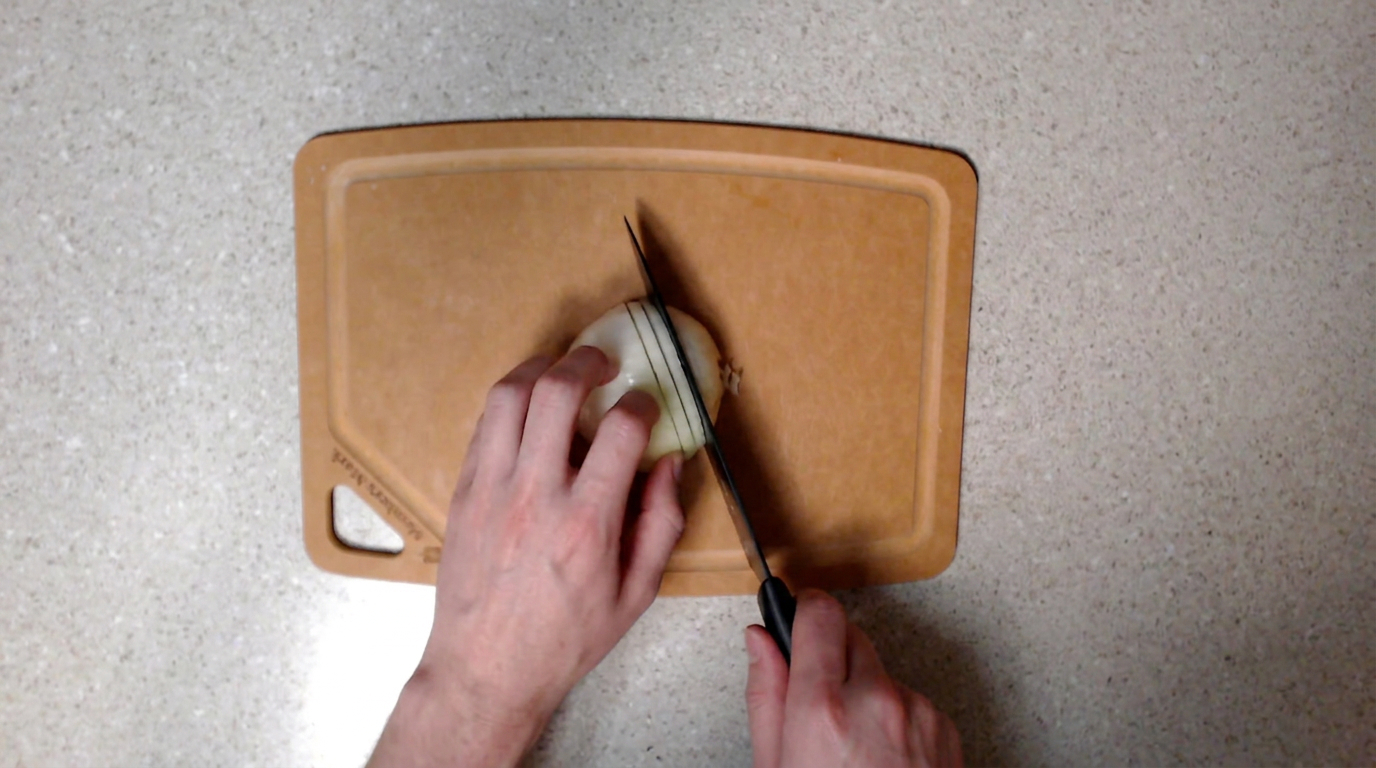}{Source}{Generated} \tabularnewline
\addlinespace[4pt]
\raggedright \textbf{B9} Key-object integrity failure\newline\textit{E1, E3} & \raggedright An important object is missing, duplicated, merged, fragmented, deformed, or loses stable identity.\par\smallskip\textit{Example:} Sandwich assembly, step 2 (Gemini 3.1 Flash Image). An additional knife appears while a spreading utensil is held over the bread. & \raggedright \probePair{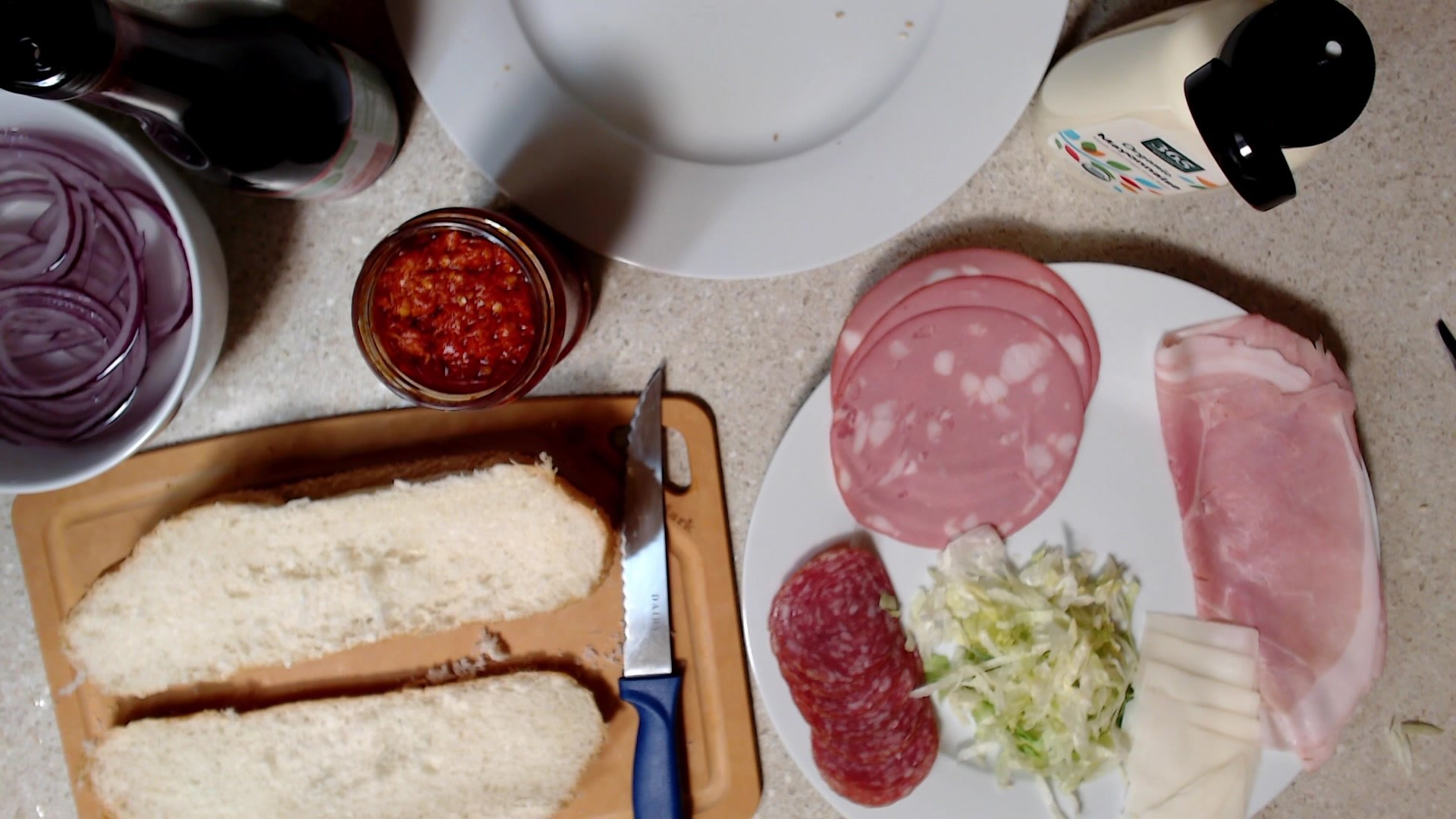}{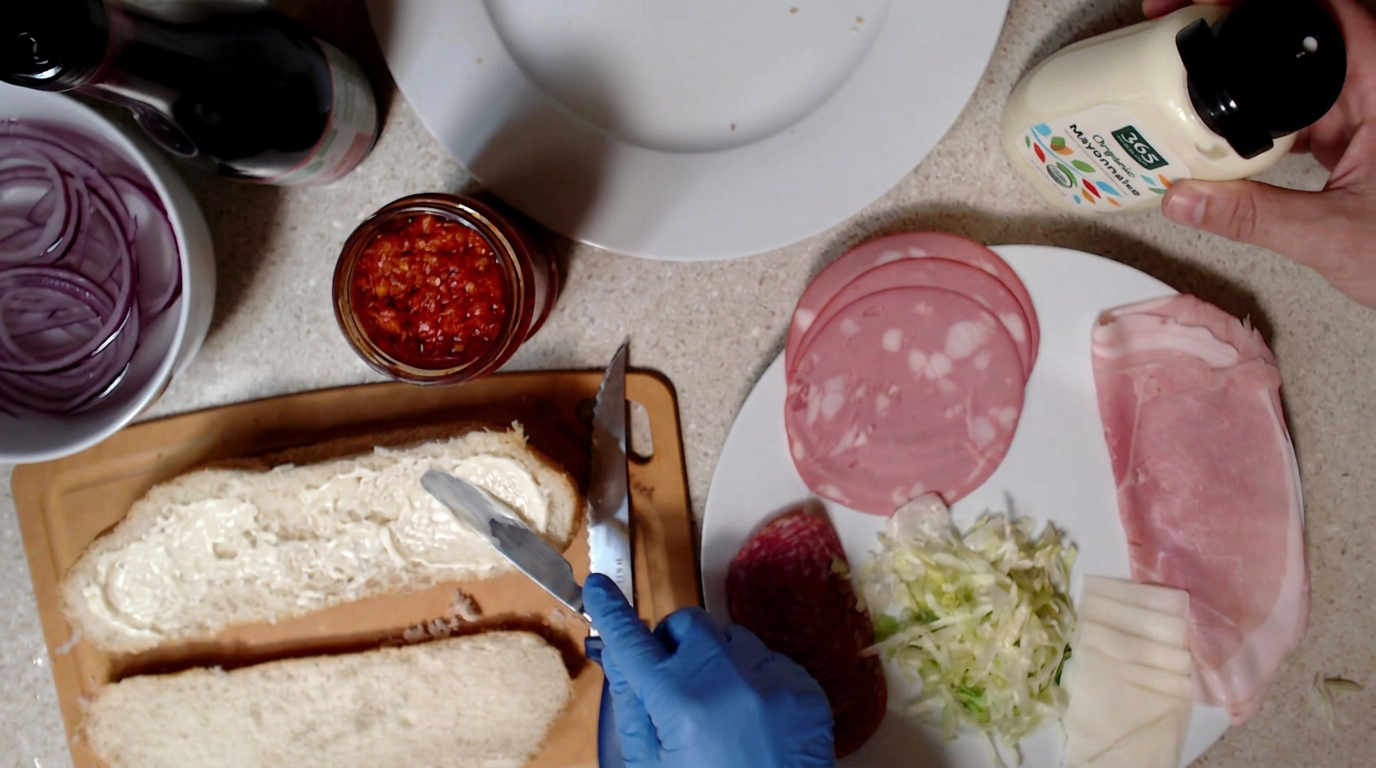}{Source}{Generated} \tabularnewline
\addlinespace[4pt]
\midrule
\multicolumn{3}{@{}l}{\textbf{C. Contextual and physical coherence}} \\*
\raggedright \textbf{C1} Noncritical attribute drift\newline\textit{E3} & \raggedright A nonessential property, such as color, texture, clothing detail, or finish, changes.\par\smallskip\textit{Example:} Gift wrapping, step 9 (Gemini 3.1 Flash Image). The printed Amazon/Prime markings disappear from the box during repositioning. & \raggedright \probePair{figures/codebook-figures/gift-wrapping-step-09-source.jpg}{figures/codebook-figures/gift-wrapping-step-09-generated.jpg}{Source}{Generated} \tabularnewline
\addlinespace[4pt]
\raggedright \textbf{C2} Peripheral context drift or removal\newline\textit{E3} & \raggedright A peripheral background object or environmental feature changes, disappears, or is replaced.\par\smallskip\textit{Example:} LEGO, step 1 (Gemini Omni Flash Preview). Loose LEGO pieces surrounding the work disappear as the clip progresses. & \raggedright \probePair{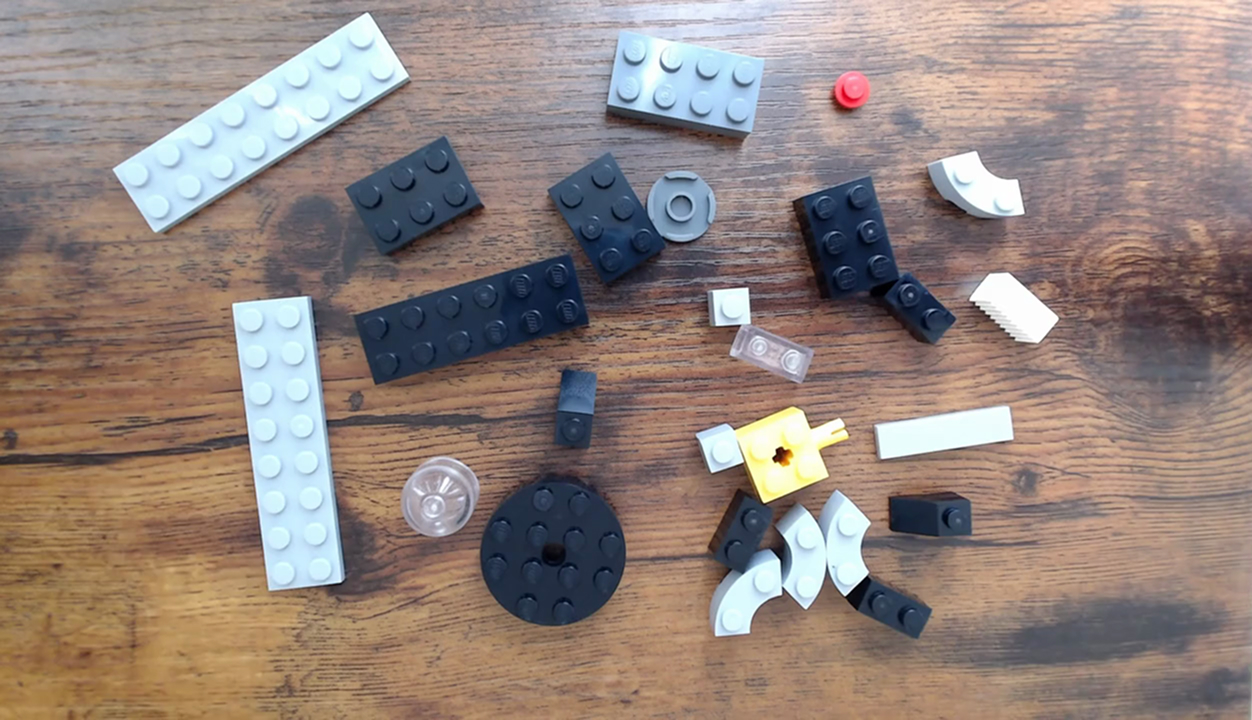}{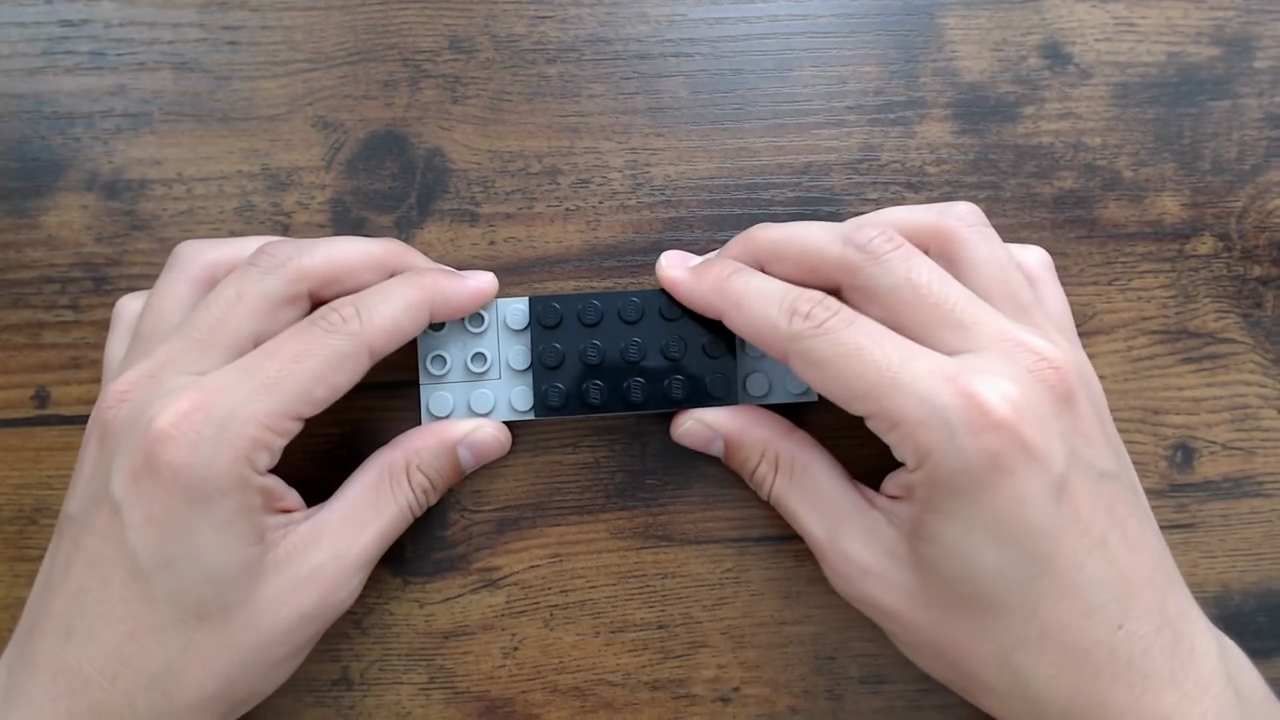}{0.00s/source}{4.91 s} \tabularnewline
\addlinespace[4pt]
\raggedright \textbf{C3} Previous-state inconsistency\newline\textit{E2, E3} & \raggedright An established task state resets, reverts, changes topology, or otherwise loses continuity.\par\smallskip\textit{Example:} Crochet, step 3 (Gemini 3.1 Flash Image). The established yarn-loop arrangement changes while the thumb is being inserted. & \raggedright \probePair{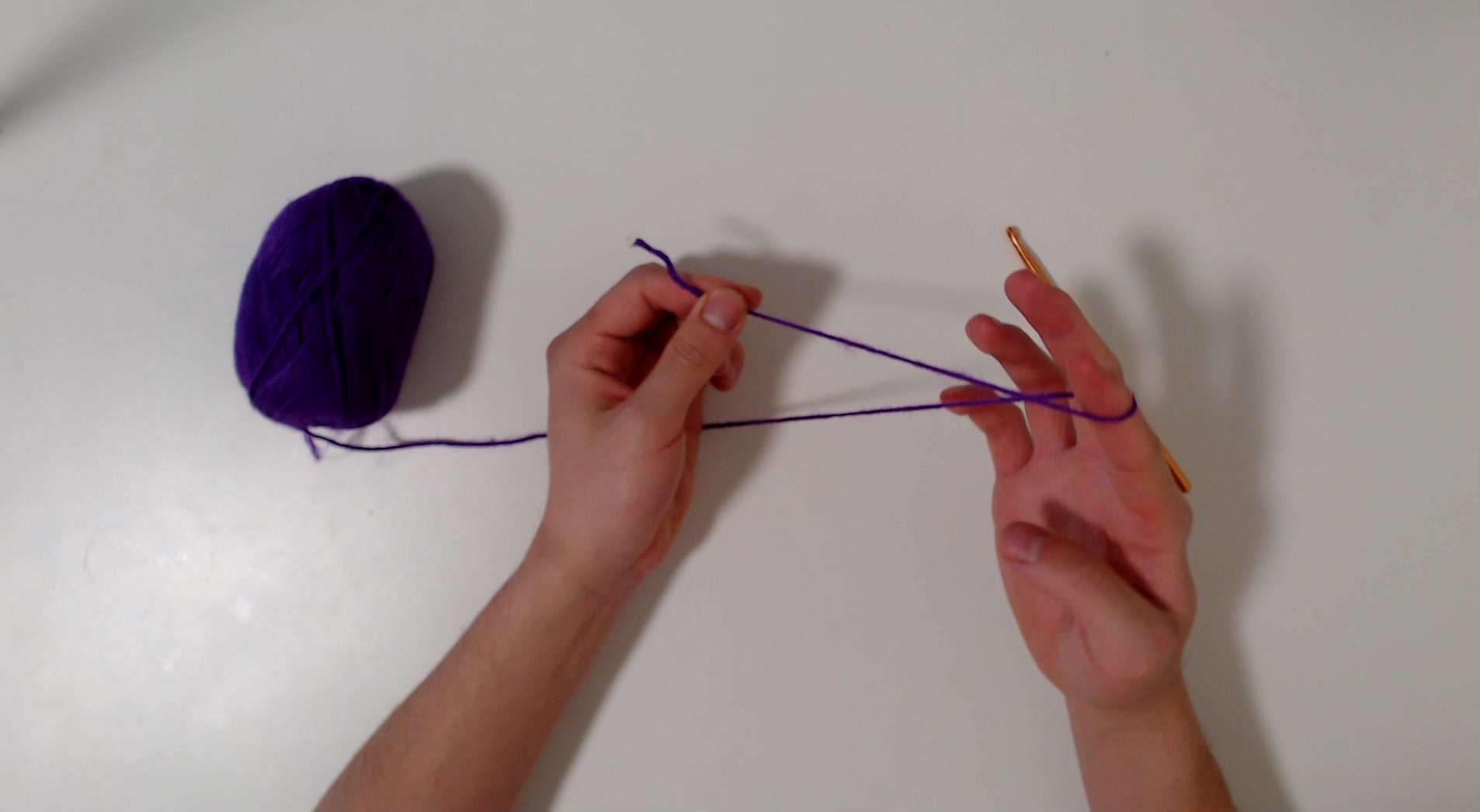}{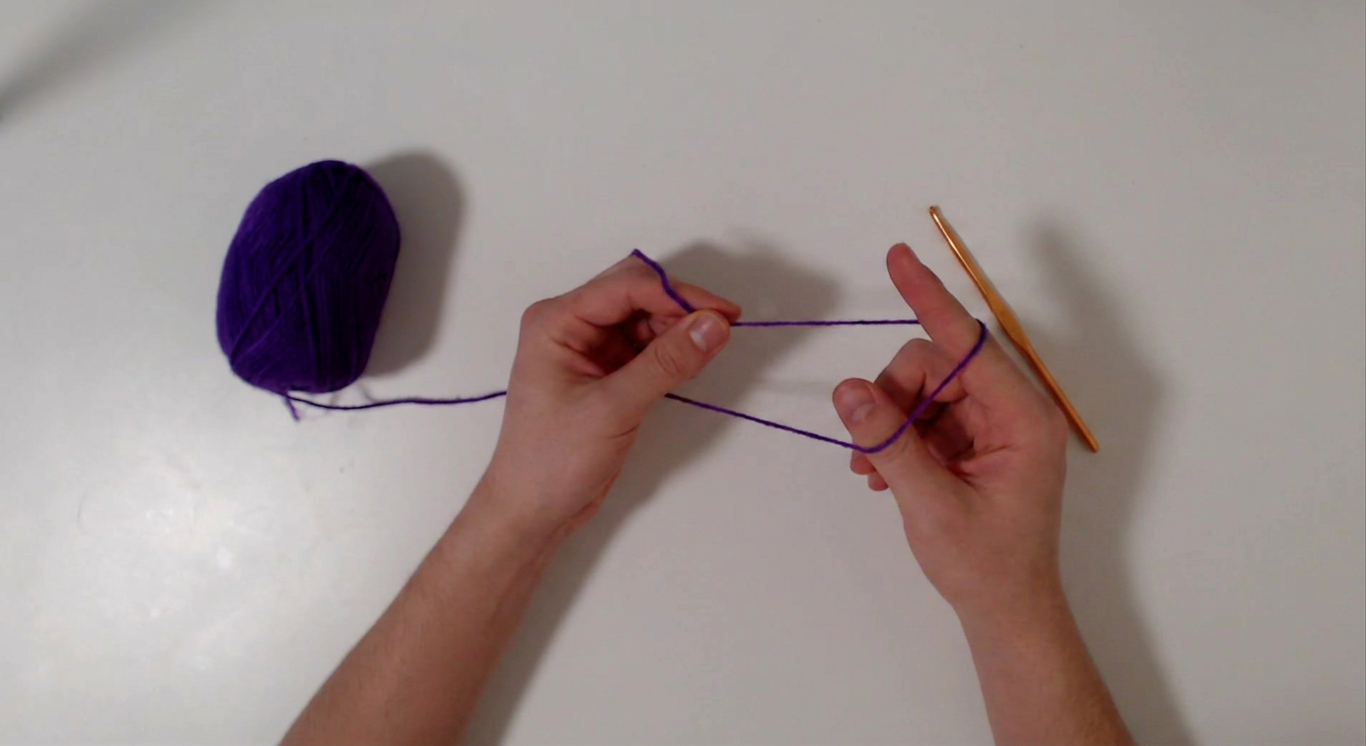}{Source}{Generated} \tabularnewline
\addlinespace[4pt]
\raggedright \textbf{C4} Missing progress or causal trace\newline\textit{E2, E6} & \raggedright Intermediate evidence needed to understand how the initial state becomes the result is absent.\par\smallskip\textit{Example:} Matcha latte, step 1 (Gemini 3 Pro). Powder appears dispersed in the water without the transfer or an intermediate causal trace. & \raggedright \probePair{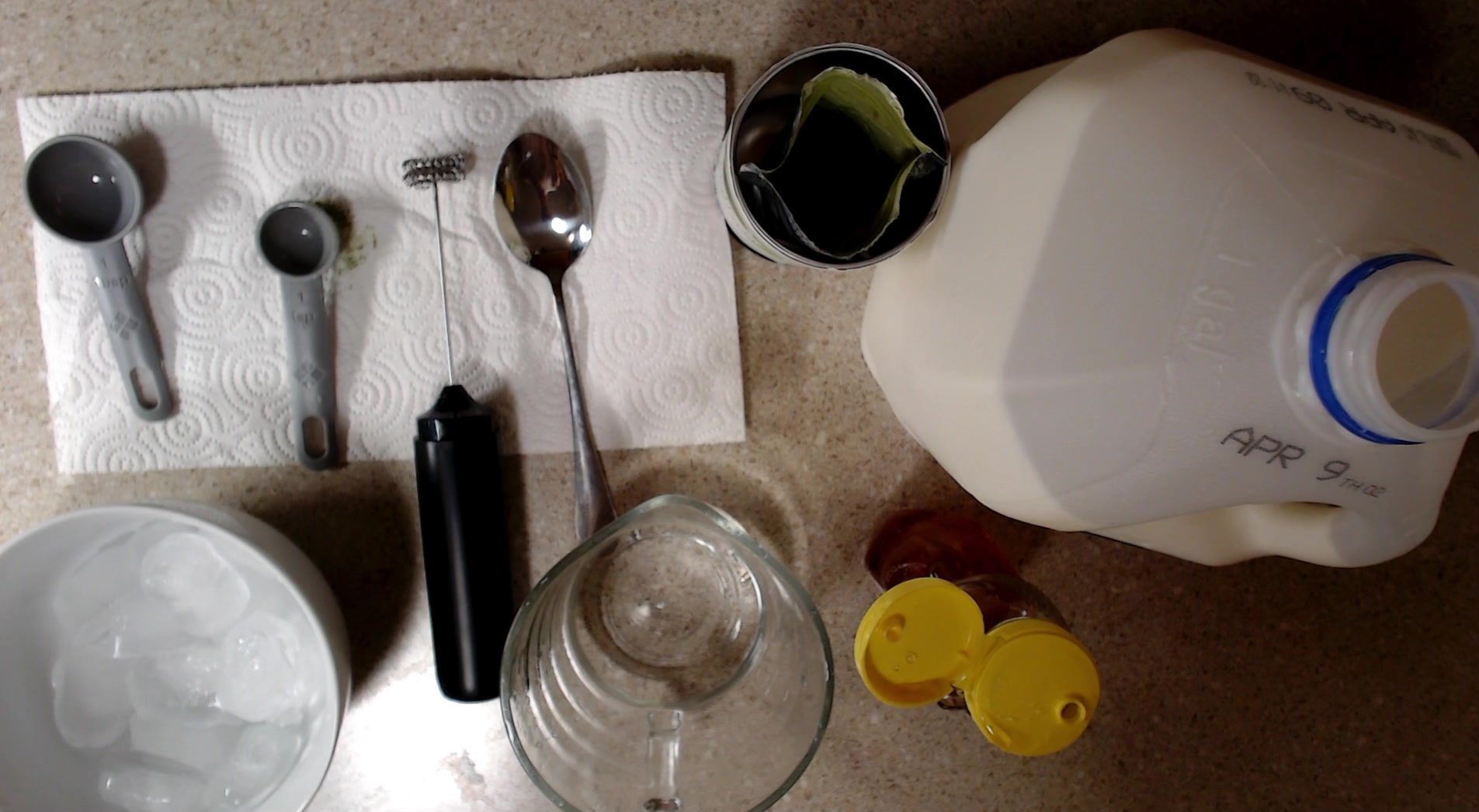}{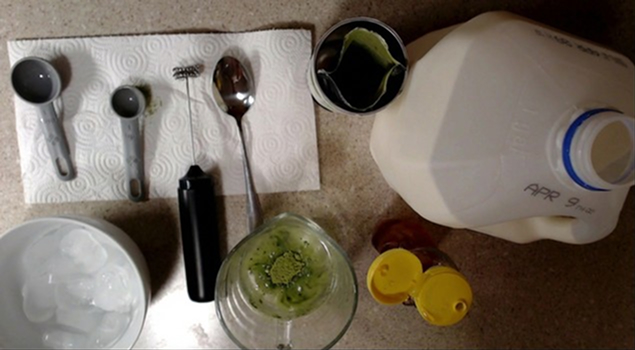}{Source}{Generated} \tabularnewline
\addlinespace[4pt]
\raggedright \textbf{C5} Impossible or malformed tool\newline\textit{E1, E2} & \raggedright A tool is visibly malformed, incorrectly constructed, or impossible to operate as depicted.\par\smallskip\textit{Example:} Furniture assembly, step 6 (Gemini 3.1 Flash Image). The fastening tool changes into an implausible form at the bolt. & \raggedright \probePair{figures/codebook-figures/furniture-assembly-step-06-source.jpg}{figures/codebook-figures/furniture-assembly-step-06-generated.jpg}{Source}{Generated} \tabularnewline
\addlinespace[4pt]
\raggedright \textbf{C6} Affordance violation\newline\textit{E2} & \raggedright A plausible object or tool is used incompatibly with its physical affordance.\par\smallskip\textit{Example:} Matcha latte, step 7 (Veo 3.1). The mixer is applied among ice pieces in a manner coded as incompatible with its affordance. & \raggedright \probeTriple{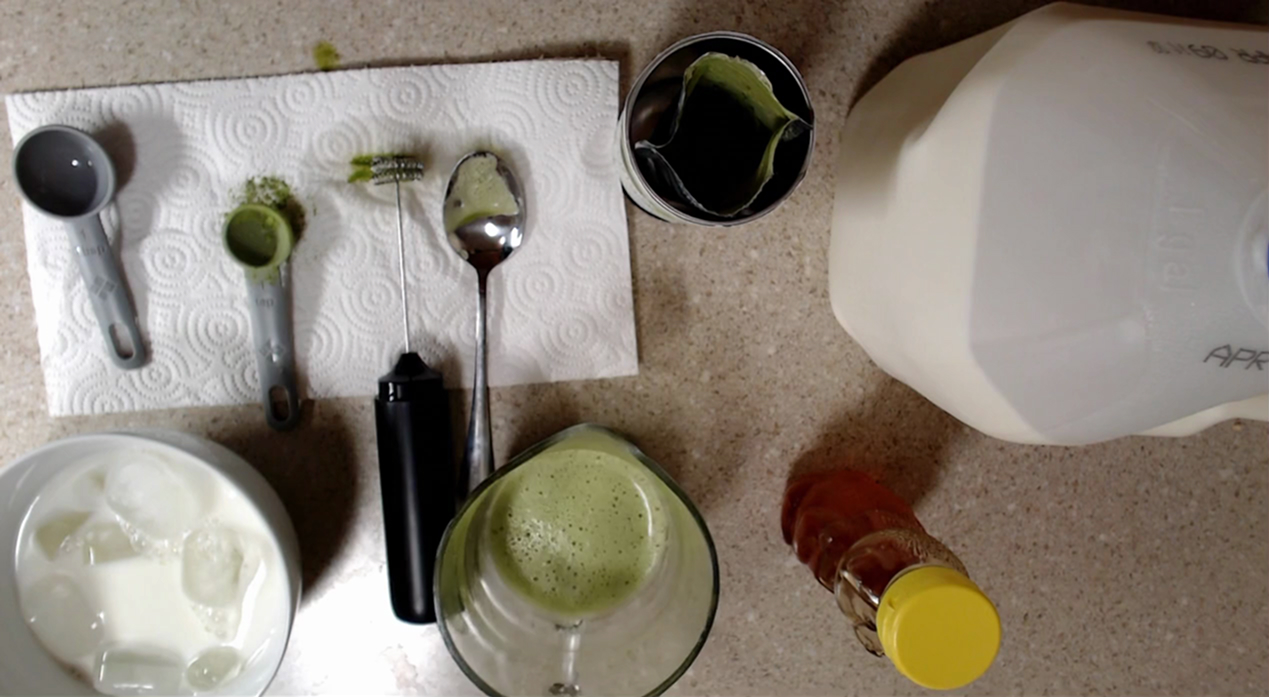}{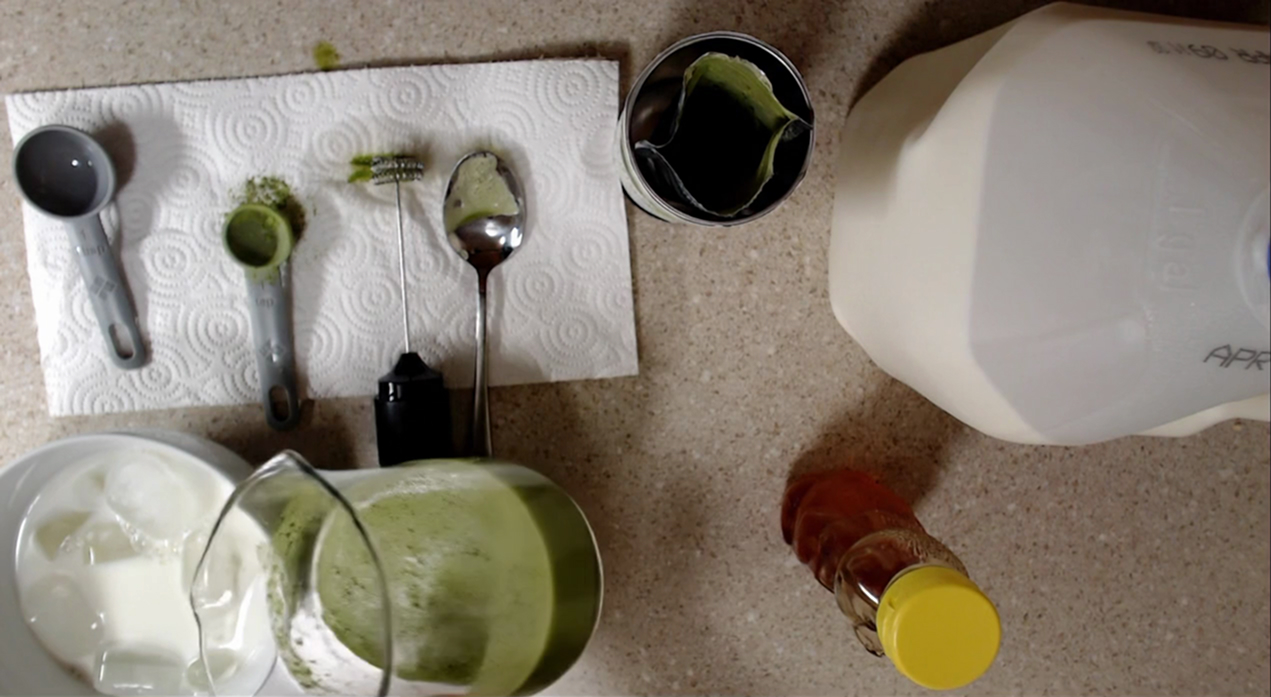}{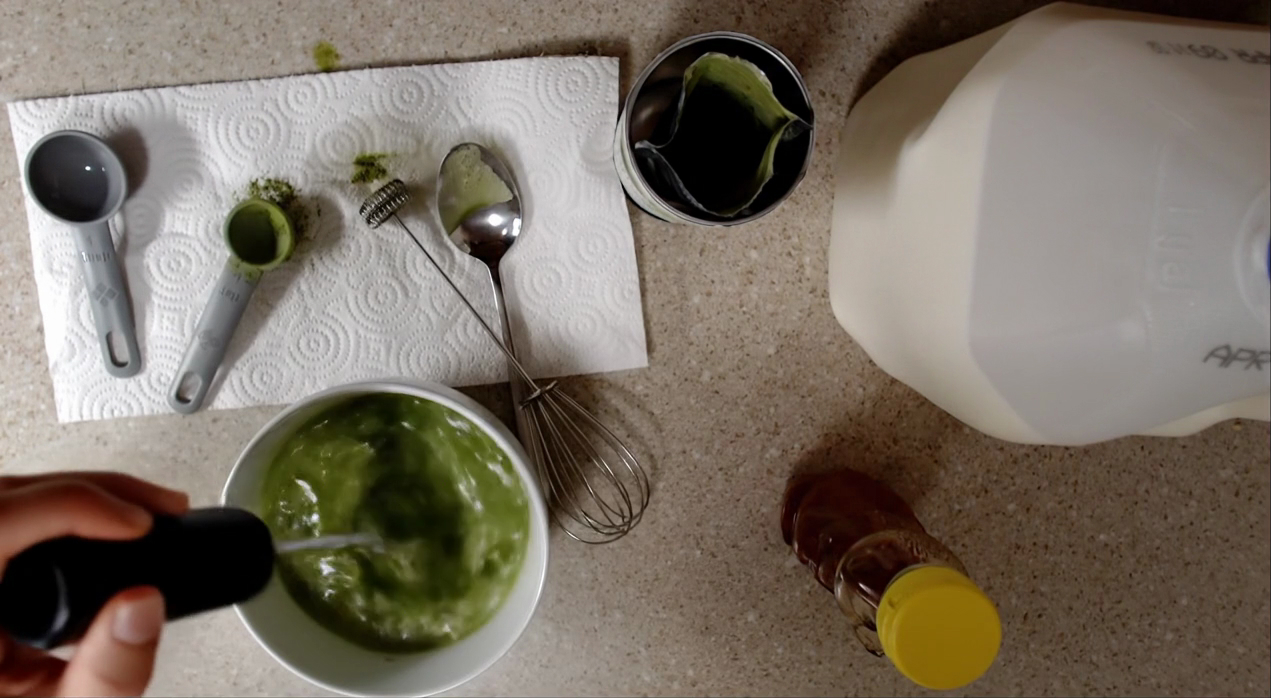}{0.00s/source}{1.00 s}{3.00 s} \tabularnewline
\addlinespace[4pt]
\raggedright \textbf{C7} Impossible physical state or transformation\newline\textit{E1, E2} & \raggedright An object, material, or body undergoes a physically implausible state or transformation.\par\smallskip\textit{Example:} Gift wrapping, step 20 (Gemini 3.1 Flash Image). The flap becomes a closed triangular form without a physically coherent paper configuration. & \raggedright \probePair{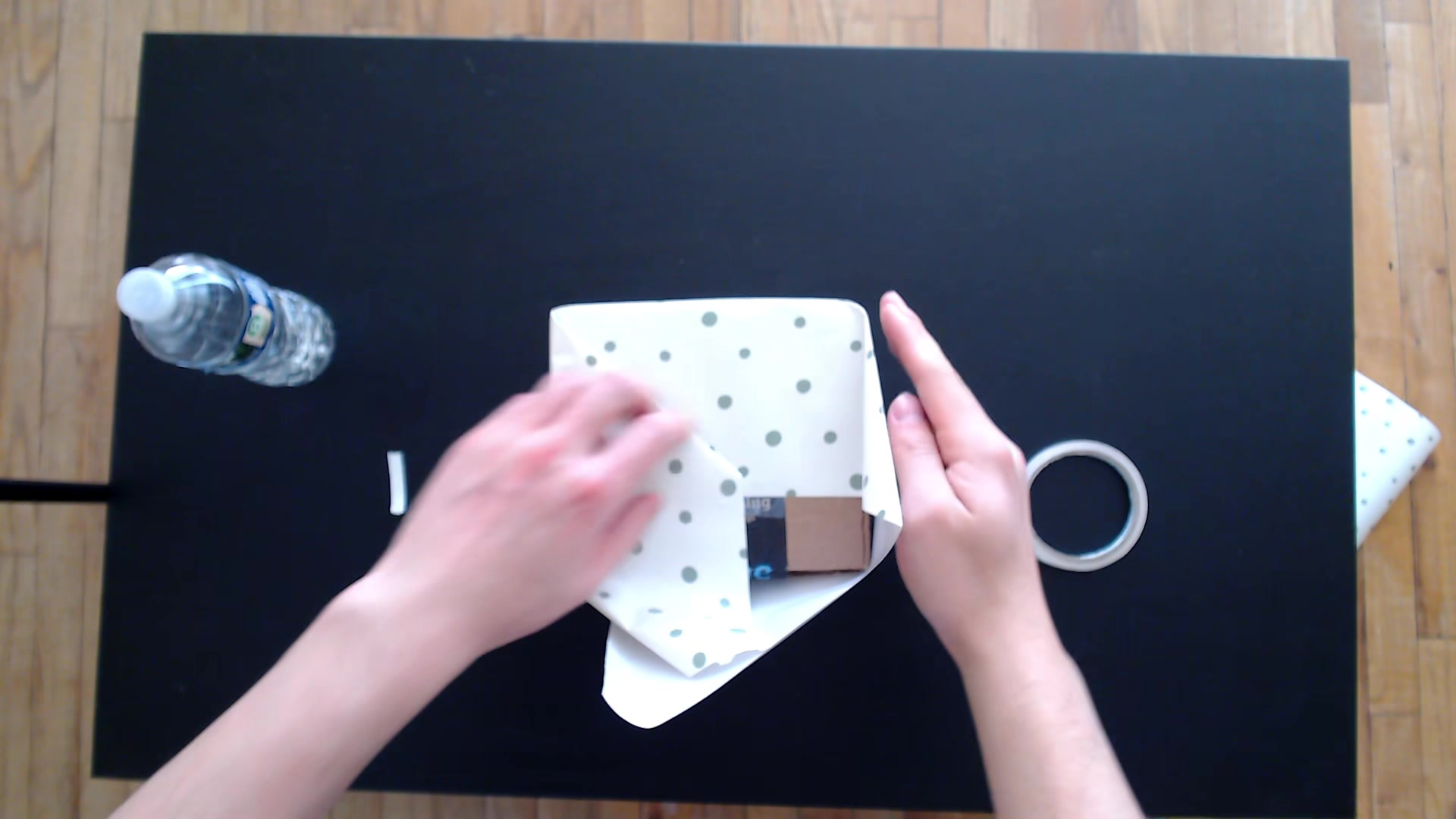}{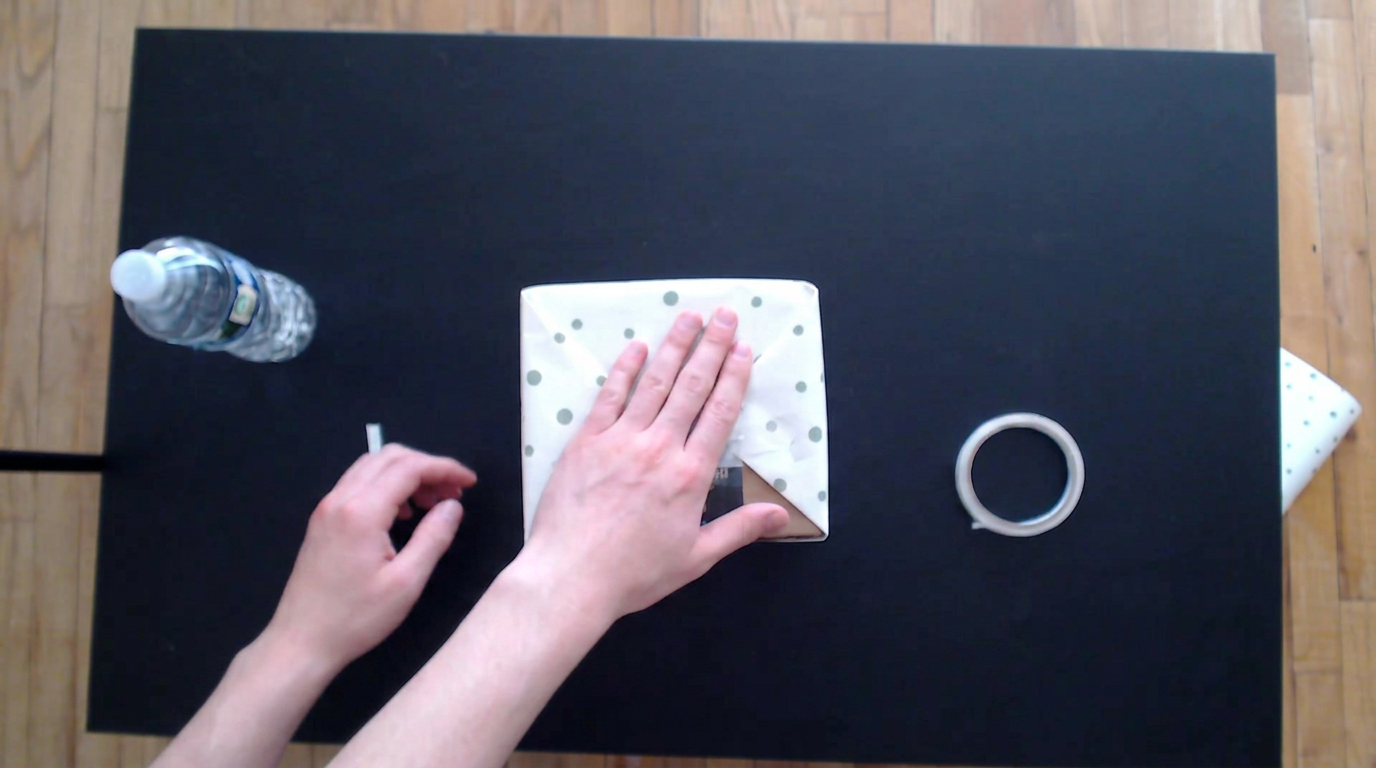}{Source}{Generated} \tabularnewline
\addlinespace[4pt]
\midrule
\multicolumn{3}{@{}l}{\textbf{D. Presentation and temporal legibility}} \\*
\raggedright \textbf{D1} Viewpoint or framing drift\newline\textit{E3, E5} & \raggedright Camera angle, distance, scale, or viewpoint changes unexpectedly and reduces continuity or comparability.\par\smallskip\textit{Example:} Packing a box, step 7 (Gemini Omni Flash Preview). The view moves from the original oblique framing to a closer overhead view of the box. & \raggedright \probePair{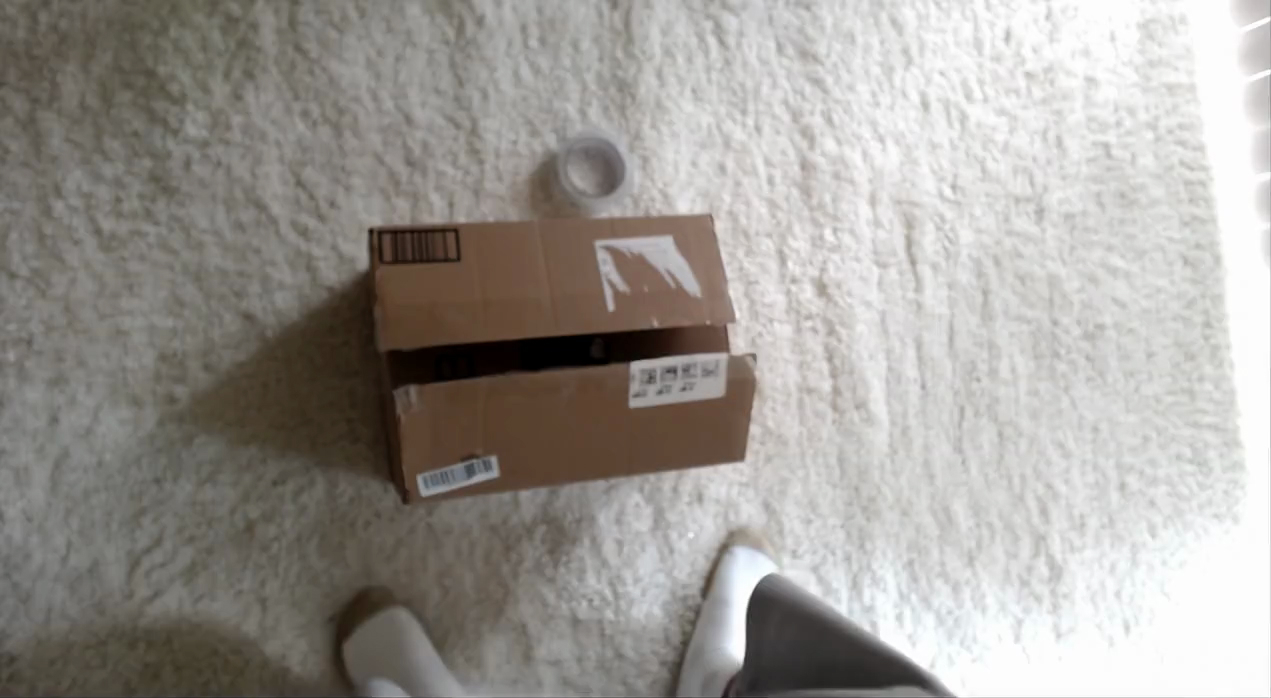}{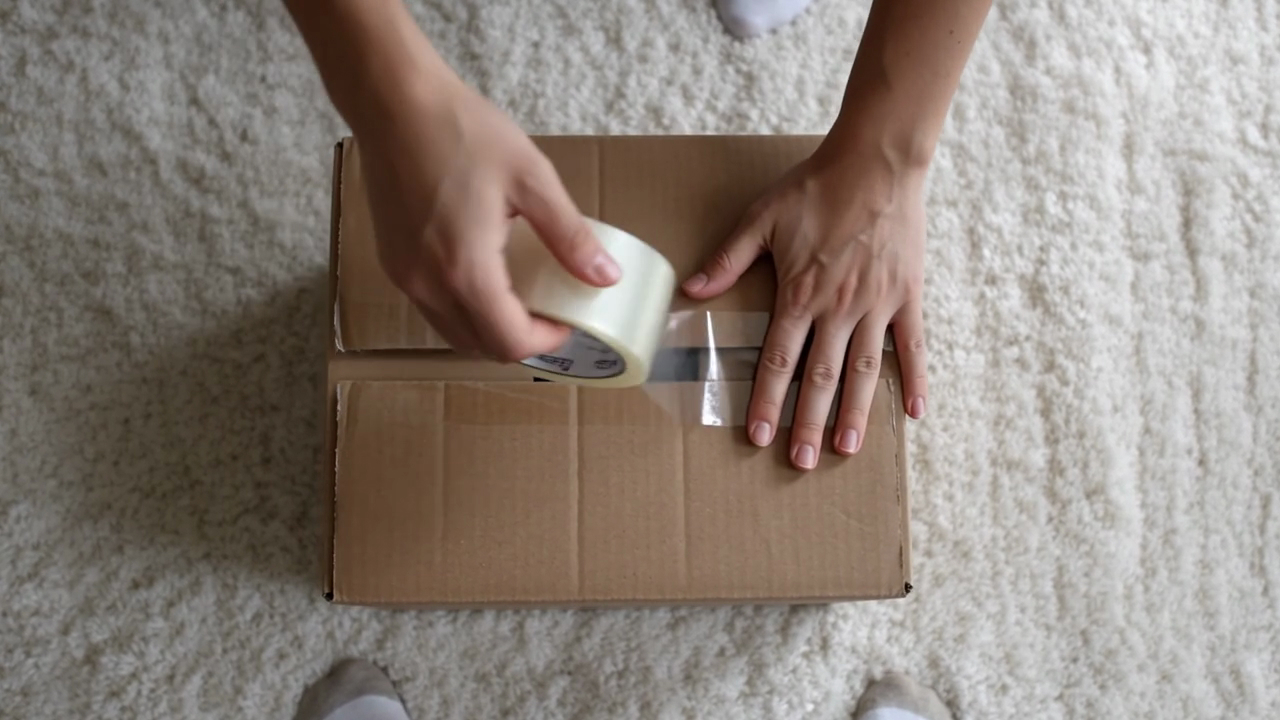}{0.00s/source}{4.91 s} \tabularnewline
\addlinespace[4pt]
\raggedright \textbf{D2} Unrequested text or graphic overlay\newline\textit{E1, E5} & \raggedright Unrequested text, arrows, labels, diagrams, or overlays obscure or distract from task evidence.\par\smallskip\textit{Example:} Tennis swing, step 4 (Gemini 3.1 Flash Image). Labels and a large curved trajectory are added over multiple copies of the player. & \raggedright \probePair{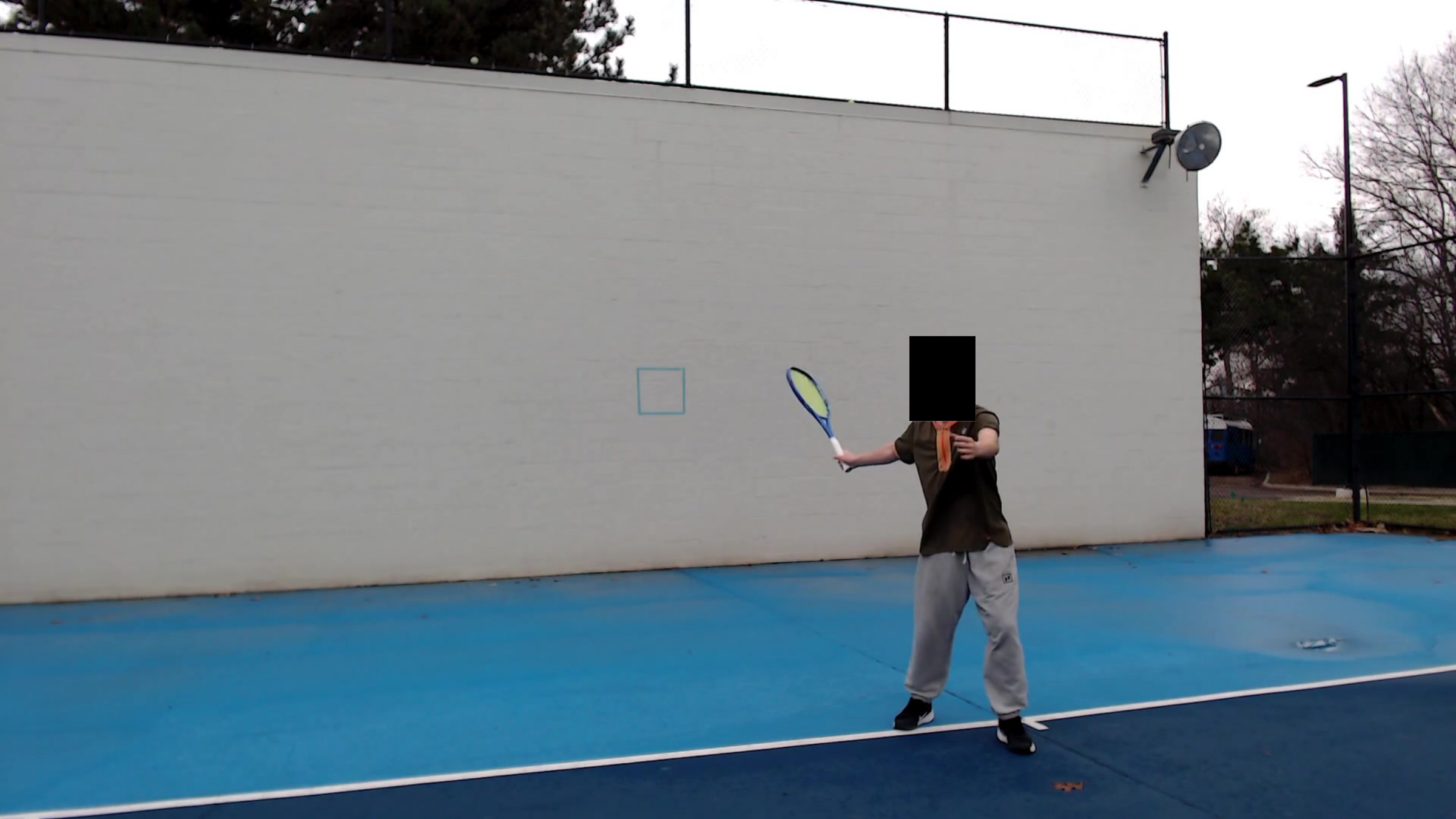}{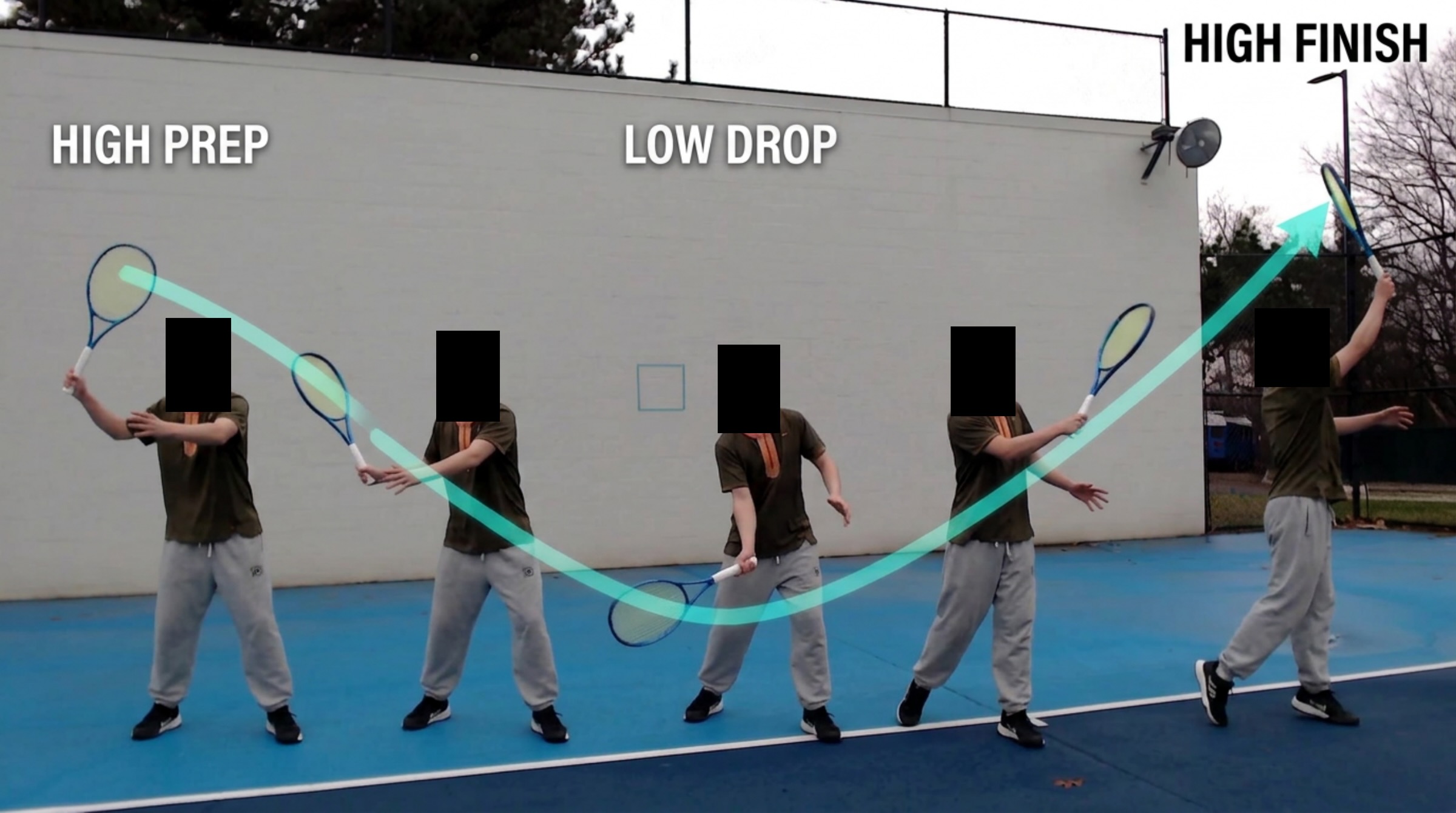}{Source}{Generated} \tabularnewline
\addlinespace[4pt]
\raggedright \textbf{D3} Key content outside the frame or time range\newline\textit{E1, E2, E5} & \raggedright A task-critical hand, tool, object, body part, action, or required moment is omitted.\par\smallskip\textit{Example:} Circuit wiring, step 1 (Gemini Omni Flash Preview). The final insertion is hidden beneath the fingers, so the specified LED rows cannot be checked. & \raggedright \probeTriple{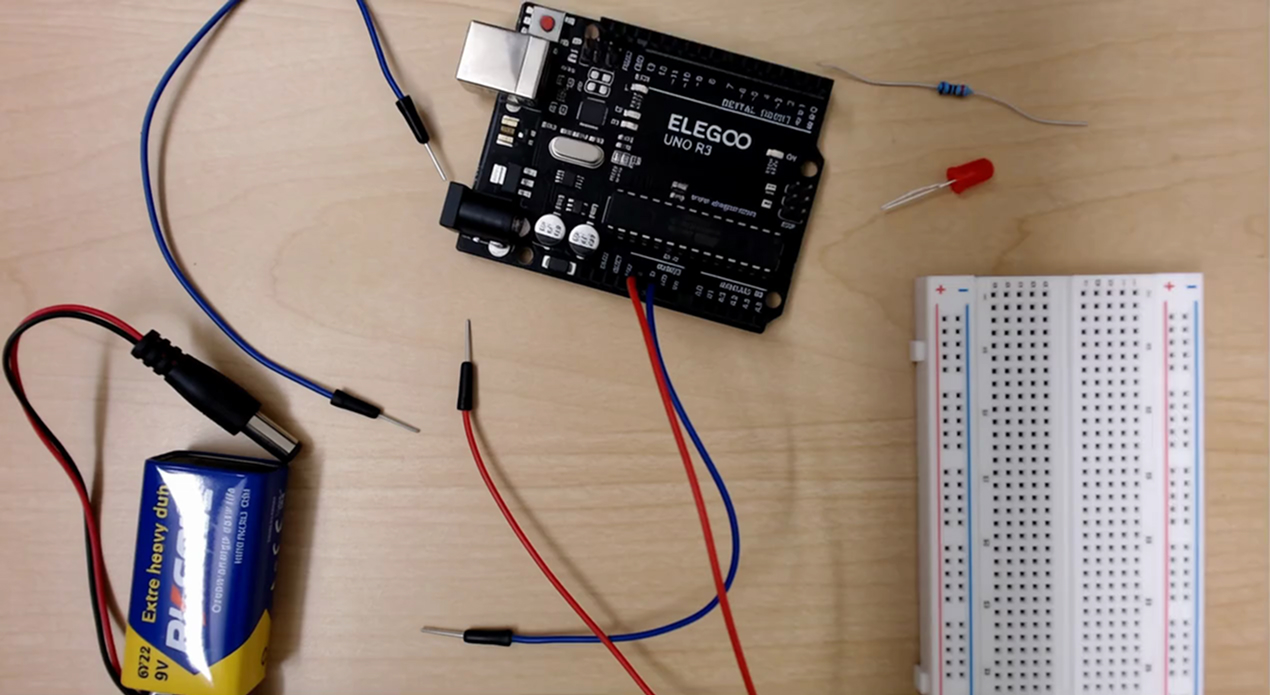}{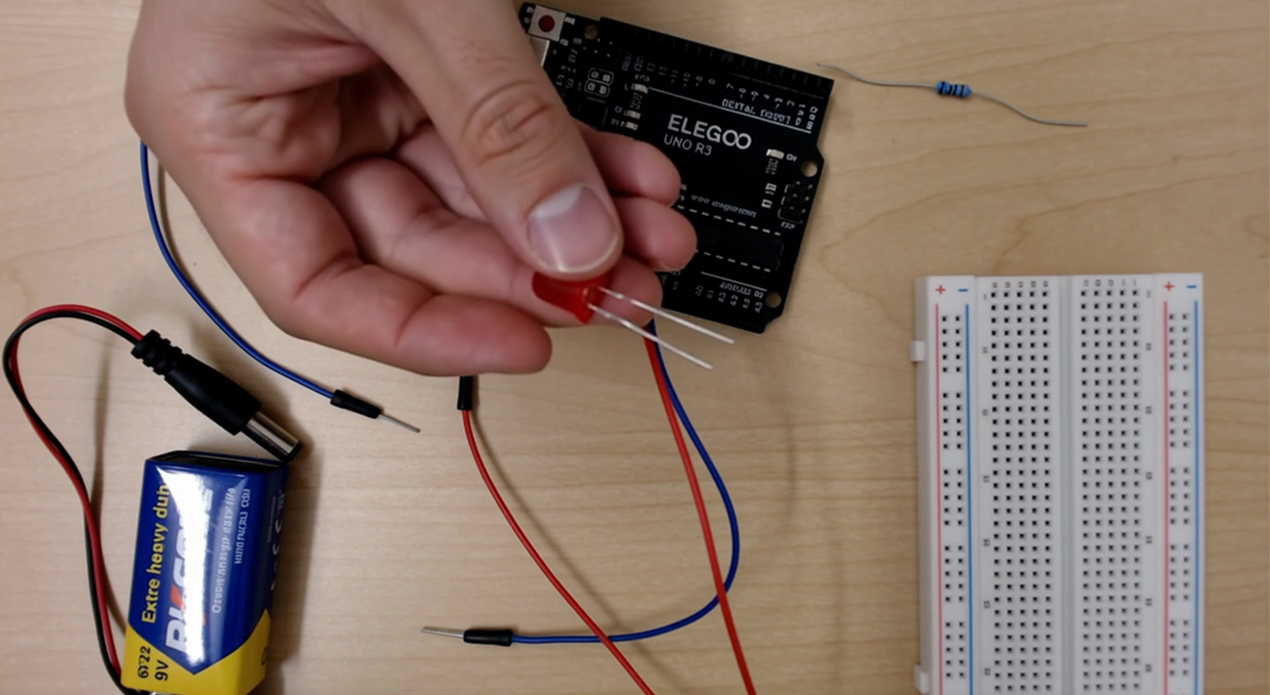}{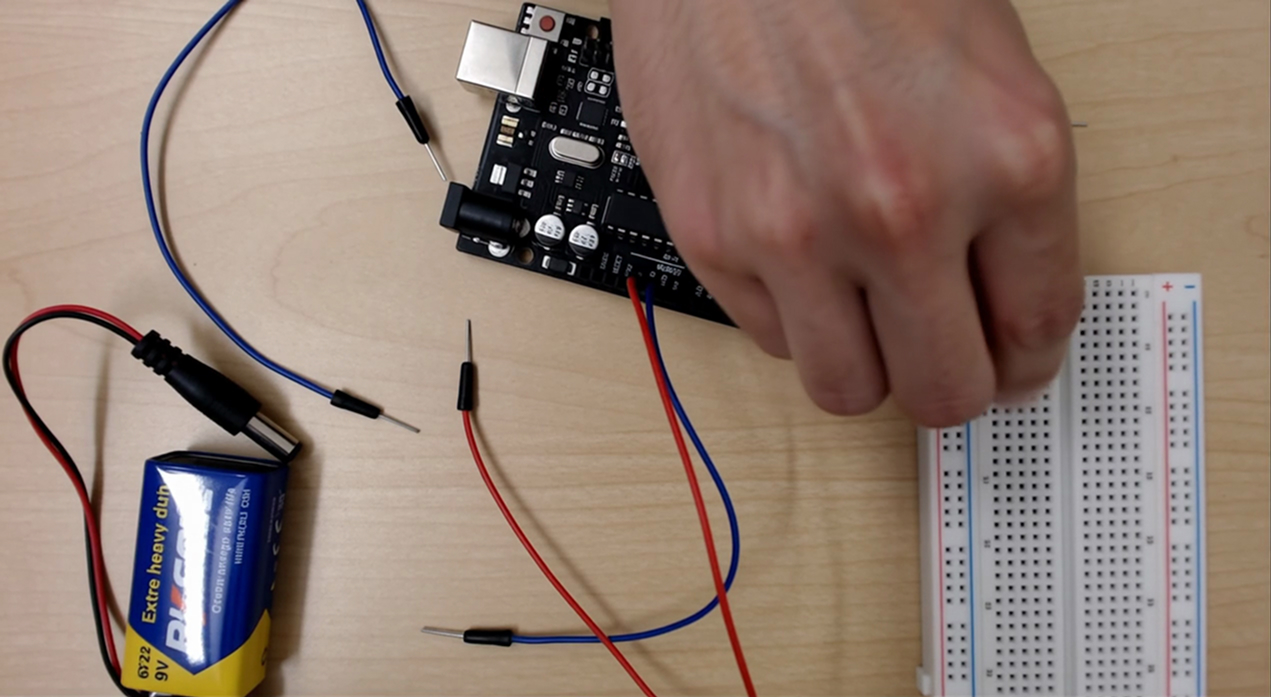}{0.00s/source}{2.51 s}{4.91 s} \tabularnewline
\addlinespace[4pt]
\raggedright \textbf{D4} Object or background jitter\newline\textit{E2, E3} & \raggedright Objects, people, or environmental features jitter inconsistently across video frames.\par\smallskip\textit{Example:} Crochet, step 4 (Veo 3.1). The hand/yarn configuration jitters between nearby moments; the clip reveals the instability. & \raggedright \probeQuad{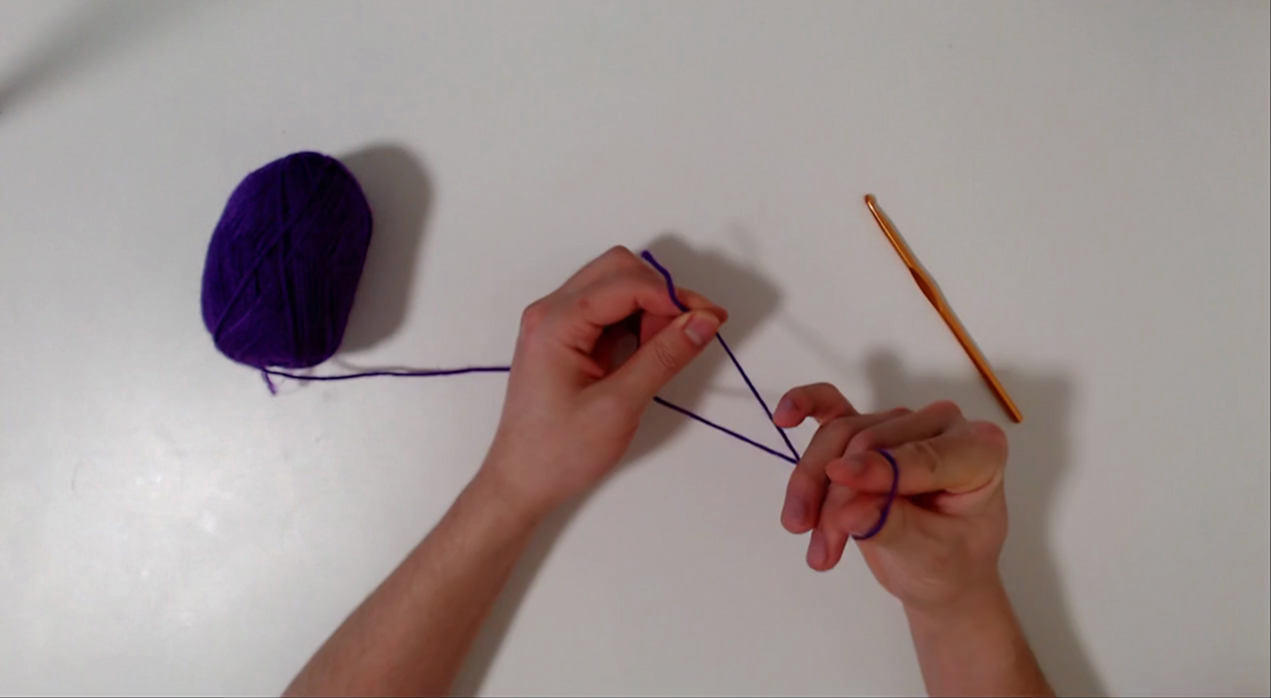}{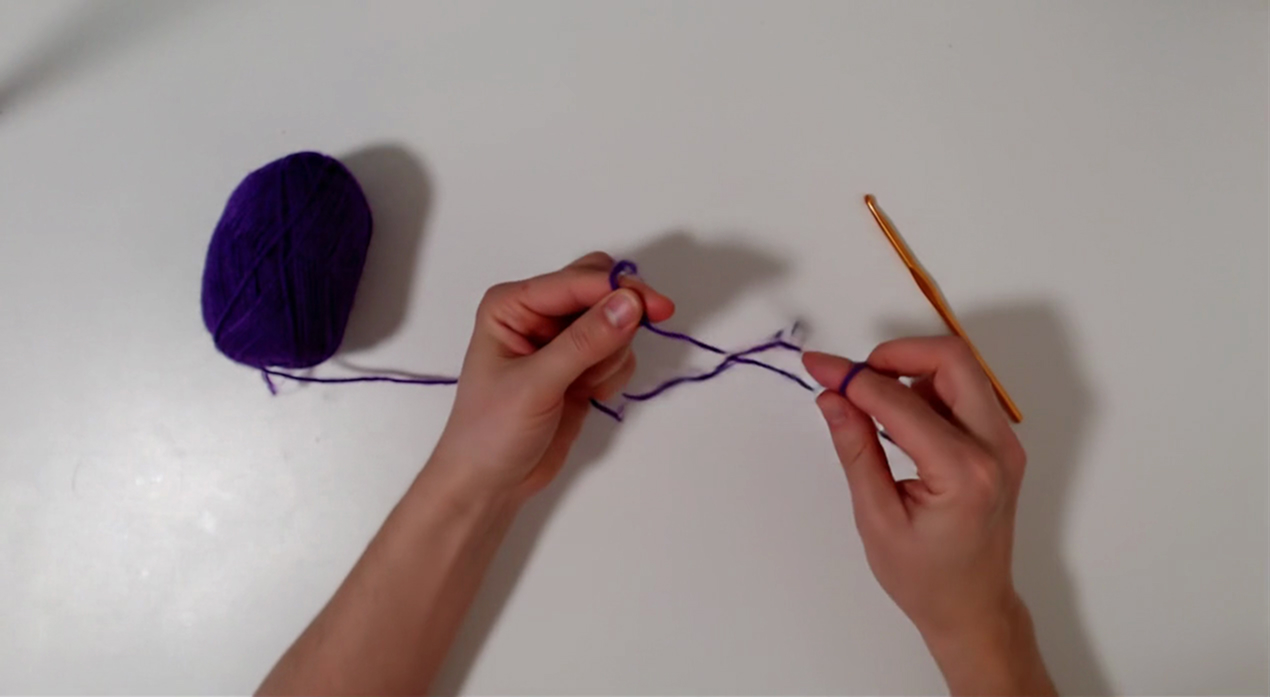}{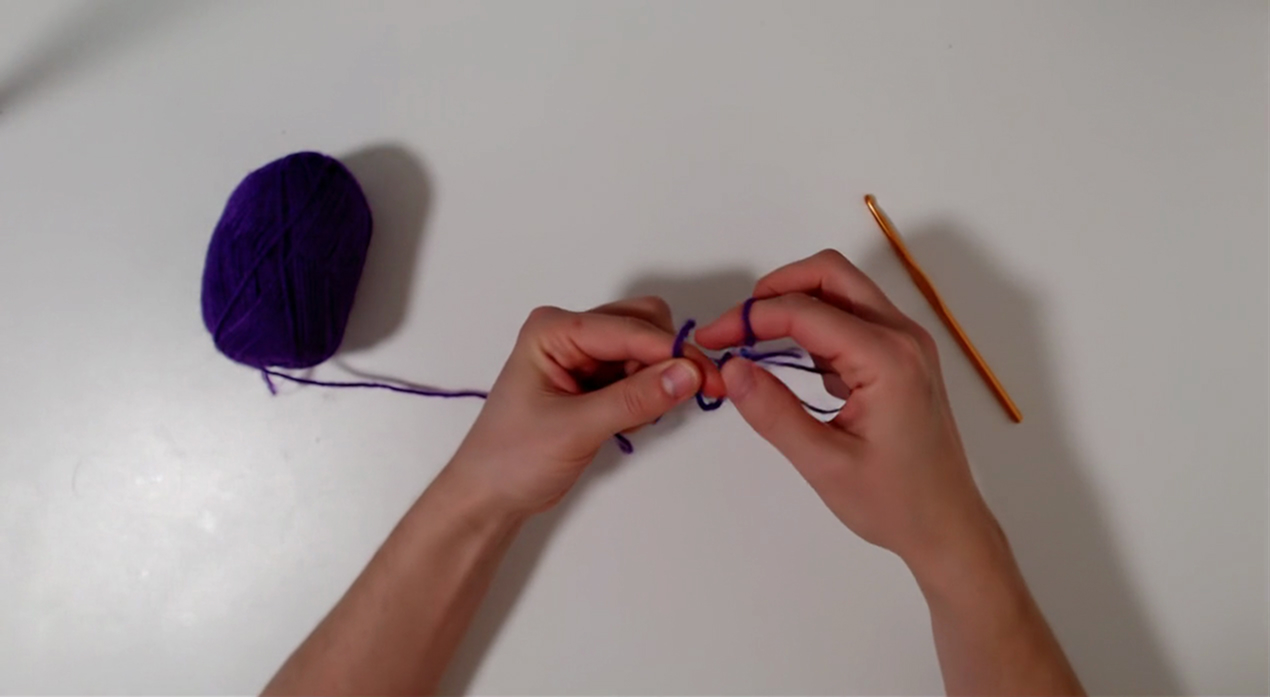}{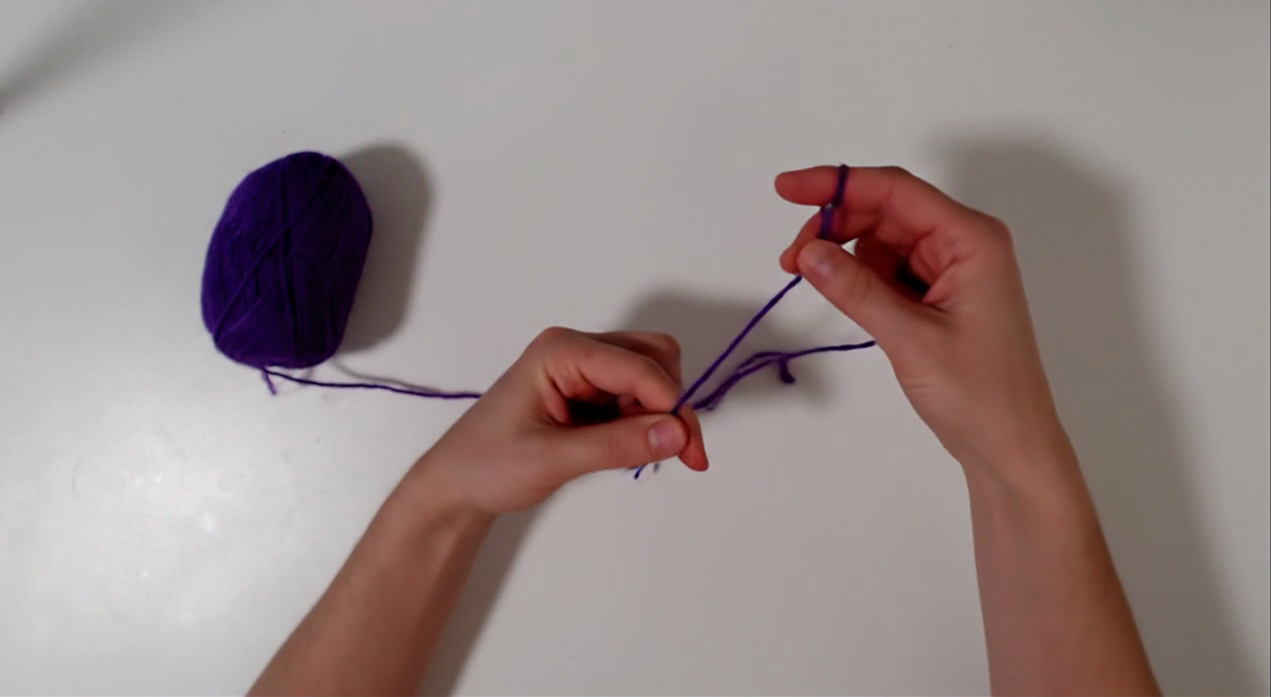}{0.00s/source}{1.00 s}{2.00 s}{3.00 s} \tabularnewline
\addlinespace[4pt]
\raggedright \textbf{D5} Abrupt or excessive motion\newline\textit{E2} & \raggedright Motion is unnaturally large, abrupt, fast, or erratic enough to impair comprehension.\par\smallskip\textit{Example:} Crochet, step 4 (Gemini Omni Flash Preview). The yarn manipulation is too abrupt to follow reliably; the sampled frames show successive moments. & \raggedright \probeQuad{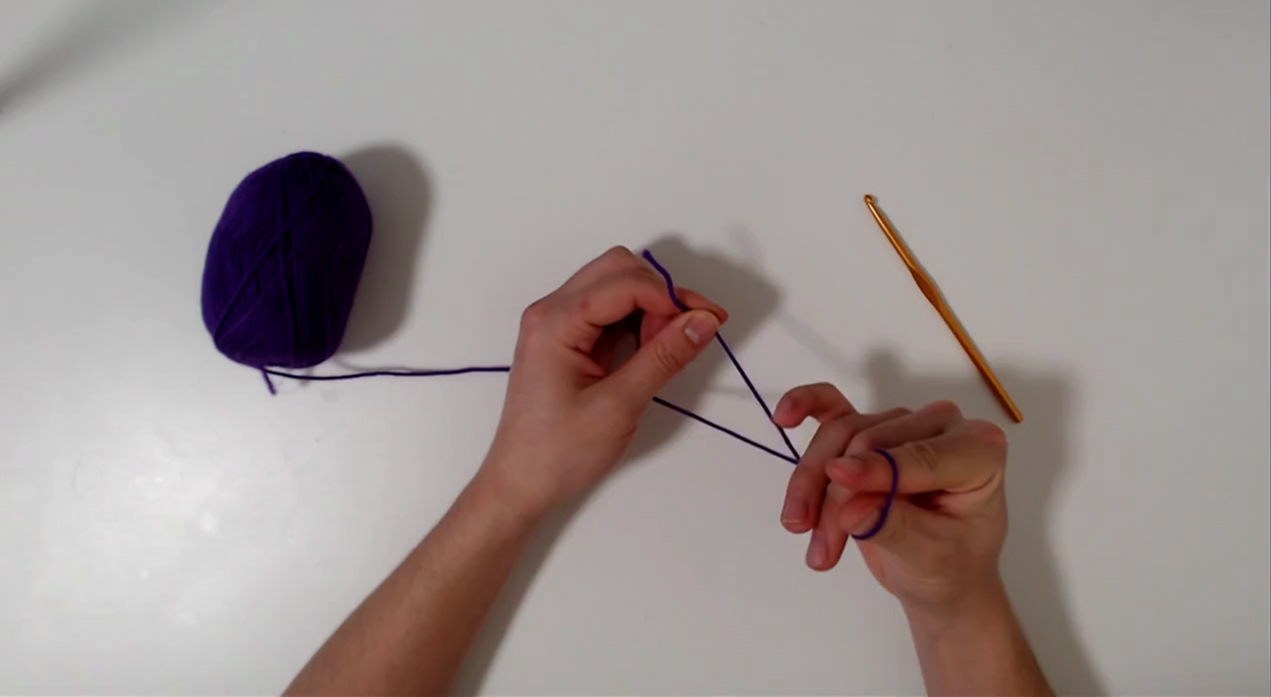}{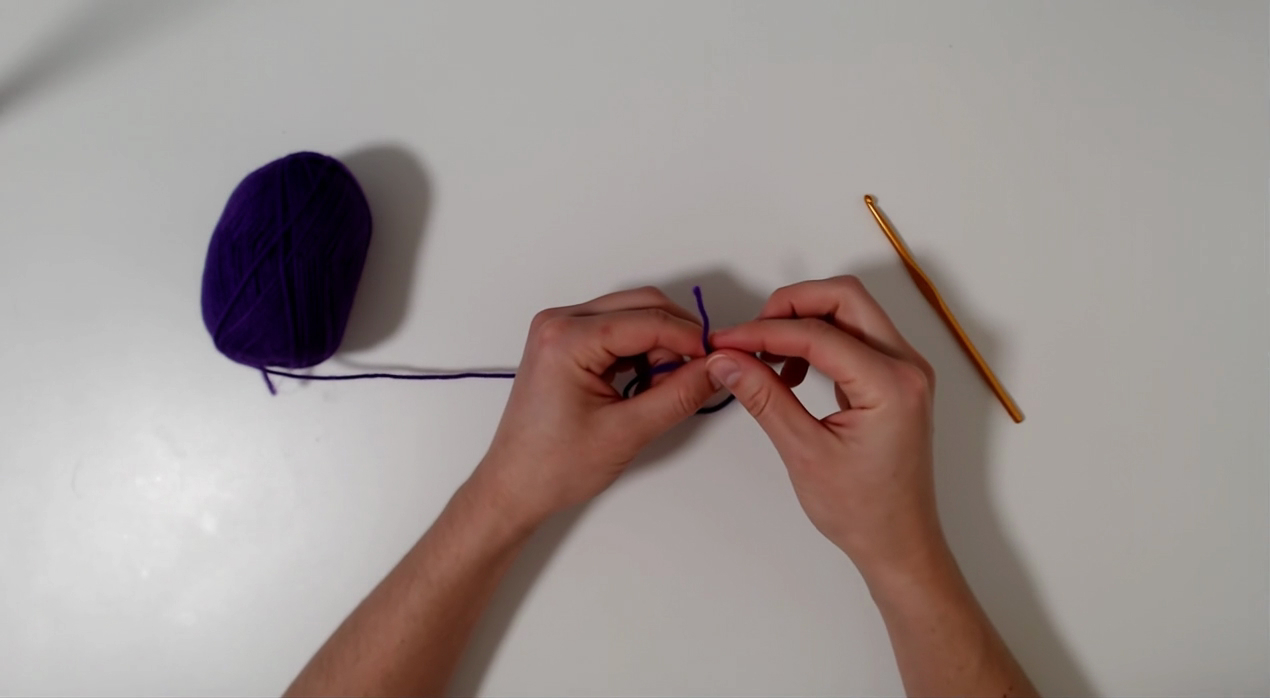}{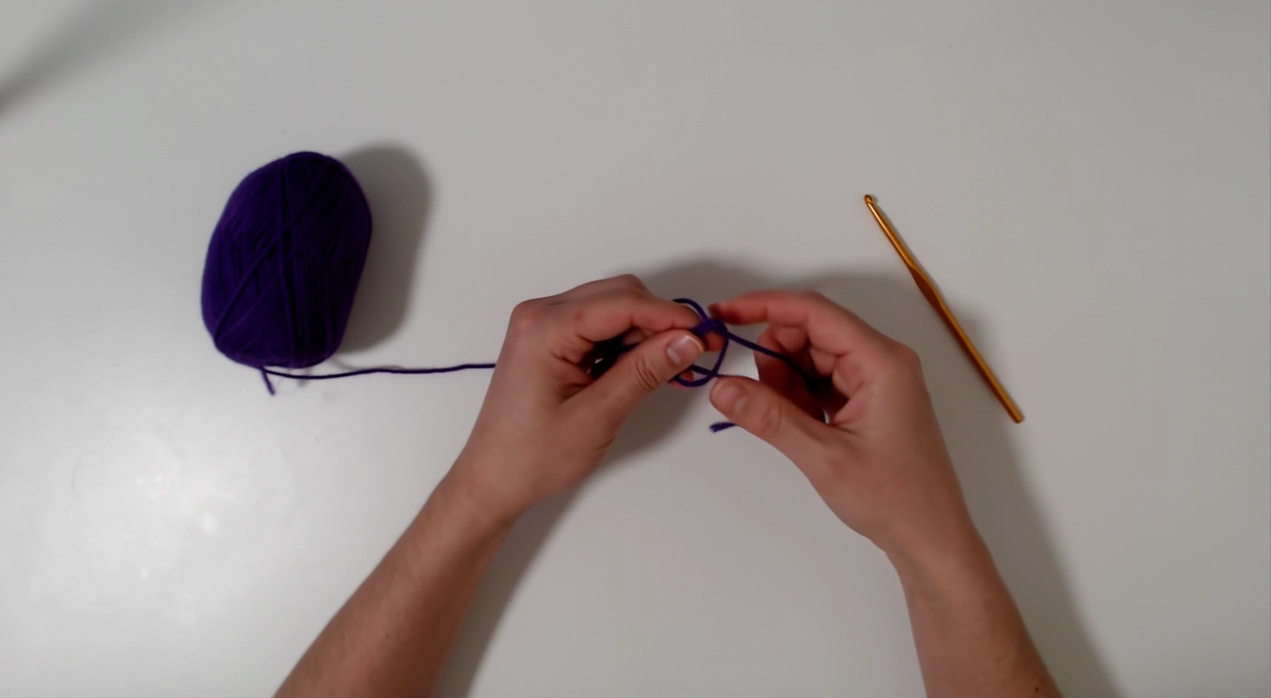}{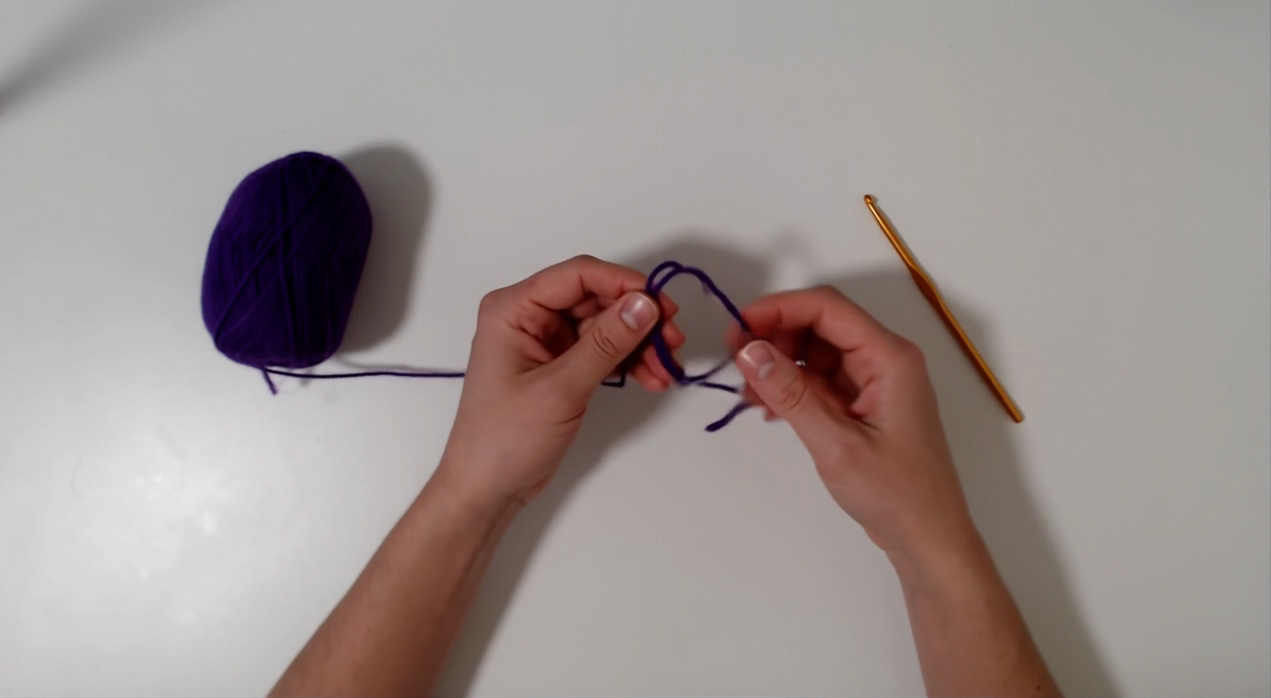}{0.00s/source}{1.25 s}{2.51 s}{3.76 s} \tabularnewline
\addlinespace[4pt]
\raggedright \textbf{D6} Temporal discontinuity\newline\textit{E2, E6} & \raggedright The video jumps between moments, poses, viewpoints, or locations without a coherent transition.\par\smallskip\textit{Example:} Sandwich assembly, step 2 (Gemini Omni Flash Preview). The clip switches from dispensing mayonnaise to spreading it without a coherent transition. & \raggedright \probeTriple{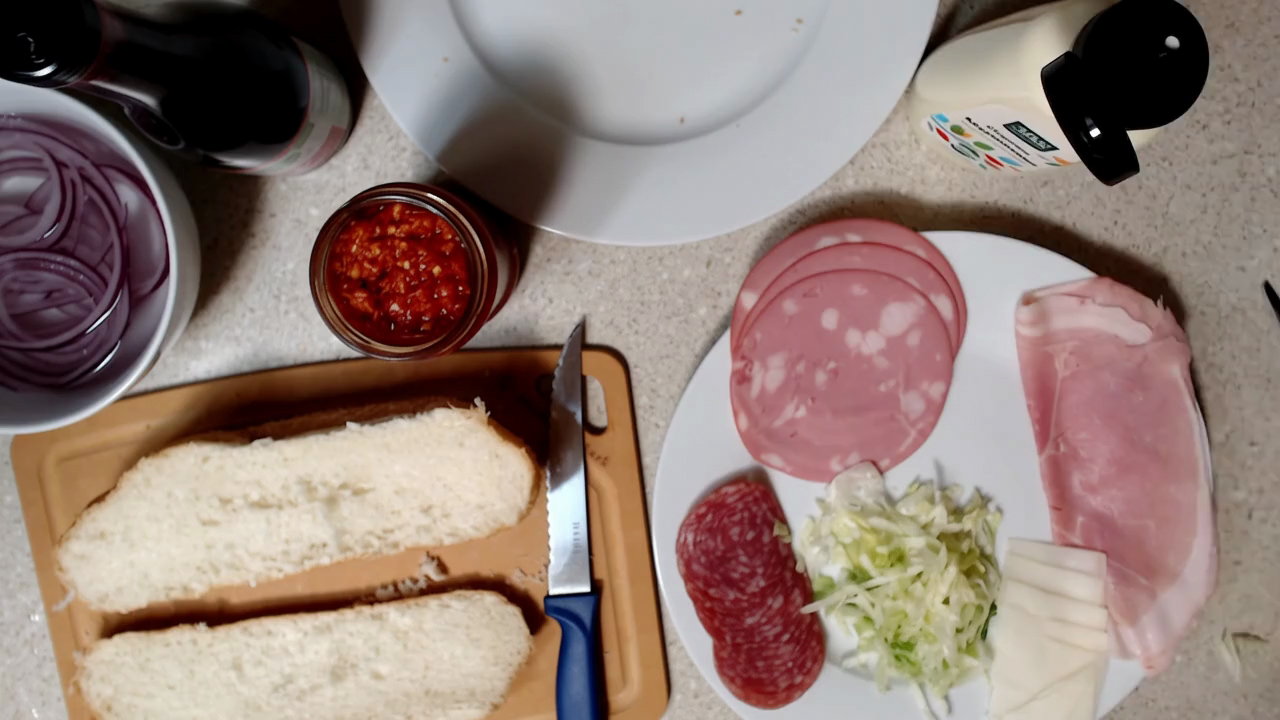}{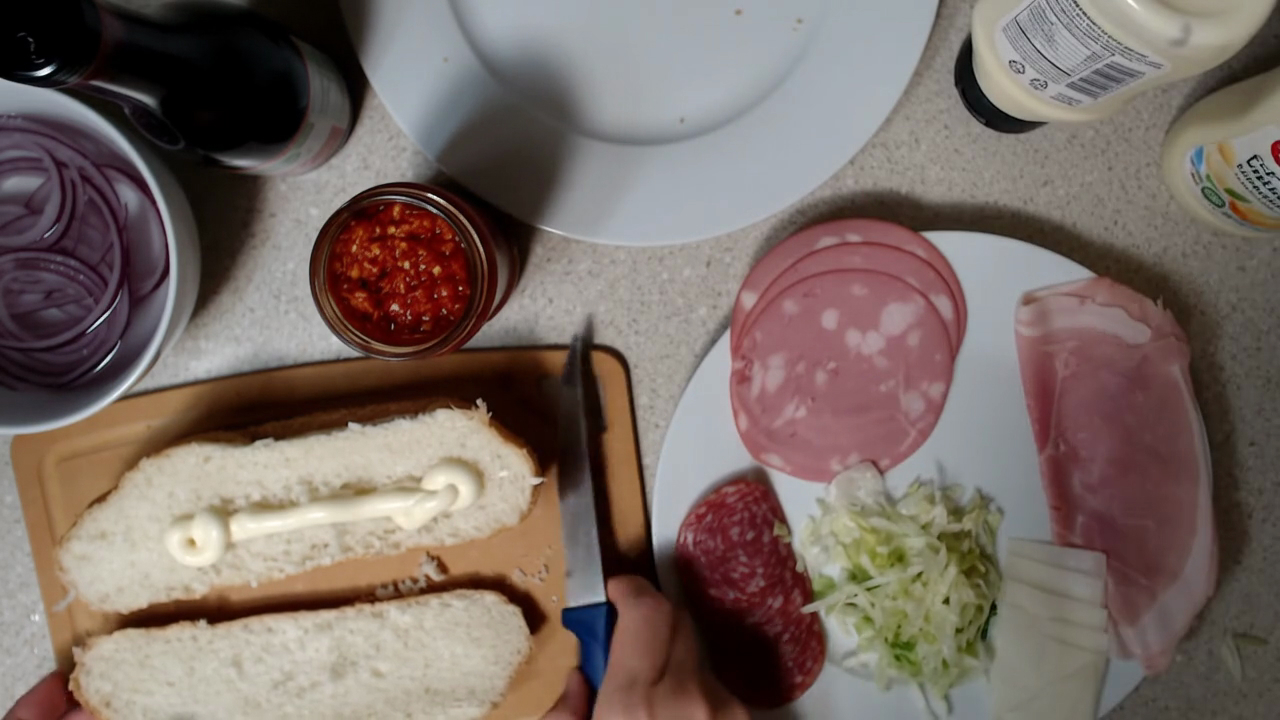}{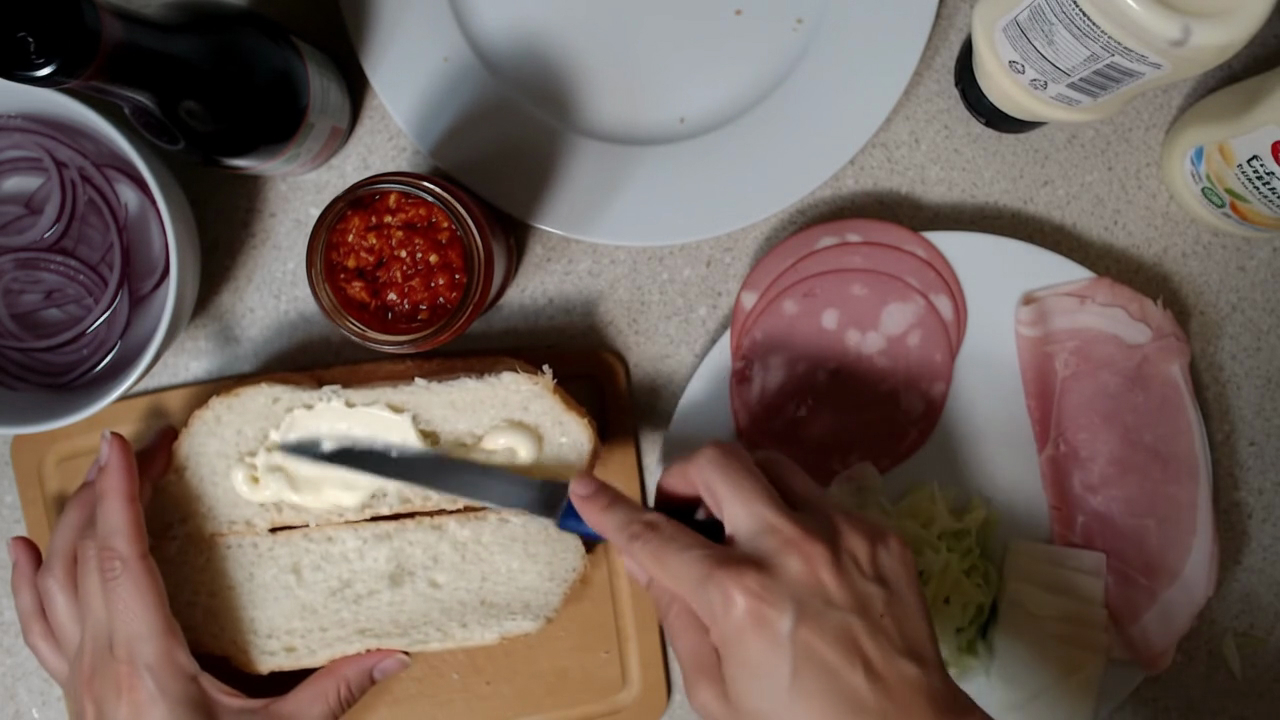}{0.00s/source}{2.51 s}{3.76 s} \tabularnewline
\addlinespace[4pt]
\raggedright \textbf{D8} Inappropriate modality\newline\textit{E1, E2} & \raggedright The medium is ill-suited to the instruction, such as a still that cannot convey essential dynamics or a video that adds unnecessary motion.\par\smallskip\textit{Example:} Tying shoelaces, step 4 (Gemini 3.1 Flash Image). A single final view cannot communicate the tuck-under and pull-tight sequence. & \raggedright \probePair{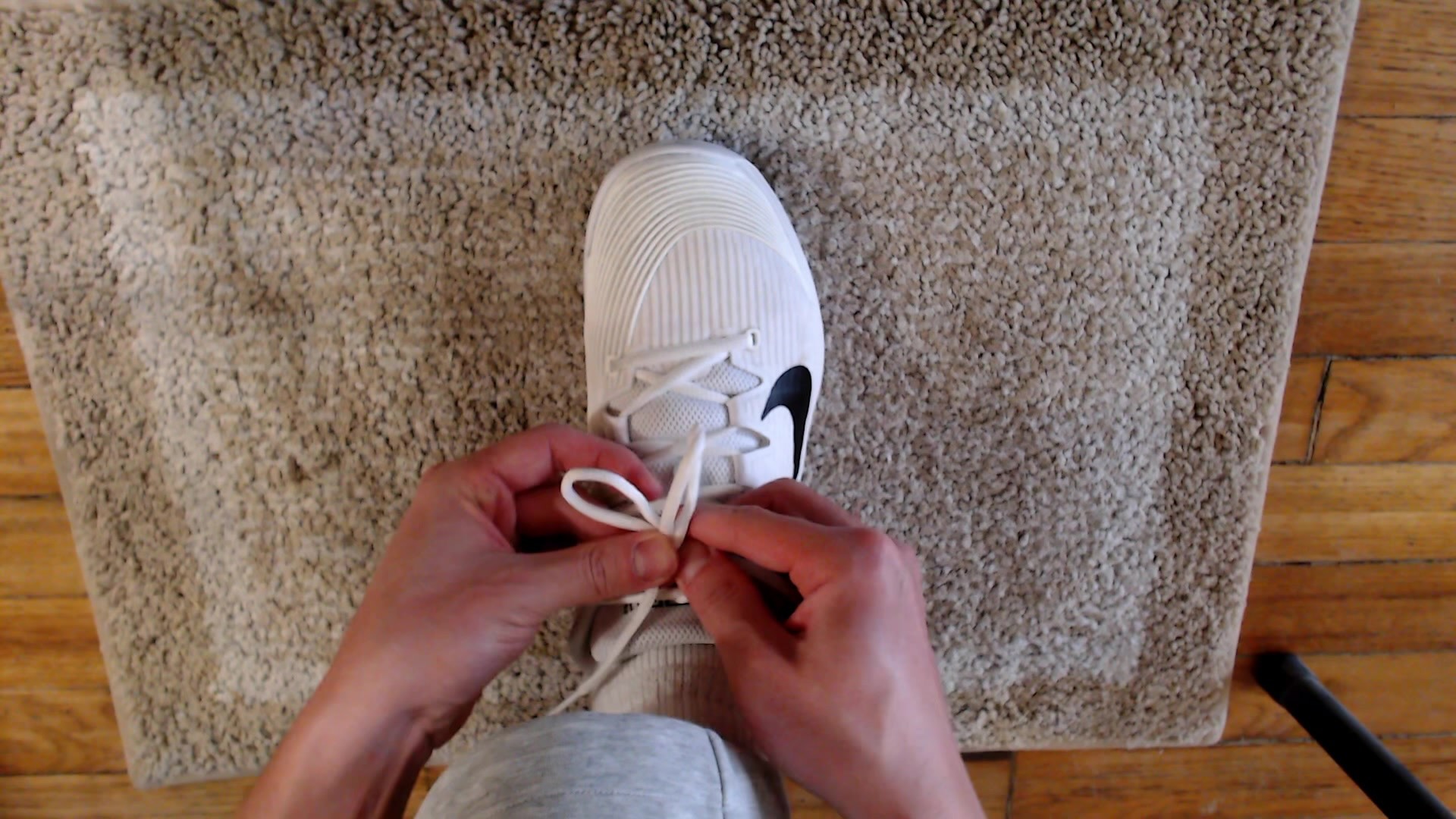}{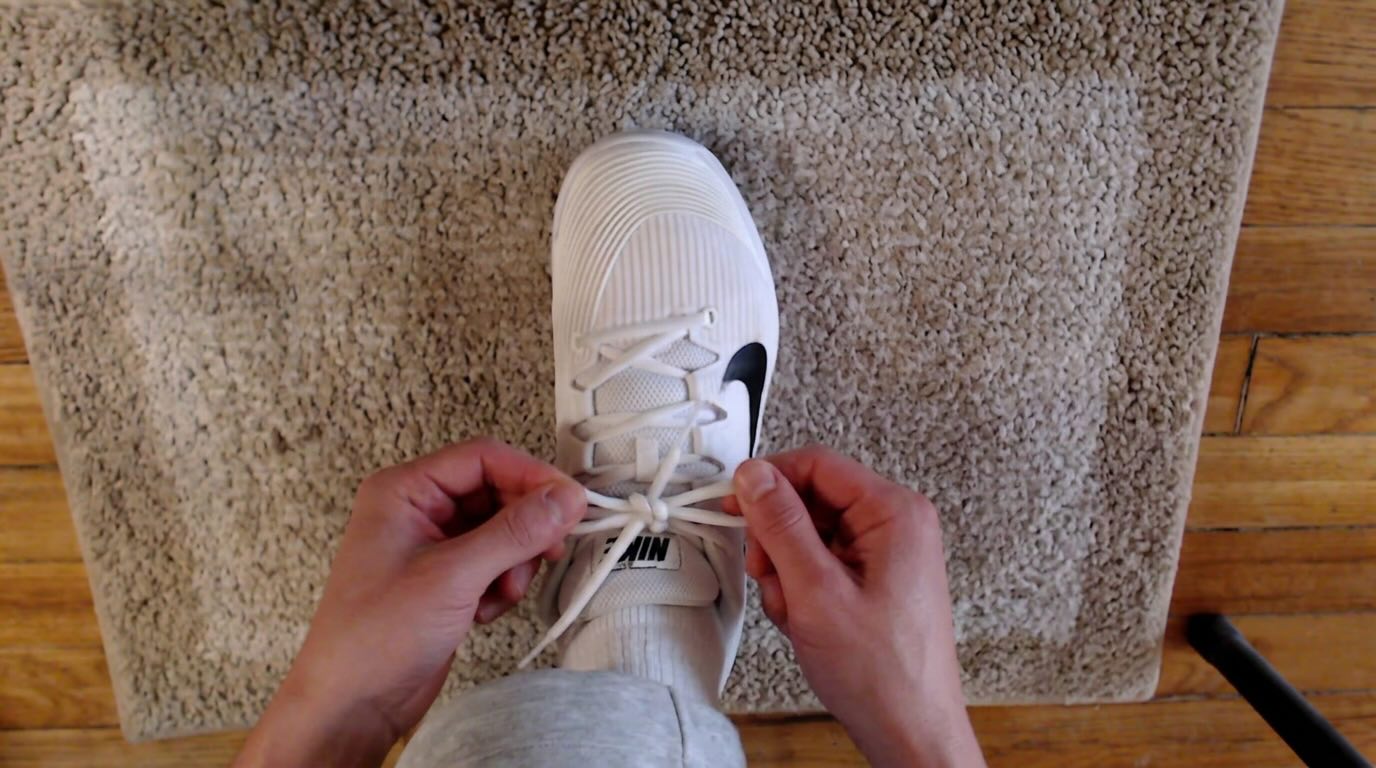}{Source}{Generated} \tabularnewline
\addlinespace[4pt]
\midrule
\multicolumn{3}{@{}l}{\textbf{E. Usefulness dimensions (analytic mappings)}} \\*
\raggedright \textbf{E1} Target-state legibility & \multicolumn{2}{p{\dimexpr.81\linewidth-\tabcolsep\relax}@{}}{The intended result, placement, configuration, grip, or posture is clear enough for learner comparison.} \tabularnewline
\addlinespace[4pt]
\raggedright \textbf{E2} Action-sequence legibility & \multicolumn{2}{p{\dimexpr.81\linewidth-\tabcolsep\relax}@{}}{The temporal order and mechanics of the instructed action are understandable.} \tabularnewline
\addlinespace[4pt]
\raggedright \textbf{E3} Source-context continuity & \multicolumn{2}{p{\dimexpr.81\linewidth-\tabcolsep\relax}@{}}{Relevant people, objects, environment, and spatial layout remain grounded in the source.} \tabularnewline
\addlinespace[4pt]
\raggedright \textbf{E4} Step-scope discipline & \multicolumn{2}{p{\dimexpr.81\linewidth-\tabcolsep\relax}@{}}{The artifact stays within the requested unit action.} \tabularnewline
\addlinespace[4pt]
\raggedright \textbf{E5} Detail-focused framing & \multicolumn{2}{p{\dimexpr.81\linewidth-\tabcolsep\relax}@{}}{Framing exposes a task-critical detail that materially helps understanding or assessment.} \tabularnewline
\addlinespace[4pt]
\raggedright \textbf{E6} Multi-state scaffolding & \multicolumn{2}{p{\dimexpr.81\linewidth-\tabcolsep\relax}@{}}{Multiple states or an explicit trajectory meaningfully clarify the transition to the target state.} \tabularnewline
\addlinespace[4pt]
\end{longtable}
\endgroup
\endgroup

\lstdefinestyle{promptblock}{
    basicstyle=\ttfamily\scriptsize,
    breaklines=true,
    breakatwhitespace=false,
    columns=fullflexible,
    keepspaces=true,
    showstringspaces=false,
    frame=single,
    framerule=0.3pt,
    framesep=1.5mm,
    xleftmargin=2mm,
    xrightmargin=2mm,
    aboveskip=6pt,
    belowskip=6pt,
    numbers=none,
    tabsize=2
}
\section{Model Prompts}
\label{sec:appendix-prompts}

This appendix reports the optional contextualization template, physical-readiness and same-step replacement-comparison prompts, media-generation templates, and narration input.
Runtime values are replaced with angle-bracketed placeholders (for example, \texttt{<<VALUE>>}).

\subsection{Contextualized Task-model Prompt}
\label{sec:appendix-contextualization-prompt}

The controlled study uses authored task models and does not invoke this planner.

\begin{lstlisting}[style=promptblock]
Create a task model for the user's goal using the supplied workspace image and context.

Goal: <<HIGH_LEVEL_GOAL>>
Workspace context: <<WORKSPACE_CONTEXT_JSON>>
Prior model (if any): <<PRIOR_MODEL_JSON>>

Represent the task as concrete, independently instructable actions suited to the available objects and workspace. Preserve unrelated arrangements. Reuse object and step IDs consistently. If a prior model is supplied, return only additional steps; keep existing steps and object IDs unchanged.

Return only {"steps": [...]} with these fields per step:
- step_id: unique, stable step identifier.
- step_number: one-based presentation order.
- instruction: detailed action instruction.
- short_instruction: participant-facing instruction, at most 280 characters.
- required_objects: array of objects needed for the action.
- completion_criteria: array of observable completion conditions.
- depends_on_step_ids: array of immediate prerequisite step IDs; [] for an independent action. Dependencies must be acyclic; presentation order alone does not imply a dependency.
- input_objects: array of object IDs used by the action.
- target_objects: array of object IDs affected by the action.
- output_states: array of {"object_id": "...", "description": "..."} entries describing the immediate resulting object states.
- parsed_verb: action verb.
- source: source location or object, or null if not applicable.
- destination: destination location or object, or null if not applicable.
- temporary: whether the action creates a temporary arrangement.

Do not return additional fields or commentary.
\end{lstlisting}

\subsection{Physical prerequisite-readiness prompt}
\label{sec:appendix-state-query-prompt}

The graph-scheduled policy instantiates this prompt for a selected physical frame and action.

\begin{lstlisting}[style=promptblock]
Decide whether the supplied physical context is ready to generate visual guidance for one action.
Destination workspace role: <<WORKSPACE_ROLE>>.
Action context: <<ACTION_CONTEXT_JSON>>
Assess only PREREQUISITE READINESS for the generation_instruction: all task-relevant materials, tools, containers and prerequisite physical states needed for this action must have visible, usable evidence in the supplied input context. A generic workspace identifier does not waive the objects named by the instruction. The action context classifies objects into destination_prerequisites, action_materials, and action_outputs. DESTINATION PREREQUISITES are states produced by earlier actions and must be visible in the CURRENT DESTINATION FRAME. ACTION MATERIALS are physical objects/tools used by this action; they may be visible in the current frame or the STAGING MATERIALS REFERENCE. ACTION OUTPUTS describe the target placement/state created by this action and must be ignored when deciding prerequisite readiness. If an object appears in both action_materials and action_outputs, check only that the physical source object is available. Report absent earlier states in missing_destination_prerequisites and absent physical source items in missing_action_materials. Return ready_to_generate only when both lists are empty. Return not_ready when a required input is visibly absent. Return unobservable when the workspace identity or required input evidence cannot be judged. Independently assess GENERATION FRAME SUITABILITY. The CURRENT DESTINATION FRAME will be used as the source image for generation. The task-relevant work region and any visible prerequisite arrangement must be comfortably inside the image and large enough to edit. Set generation_frame_suitable=false if that region is cut off, touches an image edge, is confined mainly to the outermost 15%
\end{lstlisting}

\subsection{Same-step replacement-comparison prompt}
\label{sec:appendix-replacement-comparison}

This prompt compares a prior goal with its already-displayed replacement for the same action.

\begin{lstlisting}[style=promptblock]
Compare two generated guidance images for the SAME step in an AR physical-task tutorial. IMAGE 1 is the prior displayed baseline; IMAGE 2 is the newer candidate, which is already being shown. This verdict only chooses which image should remain.
STEP INSTRUCTION: <<STEP_INSTRUCTION>>
EXPECTED VISIBLE OUTPUT STATE: <<EXPECTED_OUTPUT_STATE>>
COMPARISON CONTRACT: <<COMPARISON_CONTRACT>>
Judge each image independently. Set baseline_acceptable=true only when IMAGE 1 visibly communicates a task-correct and usable output for this step. Set candidate_acceptable=true only when IMAGE 2 does. These two booleans are how the caller detects that neither generated image is good enough and must regenerate from validated physical context.
Set target_state_met=true when IMAGE 2 visibly communicates the requested output. Set source_consistent=true when IMAGE 2 is at least as useful and task-correct as IMAGE 1. Equal task usefulness is a tie in favor of IMAGE 2 because it reflects newer context. Ignore aesthetic realism, lighting, texture, and other cosmetic differences. Set hallucination_free=false only for a material invented, missing, duplicated, identity-swapped, or contradictory task object in IMAGE 2. A worse candidate must include a short concrete violation; an equal or better candidate must return an empty violations list.
Return JSON exactly as: {"baseline_acceptable": true/false, "candidate_acceptable": true/false, "target_state_met": true/false, "source_consistent": true/false, "hallucination_free": true/false, "violations": ["short concrete discrepancy"], "evidence": "one short comparison"}.
\end{lstlisting}

\subsection{Shared media-generation constraints}
\label{sec:appendix-shared-media-prompts}

The goal-image and demonstration-video prompts are assembled from the following reusable blocks.

\begin{lstlisting}[style=promptblock]
<<CAMERA_LOCK_IMAGE>>
CAMERA: keep the input photograph's exact camera viewpoint, framing, perspective and lens look--the output must overlay the photo pixel-for-pixel in all unchanged regions.

<<CAMERA_LOCK_VIDEO>>
CAMERA: locked-off tripod shot from the exact viewpoint of the input photograph. The camera must not move at all for the entire clip--no pan, no tilt, no zoom, no dolly, no handheld drift, no perspective change, no cuts, no reframing. The first frame is the input photograph; the last frame keeps the identical framing and field of view.

<<SCENE_IDENTITY>>
OBJECTS: use ONLY the real objects visible in the input photograph. Every plate, bowl, glass, utensil, napkin, tool and prop keeps its exact real appearance--same color, material, pattern, size, wear and count. Reposition or manipulate the real items; never redraw them as different items, never restyle them, never swap them for similar-looking ones, and never add any object that is not already in the photo. If the instruction mentions something not visible in the photo, leave it out. Each real item exists exactly ONCE: an item that was moved appears only at its new location, never also at its old one--no copies, no duplicates.

<<MOVE_SEMANTICS>>
MOVING ITEMS: when an action moves, pours or uses an item, its ORIGINAL spot counts as a changed region too--paint the empty surface (or the emptied container) where it used to be. Never leave the item, or a copy of it, at the old position. Do not invent staging props (extra bowls, trays, containers) that are not in the photo.

<<INGREDIENT_FIDELITY>>
INGREDIENTS: never swap one ingredient or liquid for another. Depict it from whatever actually holds it in the photo, never from a container the photo does not show. Ingredient containers and packaging must be the EXACT ones visible in the reference photo--same brand artwork, label text, colors, shape, size, material, and plausible orientation. When a real item is an unusual variant, depict that variant exactly as photographed rather than normalizing it to a generic form.

<<NO_OVERLAYS>>
No on-screen text, captions, subtitles, labels, arrows, diagrams, watermarks or UI overlays of any kind.

<<BRISK_PACE>>
PACE: the hands move briskly and continuously--no idle pauses, no slow reaches, no lingering--so the WHOLE action completes well within the short clip.
\end{lstlisting}

\subsection{Goal-image prompt}
\label{sec:appendix-image-prompt}

For the study configuration, the image model receives the prompt below with one selected workspace frame and any supplementary grounded frames.
The template's repeated count checks reflect implementation hardening from the proactive demo: ``minimal edit'' means that all parts of the selected unit action must be completed while unrelated regions remain unchanged, not that the model may perform only part of the action.

\begin{lstlisting}[style=promptblock]
MINIMALLY EDIT this photograph to show the exact same scene after the following action(s) have been completed: <<UNIT_ACTION>>.

ALL of it: a multi-part action is complete only when EVERY part has been carried out--each item an action moves now sits at its NEW place and its old spot shows empty surface. THE COMMON ERROR TO AVOID: doing the first named move and leaving the other named items where they were.

HARD RULE--OBJECT COUNT: match the input photo(s) object for object. Every tool, vessel, container, or other item appears in the output the SAME number of times it appears in the input: never an extra copy, and never fewer. If the input shows two or three similar items, ALL remain, each at its own spot unless this action moves it. Never merge, remove, or tidy away items because they look alike, and never add an object the photo(s) do not show.

<<OPTIONAL_KNOWN_ERRORS_TO_AVOID>>

THE SINGLE MOST COMMON ERROR TO AVOID: a tool the action used drawn TWICE--once where the action ended AND once still at its old spot. Every used tool appears exactly once, set down at one plausible resting spot, its old location showing clean empty surface.

No hands appear in the output--the action is already finished. A vessel's existing contents also stay unless this step explicitly uses them; never empty, swap, or repurpose an unrelated filled vessel for the action.

HARD RULE--SAME EXACT ITEMS: every object in the output is the SAME physical object as in the input photo(s)--identical brand packaging, label artwork and text, colors, shape, size, material, and wear. Do not redraw an ingredient's package as a cleaner, prettier, or generic version.

Change ONLY the regions those actions touch--which INCLUDES the spots moved items came from. Every other region--the surface itself, background, walls, lighting, shadows, and all unrelated objects--must remain pixel-identical to the photo.

<<MOVE_SEMANTICS>>
<<INGREDIENT_FIDELITY>>
<<OPTIONAL_MATERIALS_REFERENCE>>
<<OPTIONAL_PRESENTATION_CONVENTION>>
<<CAMERA_LOCK_IMAGE>>
<<SCENE_IDENTITY>>
<<NO_OVERLAYS>>

You must always return an image, never a text reply.

THE COMPLETED STATE OF THIS STEP looks exactly like this--nothing more, nothing less: <<OUTPUT_STATE_COMPLETION_CRITERION>>. If an action mentions objects that are not visible in the supplied frame(s), keep those objects absent and depict only the observable after-state supported by the reference scene.

FINAL CHECK before you output: count every tool, vessel, container, and other item against the input photo(s). Counts must match exactly in both directions: nothing extra, nothing missing, and no object redesigned. Also re-check every known error listed above; if any is present, fix it before answering.
\end{lstlisting}

\subsection{Demonstration-video prompt}
\label{sec:appendix-video-prompt}

The preferred video path supplies two images: the selected real or speculative context as the initial state and the generated goal image as the terminal reference.
This couples the motion demonstration to the same immediate after-state.

\begin{lstlisting}[style=promptblock]
Two images of the SAME scene from the SAME viewpoint are provided: the first is the current state, the second is the exact END STATE of this step. Animate the first image into a short, realistic first-person demonstration of: <<UNIT_ACTION>>.

A pair of realistic adult hands enters the frame and physically PERFORMS the step's action with the real objects. Show the action itself as continuous visible motion, including in-place actions such as whisking, stirring, shaking, sifting, pouring, or wiping, with the objects' contents visibly transforming while it happens. The scene ends EXACTLY matching the second image; then the hands exit and the clip holds on the completed state for a moment.

HANDS: exactly ONE pair of hands belongs to ONE person--one left hand and one right hand, never a third hand, never a duplicated or detached hand, and never two hands on the same side. Both hands leave the frame at the end.

<<OPTIONAL_KNOWN_ERRORS_TO_AVOID>>
<<OPTIONAL_TECHNIQUE_REQUIREMENT>>

The action itself must fill most of the clip--never jump or dissolve straight from the first image to the second. The second image defines only the final arrangement; never show it as a separate shot. No cuts, flashes, other rooms, or other surfaces.

OBJECTS: use only the objects visible in the two images--never add, remove, restyle, or duplicate anything. An item that moves exists exactly once, leaving its old spot empty. Every object keeps the EXACT appearance it has in the two images--same brand packaging, label artwork, colors, shape, and wear; never substitute a cleaner, prettier, or generic look-alike.

AT EVERY MOMENT, the clip contains exactly the objects of the first image and nothing else: never more copies of any item than the images show and never fewer. No object may materialize in the background or foreground. When this step legitimately carries an item into the scene, that item must visibly enter in one of the hands from the frame edge, be set down once, and remain there.

An object that appears in both images is ONE physical object in transit: never render it at its first-image and second-image positions at the same time. For countable items, the clip never shows more instances than the larger of the two images contains. Everything the hands do not touch stays pixel-identical to the first image.

<<INGREDIENT_FIDELITY>>
<<BRISK_PACE>>
<<OPTIONAL_PRESENTATION_CONVENTION>>
<<CAMERA_LOCK_VIDEO>>
<<NO_OVERLAYS>>

Also re-check every known error listed above; if any is present, fix it before answering.
\end{lstlisting}

\section{User Study}
\label{app:user-study}

\subsection{Tasks}
\label{app:user-study-tasks}

Table~\ref{tab:user-study-tasks} summarizes the materials and prescribed steps for the four study tasks. Both conditions follow the same steps. Each task begins with a material check (Step 0), followed by six action steps for latte, mochi, and bouquet, and four for table setting; the Rubik's Cube practice task is excluded. Procedures are condensed from the participant-facing instructions.

\begingroup
\footnotesize
\setlength{\tabcolsep}{5pt}
\renewcommand{\arraystretch}{1.05}
\begin{longtable}{@{}p{\dimexpr.18\textwidth-\tabcolsep\relax}p{\dimexpr.30\textwidth-2\tabcolsep\relax}p{\dimexpr.52\textwidth-\tabcolsep\relax}@{}}
\caption{User-study tasks, materials, and step sequences. Step 0 is the material check; action steps begin at 1. Task names link to the YouTube sources used for pre-authored visual guidance.}
\label{tab:user-study-tasks} \\
\toprule
\textbf{Task} & \textbf{Materials and tools} & \textbf{Prescribed steps} \tabularnewline
\midrule
\endfirsthead
\caption[]{User-study tasks, materials, and step sequences (continued).} \\
\toprule
\textbf{Task} & \textbf{Materials and tools} & \textbf{Prescribed steps} \tabularnewline
\midrule
\endhead
\midrule
\multicolumn{3}{r}{\footnotesize\itshape Continued on next page} \\
\endfoot
\bottomrule
\endlastfoot
\raggedright\href{https://www.youtube.com/watch?v=ILXyVBkb79U}{\textbf{Strawberry matcha latte}}\par 7 steps
& \raggedright Matcha powder, water, milk, strawberries, ice, sieve, bowl, bamboo whisk, two spoons, and tall serving glass.
& \raggedright 0. Gather the ingredients and tools.\par
1. Pour about a teaspoon of matcha from the bag into the sieve and sift into the bowl.\par
2. Add a little water.\par
3. Whisk side to side until evenly foamed.\par
4. Reserve a garnish berry; remove the other berries' tops and mash them in the glass.\par
5. Add ice and milk to three-quarters full, keeping the layers separate.\par
6. Pour matcha over a spoon to form a green layer; garnish and dust with matcha. \tabularnewline
\addlinespace[8pt]

\raggedright\href{https://www.youtube.com/watch?v=K8Li3BjA4kY}{\textbf{Red-bean mochi}}\par 7 steps
& \raggedright Red-bean paste, cornstarch in its yellow box, one prepared cling-film square, fully cooked mochi dough in a bowl, scraper or knife, filling tray, and serving plate.
& \raggedright 0. Gather materials; wait for the experimenter to confirm the dough is safe to handle.\par
1. Roll one red-bean filling ball.\par
2. Wrap it in the prepared cling film.\par
3. Dust the dough and surface with cornstarch; stretch the dough into a cylinder.\par
4. Divide into four equal portions and roll four dough balls.\par
5. Flatten one dough ball into a wrapper.\par
6. Unwrap the filling and set the film aside; center the filling, enclose it, and pinch and fold the edge inward. \tabularnewline
\addlinespace[8pt]

\raggedright\href{https://www.youtube.com/watch?v=p9mzBckf3G4}{\textbf{Western-style table setting}}\par 5 steps
& \raggedright Twelve pieces: charger, dinner plate, bread plate, three forks (dinner, salad, dessert), two long knives, butter spreader, two spoons (soup, dessert), and drinking glass. Staging surface and dining table.
& \raggedright 0. Check all twelve pieces at the staging surface.\par
1. Place the charger at the table's near edge.\par
2. Place forks on the left, knives and soup spoon on the right; align handles and turn blades inward.\par
3. Place dessert spoon and fork beyond the charger, with handles right and left, respectively.\par
4. Add the dinner plate, bread plate with spreader, and glass, preserving earlier placements. \tabularnewline
\addlinespace[8pt]

\raggedright\href{https://www.youtube.com/watch?v=n58FtBmeJ_0}{\textbf{Spiral bouquet}}\par 7 steps
& \raggedright Four artificial stems: one dark foliage stem, two cream/white roses, and one pink focal stem; one elastic loop.
& \raggedright 0. Check the four stems and elastic.\par
1. Hold the foliage stem in the left hand.\par
2. Cross one cream/white rose over it to form an X.\par
3. Add the second rose on the opposite side at the same crossing angle.\par
4. Turn the held bundle a quarter turn.\par
5. Cross the pink stem with the bundle; do not turn again.\par
6. Hook, wrap, and anchor the elastic around all four stems. \tabularnewline
\end{longtable}
\endgroup

\noindent For mochi, the experimenter supplies fully cooked dough; dough preparation and heating are not participant steps. Participants prepare one filling ball and finish one mochi, although the dough is divided into four portions. Table-setting directions use the diner's viewpoint.

\providecommand{\surveyq}[2]{%
  \par\begingroup\emergencystretch=1em\noindent\hangindent=2.4em\hangafter=1%
  \makebox[2.4em][l]{\bfseries #1}#2\par\endgroup}
\providecommand{\surveyopts}[1]{\textit{[#1]}}
\providecommand{\surveyscaleSeven}{\surveyopts{1 = strongly disagree; 7 = strongly agree}}
\providecommand{\surveyscaleFive}{\surveyopts{1 = strongly disagree; 5 = strongly agree}}
\providecommand{\surveyfree}{\surveyopts{free response}}

\subsection{Questionnaires and Interview Guide}
\label{sec:appendix-study-questionnaires}

This appendix reproduces the background, post-task, and exit questions used to characterize participants and their experience. Demographic questions and administrative metadata fields, such as participant, trial, task, and condition identifiers, are not included. Question numbers are local to each instrument; the same SUS items are used for both conditions. We adopt Q5 from Stanford Sleepiness Scale~\cite{shahid2012stanford}.

\Needspace{4\baselineskip}
\subsubsection{Pre-Questionnaire}\mbox{}\par\nopagebreak
\label{app:pre-questionnaire}
\surveyq{Q1}{How often have you used a VR/AR headset before today? \surveyopts{never; a few times ever; a few times a month; weekly or more}}
\surveyq{Q2}{How often have you used AI tools to generate or edit images or videos before today? \surveyopts{never; a few times ever; a few times a month; weekly or more}}
\surveyq{Q3}{Have you ever used an AR app that guides you through a physical task (assembly, cooking, repair, \ldots)? If so, can you briefly describe it (e.g., what app/task)? \surveyopts{short free response}}
\surveyq{Q4}{Have you ever used AI tools to get guidance for a physical task (e.g., cooking, assembly, or repair)? If yes, briefly describe the tool, task, and type of guidance (text, images, or video). If not, write ``No''. \surveyopts{short free response}}
\surveyq{Q5}{Pick what best represents how you are feeling right now. \surveyopts{feeling active, vital, alert, or wide awake; functioning at high levels, but not fully awake; awake, but relaxed, responsive but not fully alert; somewhat foggy, let down; foggy, losing interest in remaining awake, slowed down; sleepy, woozy, fighting sleep, prefer to lie down; no longer fighting sleep, sleep onset soon, having dream-like thoughts}}

\Needspace{4\baselineskip}
\subsubsection{Post-Task Survey}\mbox{}\par\nopagebreak
\label{app:post-condition-survey}
Participants answer about the task they have just completed, after every task in either condition. NASA Task Load Index questions preserve the original scale and omit the pairwise comparisons suggested by Lee~\etal~\cite{lee2026nasa}.

\surveyq{Q1}{Mental Demand --- How mentally demanding was the task? \surveyopts{0 = very low; 100 = very high; increments of five}}
\surveyq{Q2}{Physical Demand --- How physically demanding was the task? \surveyopts{0 = very low; 100 = very high; increments of five}}
\surveyq{Q3}{Temporal Demand --- How much time pressure did you feel because of the task pace? \surveyopts{0 = very low; 100 = very high; increments of five}}
\surveyq{Q4}{Performance --- How successful were you in accomplishing what you were asked to do? \surveyopts{0 = perfect; 100 = failure; increments of five}}
\surveyq{Q5}{Effort --- How hard did you have to work, mentally and physically, to accomplish your level of performance? \surveyopts{0 = very low; 100 = very high; increments of five}}
\surveyq{Q6}{Frustration --- How insecure, discouraged, irritated, stressed, or annoyed did you feel during the task? \surveyopts{0 = very low; 100 = very high; increments of five}}
\surveyq{Q7}{The visual guidance matched the task instruction. \surveyscaleSeven}
\surveyq{Q8}{The visual guidance matched my physical workspace. \surveyscaleSeven}
\surveyq{Q9}{The visual guidance was easy to understand. \surveyscaleSeven}
\surveyq{Q10}{The visual guidance made me confident that I was performing the steps correctly. \surveyscaleSeven}
\surveyq{Q11}{I relied on the visual guidance when deciding what to do next. \surveyscaleSeven}
\surveyq{Q12}{I could view the visual guidance and the relevant workspace at the same time without interference. \surveyscaleSeven}
\surveyq{Q13}{The visual guidance distracted me from the task. \surveyscaleSeven}
\surveyq{Q14}{The visual guidance was available when I needed it. \surveyscaleSeven}
\surveyq{Q15}{I feel that I could perform this task again without guidance. \surveyscaleSeven}
\surveyq{Q16}{I would want this kind of visual guidance for similar everyday tasks. \surveyscaleSeven}
\surveyq{Q17}{How helpful was each element? \surveyopts{rows: the text instruction; the voice narration; the goal images; the demonstration video. For each: 1 = not at all helpful to 5 = extremely helpful; not present / didn't use}}
\surveyq{Q18}{Did you notice guidance that seemed incorrect, impossible, outdated, or mismatched with the workspace? \surveyopts{yes; no; not sure}}
\surveyq{Q19}{If yes or not sure, describe what seemed wrong and at which step. \surveyopts{optional free response}}

\Needspace{4\baselineskip}
\subsubsection{Condition-Block System Usability Scale}\mbox{}\par\nopagebreak
Participants answer the following ten items at the end of each two-task condition block (trials 2 and 4).
\surveyq{Q1}{I think that I would like to use this system frequently. \surveyscaleFive}
\surveyq{Q2}{I found the system unnecessarily complex. \surveyscaleFive}
\surveyq{Q3}{I thought the system was easy to use. \surveyscaleFive}
\surveyq{Q4}{I think that I would need the support of a technical person to be able to use this system. \surveyscaleFive}
\surveyq{Q5}{I found the various functions in this system were well integrated. \surveyscaleFive}
\surveyq{Q6}{I thought there was too much inconsistency in this system. \surveyscaleFive}
\surveyq{Q7}{I would imagine that most people would learn to use this system very quickly. \surveyscaleFive}
\surveyq{Q8}{I found the system very cumbersome to use. \surveyscaleFive}
\surveyq{Q9}{I felt very confident using the system. \surveyscaleFive}
\surveyq{Q10}{I needed to learn a lot of things before I could get going with this system. \surveyscaleFive}

\Needspace{4\baselineskip}
\subsubsection{Exit Survey}\mbox{}\par\nopagebreak
\label{app:exit-survey}
\surveyq{Q1}{Overall, which style of guidance did you prefer? \surveyopts{pre-made instructions; generated guidance (on your workspace)}}
\surveyq{Q2}{Why? Please explain in a few sentences. \surveyopts{free response; verbal answer permitted}}
\surveyq{Q3}{Compared with pre-made guidance, generated guidance helped me work faster. \surveyscaleSeven}
\surveyq{Q4}{Compared with pre-made guidance, generated guidance helped me make fewer mistakes. \surveyscaleSeven}
\surveyq{Q5}{I trusted generated guidance more than pre-made guidance. \surveyscaleSeven}
\surveyq{Q6}{Compared with pre-made guidance, generated guidance required less effort to follow. \surveyscaleSeven}

\Needspace{4\baselineskip}
\subsubsection{Interview}\mbox{}\par\nopagebreak
\label{app:interview-guide}

\surveyq{Q1}{What did you like and dislike about each guidance style? Please comment on both pre-made and generated guidance.}
\surveyq{Q2}{Describe a specific moment when you followed, double-checked, or ignored the generated guidance. What happened, and what influenced your decision?}
\surveyq{Q3}{In the generated guidance system, how would you compare (1) text instruction, (2) audio narration, (3) generated goal image, and (4) generated video demonstration in terms of helping task performance? Any reasons?}
\surveyq{Q4}{What changes would you make to the generated guidance system and why?}
\surveyq{Q5}{For what kinds of tasks or situations would generated guidance be most useful? When would you prefer pre-made guidance instead?}
\surveyq{Q6}{Anything else you want to tell us?}

\subsection{Task-Performance Grading Rubrics}
\label{app:study-rubrics}

The task-specific checklists below assess observable execution and outcomes of the prescribed steps. The same criteria apply to both conditions. The checklists contain 43 criteria across 22 steps; the initial material-check step (Step 0) in each task is not scored. Item identifiers combine the task (A--D), step number, and criterion number within that step, corresponding to the step sequences in Table~\ref{tab:user-study-tasks}.

Each criterion is scored \textbf{1} when visibly met and \textbf{0} when visibly not met. Obscured, missing, ambiguous, or incomplete evidence is left unrated rather than treated as failure. Outcome criteria refer to the final observable state within the step; technique and ordering criteria require review of the action sequence. A thumbs-up serves as a timing cue, not evidence of success. Log-derived boundaries must be corrected if they omit the relevant action. The item-specific guidance below specifies acceptable variation and limits on what can be judged from the recordings.

\providecommand{\studyrubricitem}[3]{%
  \par\addvspace{4pt}\begingroup\emergencystretch=1em
  \noindent\hangindent=3.5em\hangafter=1
  \makebox[3.5em][l]{\bfseries #1}\textbf{#2}\par\nopagebreak
  \noindent\hangindent=3.5em\hangafter=0 #3\par\endgroup}
\begingroup\small

\Needspace{4\baselineskip}
\subsubsection{A. Strawberry Matcha Latte}\mbox{}\par\nopagebreak

\studyrubricitem{A1.1}{Was matcha poured directly from the bag into the sieve over the bowl, without a spoon?}{1: powder travels from bag through the sieve into the bowl. 0: a spoon is used to transfer powder, or the sieve is bypassed. Do not estimate a teaspoon from the camera view.}

\studyrubricitem{A2.1}{Was water added to wet the sifted matcha in the bowl?}{1: the pour is visible and a shallow pool wets the powder. 0: no water is added in a fully observed step, or water goes into the wrong vessel. Do not infer temperature or exact volume.}

\studyrubricitem{A3.1}{Was the bamboo whisk used in short, rapid side-to-side strokes with its tines in the liquid?}{1: the main whisking passage shows the specified motion. 0: only slow circular stirring or repeated lifting out of the liquid is visible. Brief setup strokes do not determine the score; review the motion, not one still.}

\studyrubricitem{A3.2}{Did a visible foam layer cover the matcha surface by the end of whisking?}{1: foam visibly covers most of the exposed surface. 0: the visible surface remains mostly unfoamed. Leave blank when the bowl interior is hidden; do not grade microscopic bubble size from this recording.}

\studyrubricitem{A4.1}{Was one whole strawberry with its leafy top reserved outside the glass?}{1: a capped whole strawberry is visibly kept aside for later. 0: all berries go into the glass, or the reserved berry has its cap removed. Do not require a particular table location.}

\studyrubricitem{A4.2}{Were the leafy tops removed from the strawberries placed in the glass?}{1: the berries entering the glass are trimmed and no leafy cap remains in the base. 0: a visibly capped berry or detached cap is left in the glass. If insertion is obscured and contents cannot be checked, leave blank.}

\studyrubricitem{A4.3}{Were the strawberries mashed into a pulp layer at the bottom of the glass?}{1: the spoon visibly crushes the berries into a spread-out base. Some small chunks are acceptable. 0: mostly intact berries or isolated whole pieces remain. Do not require a perfectly smooth puree.}

\studyrubricitem{A5.1}{Was ice added above the strawberry base before the milk?}{1: visible order is strawberry base, then ice, then milk. 0: milk is added before ice, or ice is omitted in a fully observed step. An end frame alone cannot establish order.}

\studyrubricitem{A5.2}{Was the glass filled to roughly three-quarters with milk and ice, leaving room for matcha?}{1: level is visibly near three-quarters and substantial top space remains. 0: glass is near full, overflowing, or clearly far below the target. Borderline perspective-distorted levels remain blank; no ruler-based cutoff is implied.}

\studyrubricitem{A5.3}{Did the red strawberry base remain distinguishable below the white milk?}{1: a recognizable red base and white upper region remain. A thin pink transition is acceptable. 0: the contents are predominantly mixed into one pink region. Check the glass wall rather than the top surface.}

\studyrubricitem{A6.1}{Was the matcha poured over a spoon held just above the milk, without stirring the drink?}{1: the spoon visibly receives the pour above the milk and no subsequent stirring occurs in this step. 0: the spoon is bypassed or the drink is stirred. Leave blank if the pour path is obscured.}

\studyrubricitem{A6.2}{Are red, white, and green regions visibly distinguishable in the finished drink, from bottom to top?}{1: all three regions are identifiable; narrow blended boundaries are acceptable. 0: a region is absent or the drink is mostly blended. Do not require perfectly straight interfaces or equal layer heights.}

\studyrubricitem{A6.3}{Is the reserved whole strawberry placed on the rim of the glass?}{1: strawberry visibly rests on the rim. 0: it is absent, inside the drink, or left beside the glass at the end. Fine matcha dusting is not separately scored because these views do not reliably resolve it.}

\Needspace{4\baselineskip}
\subsubsection{B. Red-Bean Mochi}\mbox{}\par\nopagebreak

\studyrubricitem{B1.1}{Was one cohesive, roughly round red-bean filling ball formed?}{1: one compact rounded portion, even if its surface is slightly irregular. 0: multiple portions, loose paste, or a clearly flattened/elongated mass. No target diameter or mass is specified; do not invent one.}

\studyrubricitem{B2.1}{Was the filling ball enclosed in the prepared cling-film square?}{1: film surrounds the ball with no visibly bare filling. Loose film tails are acceptable. 0: the ball is left bare or part of it remains visibly uncovered. Transparent film needs a clear view; hand occlusion is not a failure.}

\studyrubricitem{B3.1}{Were both the dough and the working surface dusted with cornstarch?}{1: powder application or residue is visible on both. 0: one or both are clearly undusted throughout a fully observed step. Do not grade brand label, powder thickness, or neatness.}

\studyrubricitem{B3.2}{Was the dough formed into one elongated cylinder?}{1: one connected roll is visibly longer than it is wide. Slight tapering and roughness are acceptable. 0: dough remains a compact lump, flat sheet, or disconnected pieces. No exact length is specified.}

\studyrubricitem{B4.1}{Was all the dough divided into exactly four portions?}{1: four portions and no substantial unportioned cylinder remain. Ignore tiny scraps. 0: fewer/more portions or a substantial remainder. Check all four before a later step flattens one.}

\studyrubricitem{B4.2}{Are the four portions approximately equal in size?}{1: no portion is conspicuously larger or smaller than the others in a comparable view. 0: an obvious size imbalance remains. Small differences are acceptable; no numerical size tolerance is imposed.}

\studyrubricitem{B4.3}{Were all four portions rolled into cohesive, roughly round balls?}{1: all four are rounded, with minor creases acceptable. 0: at least one remains an obvious slab, cut chunk, or elongated piece. Do not demand smooth commercial mochi surfaces.}

\studyrubricitem{B5.1}{Was one dough ball flattened into a roughly round wrapper disc?}{1: one piece is visibly flat and broadly round. 0: it remains ball-shaped, is strongly elongated, or several pieces are combined. Minor edge irregularities are acceptable.}

\studyrubricitem{B5.2}{Is the wrapper continuous, without an open hole or a tear dividing it?}{1: a continuous sheet is visible; uneven thickness is acceptable. 0: an open hole or major tear remains. Do not infer hidden underside tears or score thickness without a clear view.}

\studyrubricitem{B6.1}{Was the filling unwrapped and the empty film kept away from the dough?}{1: bare filling enters the wrapper and the film is separate from it. 0: visible film is enclosed with the filling or left against the dough. Do not require a pixel-exact top-edge location.}

\studyrubricitem{B6.2}{Were the wrapper edges brought around the filling, pinched and folded together at the center?}{1: edges meet and enclose the filling; a gathered seam is acceptable. 0: a substantial gap exposes filling or the edges remain open. Surface smoothness and seam-down presentation are not required by this step.}

\Needspace{4\baselineskip}
\subsubsection{C. Western-Style Dinner Table}\mbox{}\par\nopagebreak

\studyrubricitem{C1.1}{Is the charger flat near the diner's table edge and centered on the intended place setting?}{1: a flat charger defines one usable setting in front of the diner. 0: it is upright, stacked on another plate, or clearly outside that setting. Use the diner's position, not the camera's center; no centimeter spacing requirement.}

\studyrubricitem{C2.1}{Are two forks on the diner's left and two long knives followed by the soup spoon on the right?}{1: from charger outward, two forks left; two long knives then soup spoon right. 0: wrong count, side, or utensil order. Visually identical forks/knives are assigned inner/outer roles by position, not an assumed size hierarchy.}

\studyrubricitem{C2.2}{Do both long knife cutting edges face the charger?}{1: both inward edges are visibly identifiable. 0: at least one blade visibly faces outward. If reflections or resolution prevent identifying an edge, leave blank.}

\studyrubricitem{C2.3}{Are the five side utensils roughly parallel with handles toward the diner?}{1: all five handles point toward the diner and the utensils are broadly parallel. 0: a reversed or transverse utensil remains. Small spacing or handle-height differences are acceptable.}

\studyrubricitem{C3.1}{Is the dessert spoon horizontal beyond the charger with its handle to the diner's right?}{1: both location and handle direction match. 0: beside the side utensils, vertical, or handle reversed. Interpret left/right from the diner's seat.}

\studyrubricitem{C3.2}{Is the dessert fork between the spoon and charger, horizontal with its handle to the diner's left?}{1: both relative location and orientation match. 0: fork is above the spoon, beside the charger, or reversed. Judge the supplied third fork; do not require it to look smaller.}

\studyrubricitem{C4.1}{Is the patterned dinner plate centered on the black charger?}{1: dinner plate sits within the charger with a reasonably balanced visible rim. 0: wrong plate, separate placement, or obvious off-center overhang. Minor centering differences are acceptable.}

\studyrubricitem{C4.2}{Is the small bread plate beyond the left forks, with the metal spreader diagonally on it, handle lower-right?}{1: plate location and spreader placement/direction all match from the diner's viewpoint. 0: bread plate is on the dinner plate, on the wrong side, or spreader is missing/wrongly oriented. This is one composite bread-setting criterion.}

\studyrubricitem{C4.3}{Is the regular drinking glass upright beyond the right-hand knives?}{1: the stemless drinking glass is upright in that region. 0: wrong side, tipped glass, or a substituted stemmed glass. Small spacing differences are acceptable.}

\Needspace{4\baselineskip}
\subsubsection{D. Spiral Bouquet}\mbox{}\par\nopagebreak

\studyrubricitem{D1.1}{Is the single foliage stem held in the left hand, with the other three stems still separate?}{1: foliage alone is held and no flower stem has been added. 0: wrong stem or additional stems are picked up into the bundle. Identify physical base stems, not every branch or bloom.}

\studyrubricitem{D2.1}{Was exactly one cream/white rose stem added to the foliage?}{1: bundle contains foliage and one rose. 0: wrong flower or both roses are added at this step. Count base shafts or track the pickup; do not count flower heads.}

\studyrubricitem{D2.2}{Does the rose cross the foliage diagonally to form an X at the holding point?}{1: a visible diagonal crossing, not simply parallel shafts. 0: the stems are visibly parallel or joined at separate points. Leave blank if fingers hide the crossing throughout.}

\studyrubricitem{D3.1}{Was the second cream/white rose added on the open side opposite the first?}{1: the two rose stems occupy opposing sides of the arrangement. 0: the second remains loose or is placed alongside the first on the same side. Use the held bouquet's orientation rather than screen-left/right.}

\studyrubricitem{D3.2}{Does the second rose cross at the existing holding point while maintaining the established diagonal arrangement?}{1: all three stems converge at the same grip with a diagonal arrangement. 0: a separate crossing point or clearly parallel insertion breaks the arrangement. Do not demand an exact angle from an oblique view.}

\studyrubricitem{D4.1}{Was the three-stem bundle rolled roughly a quarter turn as one unit?}{1: a visible turn about the stems' long axis approximates a quarter turn and the stems move together. 0: no turn, an obvious half/full turn, or independent rearrangement instead of rolling. Small angular differences are acceptable. Moving the arm or tilting the bouquet is not itself axial rotation; leave blank if the motion cannot be tracked.}

\studyrubricitem{D5.1}{Was the single pink focal stem added, bringing the bouquet to all four stems?}{1: foliage, both rose stems and the pink stem are together, with none left loose. 0: pink is absent or a previous stem has been dropped/left out. Count base stems rather than multiple blooms on one stem.}

\studyrubricitem{D5.2}{Does the pink stem cross the bundle diagonally through the existing binding point?}{1: it joins the common grip as a crossing shaft. 0: it is parallel beside the bundle or crosses at a separate point. Hidden shafts remain unrated; bloom placement alone does not establish a crossing.}

\studyrubricitem{D6.1}{Does the elastic wrap around all four stems at the binding point?}{1: the same band visibly surrounds the full bundle there. 0: only some stems are encircled, band is elsewhere, or no band is applied in a fully observed step. Do not infer success when the bouquet is off camera.}

\studyrubricitem{D6.2}{Is the elastic anchored around a stem so it remains in place after the wrapping hand lets go?}{1: an anchored loop stays on the bundle when the right hand releases it. The left hand may keep holding the bouquet. 0: band visibly slips off, remains unanchored, or releases the bundle. Do not require a pull test or infer tightness from a hidden final state.}

\endgroup

\end{document}